\pdfoutput=1
\RequirePackage{fix-cm}
\documentclass [12pt, fleqn, letter] {article}
\usepackage[fontsize=12pt]{scrextend}
\usepackage[english]{babel}

\usepackage{amssymb}
\usepackage{bbm}
\usepackage{caption}
\usepackage{booktabs}
\usepackage{comment}
\usepackage{subcaption}
\usepackage{dcolumn}
\usepackage{booktabs}
\usepackage{amsmath}
\usepackage[flushleft]{threeparttable}
\usepackage{array}
\usepackage{tfrupee}
\usepackage{longtable}
\usepackage{bbm}
\usepackage{graphicx}
\usepackage{comment}
\usepackage{amsthm} 
\usepackage{bm}
\usepackage{color} 
\usepackage{float}
\usepackage{graphicx,float} 
\usepackage{mathrsfs}
\usepackage{mathtools}
\usepackage[title]{appendix}
\usepackage{tikz}
\usetikzlibrary{snakes}
\usepackage[left=1in,right = 1in,top= 1in, bottom = 1in]{geometry}
\usepackage{psfrag,epsf}
\usepackage{enumerate}
\usepackage{graphicx}
\usepackage[authoryear,round]{natbib}
\usepackage{setspace}
\usepackage{mathtools}
\usepackage[hang,flushmargin]{footmisc}
\usepackage{csquotes}
\usepackage{pdflscape}
\usepackage{subcaption}
\usepackage[flushleft]{threeparttable}
\usepackage{array}
\usepackage{longtable}
\usepackage{multirow}
\usepackage{arydshln}
\usepackage{fixltx2e}
\usepackage[normalem]{ulem}
\usepackage{hyperref}
\usepackage{chngcntr}
\usepackage[title]{appendix}

\usepackage[T1]{fontenc}
\usepackage{pxfonts}  

\newcommand{\RNum}[1]{\uppercase\expandafter{\romannumeral #1\relax}} 

\hypersetup{
	colorlinks=true,
	citecolor=blue,
	linkcolor=blue,
	filecolor=blue,      
	urlcolor=blue,
}
\let\svfootnoterule\footnoterule
\renewcommand\footnoterule{\vfill\svfootnoterule}
\makeatletter
\renewcommand\footnotesize{%
	\@setfontsize\footnotesize\@ixpt{10}%
	\abovedisplayskip 8\p@ \@plus2\p@ \@minus4\p@
	\abovedisplayshortskip \z@ \@plus\p@
	\belowdisplayshortskip 4\p@ \@plus2\p@ \@minus2\p@
	\def\@listi{\leftmargin\leftmargini
		\topsep 4\p@ \@plus2\p@ \@minus2\p@
		\parsep 2\p@ \@plus\p@ \@minus\p@
		\itemsep \parsep}%
	\belowdisplayskip \abovedisplayskip
}
\makeatother

\newcolumntype{L}{>{\centering\arraybackslash}m{5cm}}
\newcolumntype{P}{>{\centering\arraybackslash}m{2cm}}
\newcolumntype{D}{>{\arraybackslash}m{5cm}}

\usepackage{collcell}
\makeatletter
\newcolumntype{G}{>{\collectcell\@gobble}c<{\endcollectcell}@{}}
\makeatother
\def\eatcell#1\unskip{}
\newcolumntype{E}{>{\eatcell}c@{}}
\newcolumntype{H}{>{\setbox0=\hbox\bgroup}c<{\egroup}@{}}
\providecommand{\sym}[1]{\ifmmode^{#1}\else\(^{#1}\)\fi}

\title{\textbf{\Large{Political Power-Sharing, Firm Entry, and Economic Growth:
Evidence from Multiple Elected Representatives}}\footnote{We thank Isha Agarwal, Sumit Agarwal, Sam Asher, Marianne Bertrand, Robin Burgess, Utpal Bhattacharya, Maxime Bonelli, Emanuele Colonnelli, Eliana La Ferrara, Stefano Fiorin, Raymond Fisman, Anand Goel, Todd Gormley, Allan Hsiao, Yan Ji, Elisabeth Kempf, Stefan M. Lewellen, Clara Mart\'inez-Toledano, Ron Masulis, Lyndon Moore, Abhiroop Mukherjee, Deniz Okat, Imran Rasul, Amit Seru, Amir Sufi, Margarita Tsoutsoura, Robert Vishny, Vikrant Vig, Christopher Woodruff, Guo Xu, Alminas Zaldokas and Luigi Zingales for helpful comments and suggestions. We are thankful to the seminar participants at the BREAD Conference on Development Economics, CEPR/IFS/UCL/BREAD/TCD Workshop in Development Economics, EFA, Finance Down Under, Hong Kong University of Science and Technology, Imperial College London, Inter-Finance PhD Seminar, NSE-NYU Conference on Indian Capital Markets, National University of Singapore, TADC, University of Chicago, WEFIDEV. We thank the NSE-NYU Stern Initiative on the Study of Indian Capital Markets for financial support. We take responsibility for all errors.}
} 

\author{\\ Harsha Dutta\footnote{Harsha Dutta is at the Imperial College London. email: \href{sdutta@imperial.ac.uk}{sdutta@imperial.ac.uk}}
\and \\
        Pulak Ghosh\footnote{Pulak Ghosh is at Indian Institute of Management Bangalore. email:  \href{pulak.ghosh@iimb.ac.in}{pulak.ghosh@iimb.ac.in}} \and \\
        Arkodipta Sarkar\footnote{Arkodipta Sarkar is at the National University of Singapore. email:  \href{asarkar@nus.edu.sg}{asarkar@nus.edu.sg}}
		 \and \\             
		Nishant Vats\footnote{Nishant Vats is at Olin Business School, Washington University - St. Louis. email: \href{vats@wustl.edu}{vats@wustl.edu}} \\
	}
\date{}

\begin{document}
\maketitle
\thispagestyle{empty}
\begin{abstract}
\singlespacing
\noindent

Unchecked political power impedes economic growth, and constraints on such power are inherently limited. We study whether increasing the number of politicians sharing authority over administration, a feature of political institutional design intended to constrain power, fosters economic growth. We implement discontinuity-based research designs, leveraging the haphazard overlap of electoral and administrative boundaries in India, which generates quasi-random variation in the number of politicians governing adjacent areas. We assemble administrative records to measure economic outcomes at high spatial resolution and find that areas governed by multiple politicians exhibit higher firm entry, higher employment, and greater economic activity. We find suggestive evidence that checks and balances operate. Economic benefits concentrate in areas governed by partisanly non-aligned politicians, who face stronger incentives to monitor one another. Politicians facing higher intensity of mutual oversight accumulate lower private returns from public office. We further find that firm entry improves along margins typically affected by unchecked political power. In particular, patronage and cronyism decline, regulatory hold‑up eases, and delivery of essential public infrastructure improves.
\\
\\
\noindent
\end{abstract}

\clearpage
\newpage

\section{Introduction}
\begin{singlespace}
\emph{``Ambition must be made to counteract ambition. \ldots{} It may be a reflection on human nature, that such devices should be necessary to control the abuses of government. \ldots{} If men were angels, no government would be necessary."}
\begin{flushright}
--- James Madison, \textit{Federalist No.\ 51} (1788)
\end{flushright}
\end{singlespace}

Political institutions such as constitutions are inherently incomplete contracts. The authority these institutions delegate leaves discretion in the hands of those in power, creating scope for abuse that generates substantial growth costs. A long-standing concern in the design of political institutions has therefore been the extent of unchecked power delegated to those in authority and the mechanisms through which that power is disciplined \citep{de1751spirit, kant1795perpetual, hayek1960constitution}. In this paper, we examine one widely used dimension of institutional design through which political power is constrained: the degree of horizontal power-sharing among elected representatives.\footnote{Institutional arrangements where multiple politicians oversee the same administrative apparatus are common but understudied. Examples include the two-senator representation of each state in the U.S. Senate and Switzerland's seven-member Federal Council.} More specifically, we study whether varying the number of politicians who hold effective authority over the same administrative apparatus affects economic outcomes.

The relationship between the number of politicians and economic growth is theoretically ambiguous. On the one hand, multiple politicians can negatively affect local economic conditions by increasing corruption due to the presence of multiple grabbing hands, higher holdup problem, and coordination failure or free-rider problems discussed in the common agency literature.\footnote{See \cite{shleifer1993corruption} for a discussion on multiple grabbing hand problem. See \cite{grossman1986costs}, \cite{hart1990property}, \cite{blanchard1997disorganization}, and \cite{olken2009simple}, among others for a discussion on the holdup problem. See \cite{bernheim1986common}, \cite{dixit1997power}, and \cite{dixit1997common}, among others for a discussion on common agency problem.} On the other hand, multiple politicians can improve local economic conditions by reducing power concentration and imposing checks and balances, leveraging diverse skills and division of labor, and reducing corruption.\footnote{See \cite{holmstrom1978incentives}, \cite{rasmusen1994cheap}, \cite{persson1997separation}, \cite{bardhan2002}, and \cite{aghion2004endogenous} among others for a discussion of these issues.} Despite the theoretical ambiguity, the empirical evidence is limited, primarily due to issues of endogeneity associated with the emergence of political institutions that foster power-sharing among politicians, as discussed in \cite{aghion2004endogenous}.

We make progress on this discussion by estimating the effect of being governed by multiple politicians on firm entry and local economic growth, using a large-scale natural experiment in India. We take advantage of the nationwide haphazard overlap of electoral and administrative boundaries, which generates quasi-random variation in whether local administrative units fall under the jurisdiction of one or multiple elected representatives. Our central finding is that being governed by multiple politicians increases firm entry and economic growth. We find this positive effect to be stronger when politicians have greater incentives to impose mutual checks and balances on each other. We further show that such constraints foster firm entry in precisely those domains where unchecked political authority typically impedes it, reducing preferential access for connected firms, limiting cronyistic arrangements, and reducing regulatory hold‑up in the issuance of permits.

The ideal research design for answering this question would require variation in the number of politicians exercising authority over otherwise similar administrative units, in a way that is orthogonal to local economic conditions. We exploit a unique setting in India that approximates such a scenario, arising from the haphazard overlap between electoral and administrative boundaries. \emph{Administrative Blocks} are the basic governance units through which local policy implementation occurs. \emph{Assembly Constituencies} are the electoral units through which representation in the state legislature is organized. Electoral boundaries are drawn by an independent, non-partisan Delimitation Commission with the objective of equalizing populations across all assembly constituencies within a state. When defining electoral boundaries, this commission does not take into account administrative boundaries at the block level. As a result, electoral boundaries may intersect with administrative blocks, leading some blocks to be divided across multiple constituencies (\emph{Split Blocks}), while others are entirely contained within a single assembly constituency (\emph{Unsplit Blocks}). This incongruence between electoral and administrative boundaries generates quasi-random variation in the number of politicians with authority over a given administrative block.

At the block level, the key administrative authority is the Block Development Officer (BDO), a bureaucrat who serves as the ``eye of the needle'' through which most public projects and administrative decisions must pass \citep{dasgupta2020political}. BDOs exercise statutory authority over a range of local approvals, including permits that are prerequisites for economic activity and public service provision. Their interpretive authority in applying these regulatory requirements results in variation in administrative outcomes across blocks. State-level politicians (Members of the Legislative Assembly, or MLAs) are elected to represent their assembly constituencies. They face electoral incentives to deliver for constituents and opportunities to extract private benefits from public office \citep{fisman2014private}. However, MLAs are legislators and lack direct executive authority over policy implementation, so achieving these objectives requires bureaucratic implementation. MLAs therefore work through BDOs, using their influence over transfers and career advancement to induce bureaucratic compliance. This politician–bureaucrat nexus determines how policy is implemented at the block level.

The MLA–BDO interactions occur through both informal channels and regular block-level meetings, which provide a structured setting where MLAs can observe BDO performance and, when multiple MLAs are present, one another’s behavior. When a single MLA’s constituency encompasses an entire block, that MLA is the sole political principal overseeing the BDO, plausibly making it easier for the two to coordinate on rent-sharing arrangements with limited scrutiny, external exposure, or electoral discipline. By contrast, when a block spans multiple constituencies, several MLAs oversee the same BDO, making their actions mutually observable to political rivals with electoral incentives to monitor and expose self-serving behavior. These differences in mutual oversight create the possibility of systematically different patterns of rent extraction and policy implementation across blocks.

This institutional design allows us to use discontinuity based designs to test whether governance by multiple elected representatives causally affects local economic outcomes. The boundaries separating split blocks from unsplit blocks generate a sharp discontinuity in the number of elected representatives exercising authority over local administration, while other fundamental determinants of economic activity are assumed to vary smoothly across the boundary. Our primary design is a cross-sectional geographic regression discontinuity that compares villages just inside split blocks to those just outside in unsplit blocks, within a narrow bandwidth around the boundary. We augment this with a second strategy that relies on the timing of the 2008 delimitation, which redrew constituency boundaries nationwide and altered the number of elected representatives exercising authority within blocks over time. This allows us to implement a differences‑in‑discontinuity design, comparing changes in outcomes over time for locations on either side of split–unsplit boundaries. The design allows us to how local economic activity responds when areas transition between single and multiple representative authority over local administration, while controlling for time‑invariant local characteristics.

We assemble rich administrative data on firm entry from the Ministry of Corporate Affairs (MCA), Government of India, which maintains the statutory registry for the near-universe of the formal private corporate sector. Firm entry is our primary outcome variable and is well-suited for our analysis for three reasons. First, new firm creation is a key determinant of productivity growth and local employment \citep{mcmillan2003central, woodruff2018addressing}, and also serves as an indicator of whether the local regulatory environment facilitates or impedes productive economic activity \citep{djankov2002regulation, djankov2009regulation}. Second, establishing a firm requires permits and approvals that fall under BDO discretion and MLA influence, making firm entry a direct reflection of how political oversight shapes the local business environment. Third, the MCA registry records precise registration addresses for each firm, which we geocode at high spatial resolution, enabling us to implement our discontinuity-based designs.

Our baseline analysis employs a geographic regression discontinuity (RD) design, comparing villages just inside split blocks to those just outside in unsplit blocks within a narrow bandwidth around the boundary. We find that villages in split blocks exhibit 3\% higher firm entry and 7\% higher nightlight intensity relative to villages in unsplit blocks. The effect increases monotonically with the number of representatives exercising authority over the block, consistent with a dose-response relationship. 

We augment the cross-sectional estimates with a differences-in-discontinuity design that leverages the timing of the 2008 delimitation. This design examines how firm entry responds when areas transition from a single representative having authority to multiple representatives having authority over local administration. More specifically, we compare blocks that switched from unsplit to split following the delimitation against blocks that always remained unsplit, and separately compare blocks that switched from split to unsplit against blocks that remained split. We find that the transition from being governed by a single to multiple politicians is associated with a 1.1–1.6\% increase in firm entry, while the reverse transition is associated with a 0.7–1.0\% reduction. This symmetry further strengthens our causal interpretation.

These effects extend to labor market outcomes. Villages in split blocks exhibit 5\% higher employment rates, consistent with the increased firm entry documented above. We further examine applications to the National Rural Employment Guarantee Scheme (NREGA), India's large-scale public workfare program, which provides a direct measure of demand for public employment assistance. We find that split blocks exhibit lower NREGA application rates. This pattern is directionally consistent with \cite{gulzar2017politicians}, who document lower NREGA disbursement in areas governed by multiple politicians and interpret this as evidence of poor program implementation. However, the observed lower disbursement is an equilibrium outcome. The data on NREGA applications at our disposal allows us to isolate demand for such assistance. The lower application rates are consistent with our findings of increased firm entry and higher private sector employment. This pattern suggests that workers in split blocks have better outside options in the private labor market, reducing their demand for public employment assistance.

The causal interpretation of the RD estimates rests on two identifying assumptions. First, the underlying determinants of economic activity are assumed to vary smoothly at block boundaries. We conduct balance tests using data from the 2001 Census to support this assumption. Villages located just within split blocks closely resemble those in adjacent unsplit blocks across multiple dimensions, including population, public infrastructure, financial access, and geography. Second, we assume boundary placement was free of selective sorting, i.e., that the delimitation process did not systematically assign more productive areas to split or unsplit blocks. \cite{iyer2013redrawing} documents that the 2008 delimitation was largely free from gerrymandering, supporting this assumption. Nevertheless, we restrict the sample to boundaries where this assumption is more plausibly satisfied: a) those following straight-line patterns, and b) those defined by natural features such as rivers and watersheds. We obtain similar estimates across these subsamples. Together, these tests support the assumptions underlying our causal estimates.

Next, we investigate the mechanism through which multiple politicians foster firm entry. We propose that multiple politicians holding effective authority over the same administrative apparatus impose mutual checks and balances that constrain rent extraction. The institutional structure facilitates such oversight: in split blocks, multiple MLAs participate jointly in block-level Panchayat Samiti meetings alongside the same BDO. This arrangement places politicians in a shared information environment where they can observe each other's interactions with bureaucrats. Politicians from rival parties face electoral incentives to monitor and expose self-serving behavior. For the BDO, the presence of multiple political principals with divergent interests makes exclusive rent-sharing arrangements difficult to sustain. We first establish that checks and balances operate by leveraging institutional features that induce variation in the intensity of mutual oversight. We then provide further evidence by examining domains where unchecked political discretion typically impedes firm entry.

We first leverage variation in the partisan composition of politicians governing split blocks. Politicians from different parties face stronger incentives to monitor one another, as they benefit electorally from exposing rivals' misconduct. Ideological differences and political animosity also create barriers to collusive arrangements. Checks and balances should therefore operate with greater intensity when split blocks are governed by non-aligned politicians. We test this by examining whether firm entry varies with partisan alignment among politicians sharing authority over the same block. We find that split blocks governed by non-aligned politicians experience significantly greater firm entry than those governed by aligned politicians. These patterns are suggestive of checks and balances in play.

We further examine whether politicians subject to greater oversight accumulate less private wealth. Politicians can extract rents through their discretionary authority over administrative processes, and such rent extraction often manifests in personal wealth accumulation. Electoral law in India requires candidates to file sworn asset affidavits, allowing us to measure changes in private wealth over a politician's term in office. Following \cite{fisman2014private}, we proxy for rent extraction through the winner's premium, the differential asset growth of election winners relative to runners-up. A politician's constituency contains multiple blocks. Haphazard boundary overlap implies that constituencies vary in the share of split blocks they contain, and hence in the degree of mutual oversight politicians face. We find that the winner's premium is lower in constituencies with a higher share of split blocks. Such constrained rent extraction in constituencies with greater mutual oversight is suggestive of checks and balances in operation.

Having established that checks and balances operate, we examine how they foster firm entry. We focus on three domains where unchecked political discretion typically impedes firm entry: preferential treatment that favors connected agents, regulatory hold-up that delays permits, and public goods provision that suffers from rent extraction. First, we examine preferential treatment. Politicians wielding unconstrained authority can facilitate firm entry based on personal connections, potentially favoring co-ethnic entrepreneurs. We identify such potential patronage connections through shared caste (jati) identity, drawing on surname-caste associations from the People of India Project \citep{singh1996communities} and applying recent advances in language to process names. We observe that connected firms experience lower entry in split blocks. Additionally, we examine firm entry in industries prone to political favoritism and find that such industries experience relatively lower entry in split blocks. These patterns suggest that checks and balances constrain preferential treatment, expanding opportunities for firms outside patronage networks.

Second, we examine regulatory hold-up. Political discretion in the issuance of regulatory permits creates scope for rent extraction, as politicians could impose costly delays on firms seeking approvals \citep{djankov2009regulation}. We find that projects in split blocks receive regulatory approvals substantially faster than those in unsplit blocks. We further exploit heterogeneity across industries and firm sizes. Industries subject to extensive regulatory requirements face higher entry barriers where discretionary delays are prevalent. We find that such regulated industries experience disproportionately higher entry in split blocks. Similarly, regulatory compliance often involves fixed costs that impose greater burdens on smaller firms. We find that smaller firms experience disproportionately higher entry in split blocks, with effects declining monotonically with firm size. These patterns suggest that checks and balances reduce costs arising from discretionary regulatory implementation.

Third, we examine public goods provision. Public infrastructure serves as a productive input for private firms, yet the discretionary nature of implementation creates scope for rent extraction, often manifesting as cost overruns and resource misallocation. We study road construction under the Pradhan Mantri Gram Sadak Yojana (PMGSY), a nationwide program where implementation is overseen by local authorities and political intervention is well documented \citep{lehne2018building}. We find that cost overruns decline substantially when blocks transition from single to multiple politician governance. We complement this with evidence on electricity: villages in split blocks are more likely to receive reliable commercial power supply. These patterns are consistent with improved efficiency in public goods provision when mutual oversight constrains rent extraction during implementation.

\textbf{Related Literature:} Our findings contribute to the understanding of the relationship between economic environment and the design of political institutions.\footnote{There is a large literature on this broad topic. See \cite{alesina1996theory}, \cite{tsebelis1999veto}, \cite{tsebelis2002veto}, \cite{glaeser2004institutions}, \cite{aghion2004endogenous}, and \cite{acemoglu2015rise}, among others.} We contribute to this literature by focusing on a certain aspect of political institutional design -- the presence of multiple politicians. Specifically, we provide well-identified empirical evidence showing that increasing the number of politicians governing an area can add economic value by imposing checks and balances on each other. The importance of checks and balances in political institutions dates back to \cite{de1751spirit} and \cite{kant1795perpetual} and has been discussed extensively in the works of \cite{barro1973control}, \cite{ferejohn1986incumbent}, \cite{persson1997separation, persson2000comparative}, \cite{bardhan2002}, \cite{aghion2004endogenous}, and \cite{acemoglu2013voters}, among others. We add to this literature by documenting that checks and balances imposed by multiple politicians can serve as a safeguard against the abuse of power. This ensures that no single entity becomes overly dominant and that the government operates in a manner that ensures accountability and transparency.

We contribute to the literature examining the impact of constraints, especially regulatory costs, on limiting firm entry.\footnote{\cite{djankov2009regulation} presents a detailed discussion of the literature examining the role of regulatory constraints in firm entry.} For instance, \cite{djankov2002regulation} and \cite{klapper2006entry} find that costly regulations restrict the creation of new firms. On a broader note, \cite{woodruff2018addressing} presents a detailed review of the literature on the constraints faced by small and growing businesses in the context of developing markets. This paper adds to the literature by documenting a relatively unexplored barrier to firm entry -- political checks and balances. We document that political checks and balances may prevent power abuse, ensuring government accountability and transparency. Consequently, this creates an economically favorable environment with lower regulatory obstacles, reduced cronyism, and improved public infrastructure, encouraging new firms to enter the market.

This study also contributes to our understanding of the role of bureaucrats in the efficient functioning of a state (see \cite{besley2022bureaucracy} for a comprehensive review). Bureaucrats are often perceived as exploitative agents who prioritize the interests of powerful industrialists \citep{tullock1965politics, peltzman1976toward, djankov2002regulation} that can potentially crowd out the entry of other firms \citep{stigler1971theory, zingales2017towards}. Moreover, politicians can exploit bureaucrats for personal gains by controlling their appointments, transfers, and promotions, often resulting in adverse economic outcomes (\cite{iyer2012traveling}, \cite{xu2018costs}, \cite{colonnelli2020patronage}, \cite{akhtari2022political}, among others). We contribute to the existing body of knowledge by demonstrating that political arrangements involving multiple politicians can act as a system of checks and balances. This arrangement enables effective monitoring of bureaucrats' performance, resulting in better provision of public goods and the establishment of an economic environment conducive to the entry of new firms and overall economic development. Essentially, our findings indicate how multiple politicians can effectively participate in overseeing bureaucrats and improving governance outcomes.

Our paper also contributes to the literature on the common agency problem. This literature has primarily focused on issues of lack of coordination, duplication of effort, and free-rider problems to hypothesize the negative outcomes of multiple principals (see \cite{bernheim1986common}, \cite{dixit1997power}, \cite{dixit1997common}, \cite{mezzetti1997common}, \cite{peters2001common}, \cite{laussel2001conflict}, \cite{bergemann2003dynamic}, among others). However, we document that multiple principals, in our case multiple politicians, can have a positive effect through better management of their subordinates, in our case local bureaucrats.


Closest to our study is the work of \cite{gulzar2017politicians}. Like them, we use a close border design to compare villages across split and unsplit blocks. \cite{gulzar2017politicians} focus on unemployment assistance from India’s \textit{National Rural Employment Guarantee Scheme} (NREGA) and find unemployment assistance is lower in areas where bureaucrats answer to multiple politicians, and they argue that this result suggests poor implementation of the program in such areas. This paper contributes by examining the relationship between multiple politicians and economic activity in the private sector -- firm entry, employment, and consequently economic development. This allows us to present an alternative perspective to \cite{gulzar2017politicians}.  Our results indicate regions with multiple politicians have greater new firm entry and higher employment, manifesting as lower demand for unemployment assistance. We directly validate the lower demand for unemployment assistance by examining the NREGA applications data. We show that 5\% lower workday claims or applications for unemployment benefits filed in split blocks relative to unsplit blocks. Therefore, our findings suggest that the development of the private sector can reduce the need for government unemployment assistance. Moreover, we add to the identification strategy of \cite{gulzar2017politicians} by using the boundary reforms in 2008 to employ a differences-in-discontinuity design allowing us to examine pre-trends.

\par
The remainder of the paper proceeds as follows: Section \ref{sec_insti} discusses the institutional background on which the empirical investigation of this paper is based. Section \ref{sec_data} provides a brief description of the data. Section \ref{sec_baseline} lays down the empirical strategy and the main results. Section \ref{mech} provides the mechanism and section \ref{conc} concludes.

\section{Institutional Details}
\label{sec_insti}

\subsection{Administrative and Political Structure in India}

India's governance system operates through a multi-layered administrative hierarchy that extends from the federal government down to the village level (see appendix Figure \ref{app_fig_six_tiers} for a visual representation of this administrative structure). National and state governments formulate policies, while implementation occurs through a three-tier local governance system: the district, the block, and the village.\footnote{Districts, headed by District Collectors, coordinate programs across blocks, while villages serve as final delivery points for policy implementation.}

Policy implementation at the local level occurs through blocks, which serve as key administrative units in India's governance system. Block Development Officers (BDOs) administer these blocks and have statutory powers for policy implementation, serving as the "eye of the needle" through which government programs must pass \citep{dasgupta2020political}. Parallel to this administrative structure is India's political system, where Members of Legislative Assembly (MLAs) represent electoral constituencies and serve as legislators in the State Legislative Assembly.

We focus on the interplay between block-level bureaucrats and politicians. Blocks occupy a unique position in the administrative hierarchy as the largest units that can either fall within a single constituency or span multiple constituencies. This creates our key variation: some blocks fall under the jurisdiction of one MLA while others fall under the jurisdiction of multiple MLAs. BDOs are accountable to all MLAs whose constituencies encompass their blocks. The quasi-random variation in political oversight forms the foundation of our research design.

In the rest of this section, we describe India's administrative and political boundaries, the formal powers held by bureaucrats and politicians at different levels, and the interaction between politicians and bureaucrats that shapes governance outcomes.

\subsection{Overlap of Electoral and Administrative Boundaries}

Local governance and policy implementation are territorially organized at the block level, while political representation operates through constituency boundaries. Political boundaries are determined by the Delimitation Commission, which redraws electoral boundaries to equalize population across all constituencies within a state. The administrative boundaries of a block are not considered during the design of constituencies. This independent boundary-drawing process results in a haphazard overlap between the constituency and block boundaries. Appendix Figure \ref{app_fig_boundary} shows a representative example of the electoral and administrative boundaries of the state of Karnataka.

A block, whether wholly or partially, inside a constituency comes under the jurisdiction of the MLA.\footnote{Note that although a block can be under the jurisdiction of multiple MLAs from different constituencies, individuals can only vote for candidates in one constituency, depending on the village or town they reside in.} As a result of this overlap, blocks are either under the jurisdiction of multiple MLAs (split blocks) or a single MLA (unsplit blocks), as shown in appendix Figure \ref{app_fig_overlap}. The local bureaucrats are accountable to all the MLAs with jurisdiction over the block, regardless of whether a block is wholly or partially included in the constituency.

\begin{figure}[H]
    \centering
    \caption{Split and Unsplit Blocks - Overlap of Administrative and Electoral Boundaries}
    \begin{subfigure}{.32\textwidth}
    \centering
    \includegraphics[width = \textwidth]{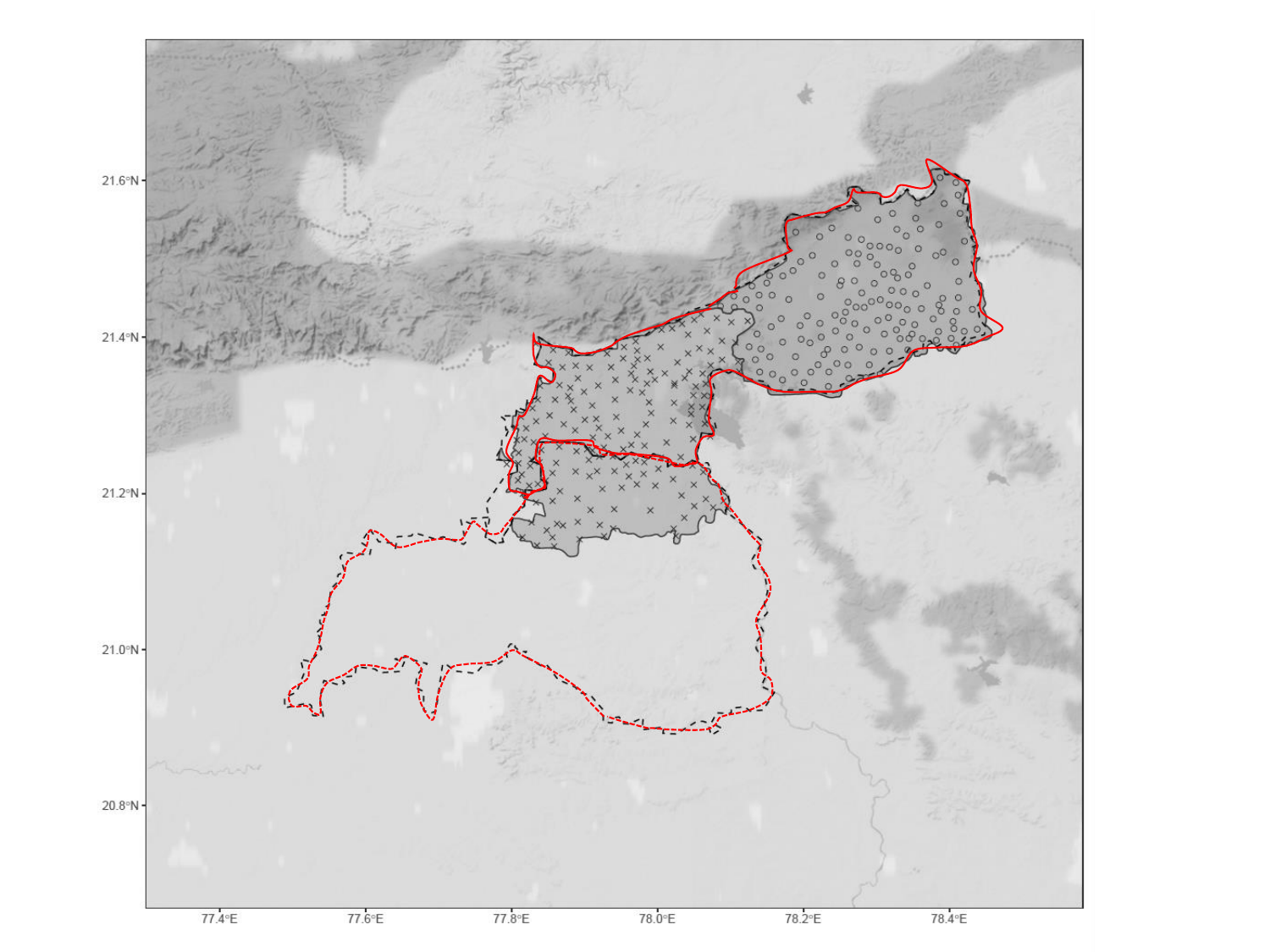}
    \caption{Electoral Boundaries}
    \label{fig_microcosm_a}
    \end{subfigure} %
    \begin{subfigure}{.32\textwidth}
    \centering
    \includegraphics[width = \textwidth]{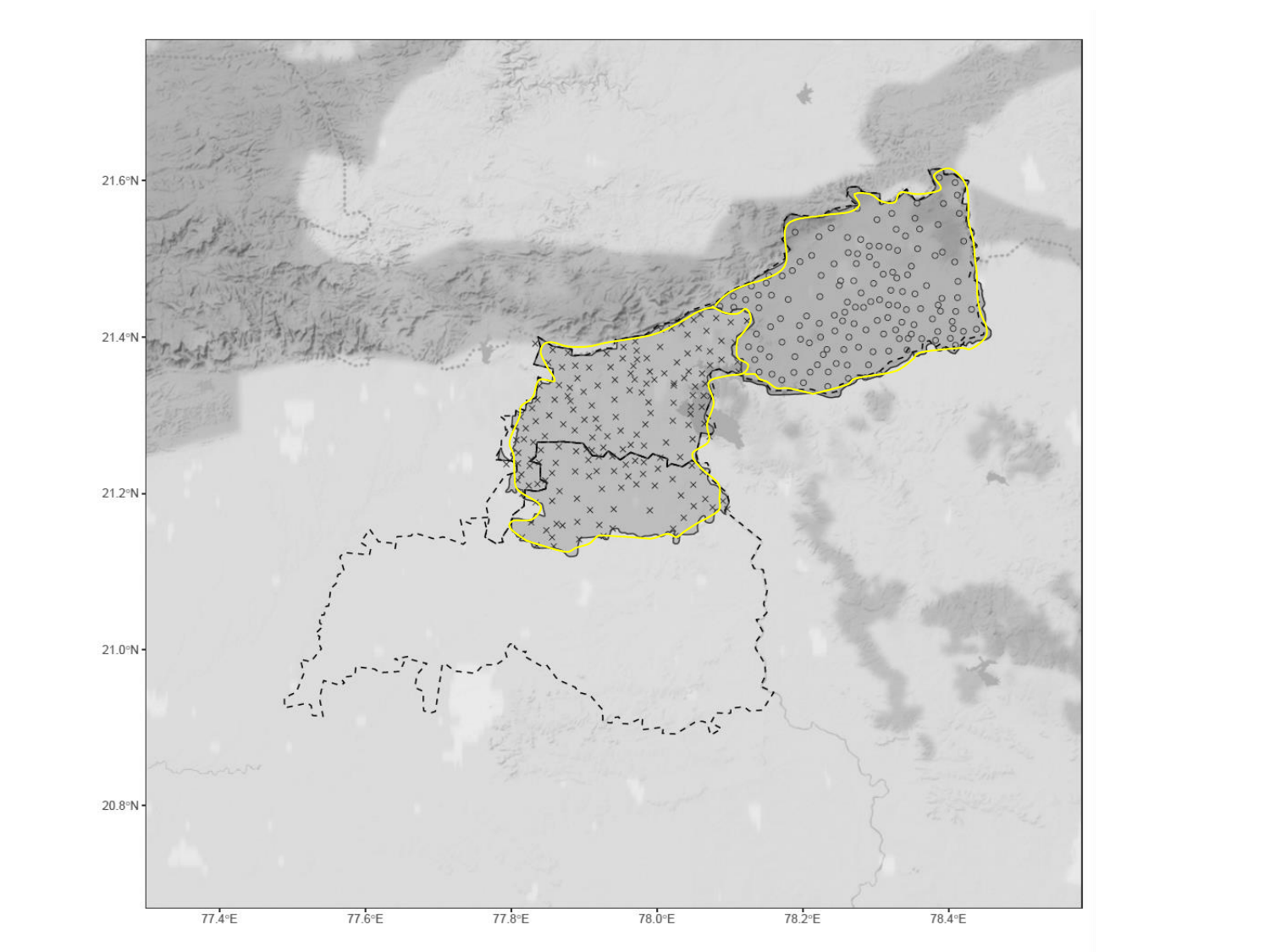}
    \caption{Administrative Boundaries (Blocks)}
    \label{fig_microcosm_b}
    \end{subfigure} %
    \begin{subfigure}{.32\textwidth}
    \centering
    \includegraphics[width = \textwidth]{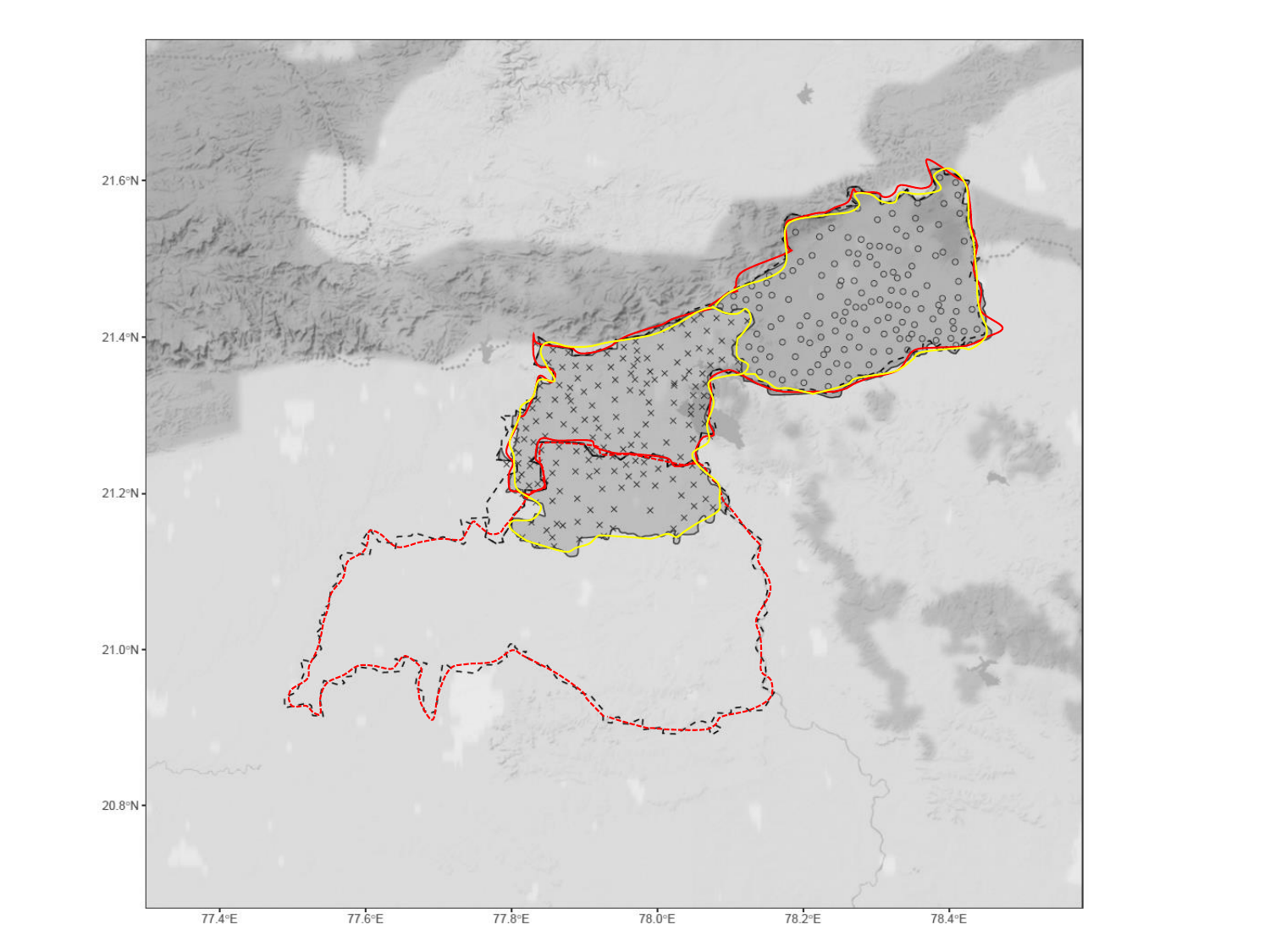}
    \caption{Overlap}
    \label{fig_microcosm_c}
    \end{subfigure} %
    \begin{subfigure}{.9\textwidth}
    \centering
    \includegraphics[width = \textwidth]{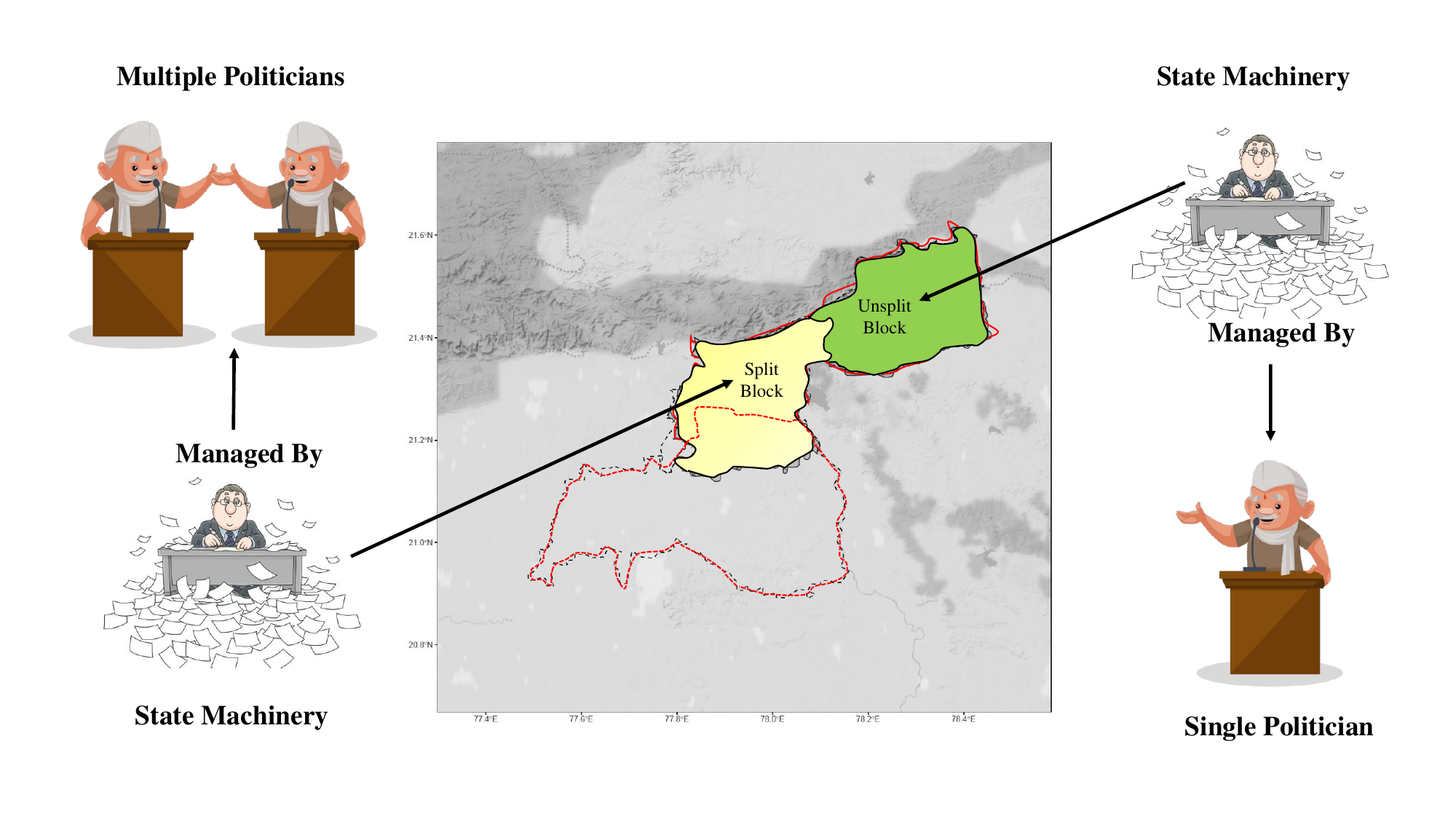}
    \caption{Split \& Unsplit Blocks}
    \label{fig_microcosm_d}
    \end{subfigure} %
    \label{fig_microcosm}
    \begin{minipage}{0.94\textwidth}
	\begin{center}
		\end{center}
		{\footnotesize 	The figure illustrates a case of adjacent split and unsplit blocks in our sample. The split block is the administrative block of Morshi, in the district of Amaravati, Maharashtra. This block is split between the two assembly constituencies of Morshi and Teosa. The neighbouring unsplit block is the administrative block of Warud, in the district of Amaravati, Maharashtra. Figure \ref{fig_microcosm_a} shows the assembly constituency boundaries of Morshi and Teosa in red. Figure \ref{fig_microcosm_b} shows the block boundaries of Morshi and Warud in yellow. Figure \ref{fig_microcosm_c} shows the overlap of assembly constituency boundaries of Morshi and Teosa in red with the block boundaries of Morshi and Warud in yellow. Figure \ref{fig_microcosm_d} presents a microcosm of our setting showing the formation of split and unsplit blocks due to the overlap of assembly and block boundaries and the state machinery being managed by multiple and single politicians, respectively. \par}
\end{minipage}
\end{figure}

Figure \ref{fig_microcosm} presents a microcosm of our setting using an example of adjacent split and unsplit blocks of Morshi and Warud in the district of Amravati, Maharashtra. The haphazard overlap of constituency and block boundaries generates the variation in the number of MLAs with jurisdiction over a block. Figure \ref{map_split_a} presents the geographic distribution of split and unsplit blocks in our sample. Split blocks constitute 40\% of blocks in our sample, and do not appear to be geographically concentrated. The variation in whether a block is split or unsplit is the primary source of heterogeneity we exploit in this paper.

\subsection{Administrative Authority at the Block Level}

Blocks represent a key implementation tier in India's governance system, where government programs and policies are implemented at the local level. Block Development Officers (BDOs), appointed by the state government, serve as the primary administrative authority at this level with discretionary power in interpreting and implementing government directives.\footnote{In Karnataka, Section 155(1) mandates that the state government appoint a Group-A Officer of the State Civil Services to serve as the Executive Officer of the Taluk Panchayat (block). Similar provisions exist across states, with variations in specific rank requirements and appointment procedures; see Appendix~\ref{app:bdo_powers} for a detailed discussion of BDO powers.} BDOs derive their authority from state legislation that establishes their position as senior gazetted officers responsible for implementing federal and state development programs.

BDOs exercise effective control over licensing and other implementation decisions through their supervisory authority over village-level administration. While many procedures, including initial permit applications, are nominally processed at the village level, BDOs hold decisive influence through budget approval powers, audit authority, and inspection directives that village institutions must comply with. Through mandatory budget approvals, BDOs can control the resources available for processing applications, while their audit powers enable them to ``disallow'' local decisions and impose personal financial liability on village officials who do not align with BDO preferences.\footnote{See Appendix~\ref{app:bdo_village_oversight} for a comprehensive description of BDO oversight powers over Grama Panchayats.} This supervisory framework ensures that BDOs function as the relevant administrative authority for decisions that ultimately affect local economic activity.

In addition to program implementation, BDOs control permits and approvals that are prerequisites for business creation, including trade licenses, factory establishment permissions, and building construction permits.\footnote{See Appendix~\ref{app:business_permits} for a detailed description of the subset of local business permits that fall under block-level administration and BDO oversight.} These permits serve as prerequisites for bank accounts, state-level clearances, and other business requirements, creating cascading dependencies through which BDO decisions determine the timing and feasibility of firm entry.

\subsection{Political Representatives and Administrative Oversight}

India operates a parliamentary democracy with elected representatives at multiple levels of government. At the state level, Members of Legislative Assembly (MLAs) are elected from single-member constituencies to serve five-year terms in their respective State Legislative Assemblies. MLAs derive their authority from India's constitutional framework, which grants them legislative power to make laws and oversee government policy. Executive authority for policy implementation is vested in the bureaucracy.\footnote{MLAs do not possess direct executive authority and cannot issue administrative orders to bureaucrats or directly implement policies. This follows the Westminster parliamentary system where the legislature and executive are constitutionally separate branches.} MLAs must therefore work through bureaucrats to achieve local governance outcomes in their constituencies.

MLAs face electoral incentives to demonstrate effective constituency service and development outcomes. In India's competitive democracy, voters evaluate MLAs based on visible improvements in infrastructure, public services, and economic opportunities, with electoral punishment for those who fail to deliver \citep{bussell2019clients}. Simultaneously, political office provides MLAs with opportunities for private benefit extraction through their position and control over bureaucratic decision-making \citep{fisman2014private}. Both electoral success and private benefit extraction require effective bureaucratic implementation. 

Despite constitutional separation of powers, the system provides MLAs with mechanisms to influence bureaucratic behavior. MLAs exercise oversight through formal legislative institutions and can affect bureaucrat career progression through the transfer and posting system. Block Development Officers and other state-level bureaucrats are subject to periodic transfers, and MLAs provide influential feedback on bureaucrat performance to state government officials who make transfer decisions \citep{iyer2012traveling}. This system enables MLAs to influence bureaucratic actions, as non-compliance with MLA preferences can result in unfavorable career consequences.

\subsection{Governance Structures and Barriers to Firm Entry}

Firm entry in India requires extensive regulatory approvals across multiple tiers of governance.\footnote{The World Bank's Ease of Doing Business Report \citep{WorldBank2009DoingBusinessIndia} provides comprehensive documentation of the regulatory procedures required for business establishment across Indian cities. These costs are substantial: building permits and utility connections in Mumbai alone cost 2,717.8\% of GDP per capita and require approximately 200 days. We refer readers to the report for the complete list of procedures.} At the block level, BDOs exercise statutory authority over the issuance of local permits essential for business establishment, including trade licensing, factory establishment permissions, and building construction permits. The BDO's interpretive authority over the issuance of regulatory permits creates differential costs for firms, with identical business applications facing different processing timelines and requirements despite uniform policy frameworks. 

State-level politicians (MLAs) have strong incentives to shape which firms are able to enter and operate in their constituencies, since influence over the allocation of business permits generates opportunities for private rent extraction while in office \citep{sukhtankar2015corruption}. However, MLAs operate as legislators and lack direct executive authority over policy implementation. \footnote{In practice, MLAs devote little time to legislative activities: only 3\% of surveyed MLAs report assembly work as their primary focus, and they spend most of their time interacting with local bureaucrats and addressing constituents \citep{chopra1996}.} Consequently, MLAs must work through the local bureaucrats who hold statutory authority over permit issuance.\footnote{This aligns with the argument put forth by \citet{chopra1996} that MLAs often act as 'fixers' who work with bureaucrats to allocate government-produced goods.} By credibly promising job security and favorable treatment to compliant bureaucrats, MLAs can influence the enforcement and interpretation of regulatory requirements. This politician–bureaucrat nexus determines the effective barriers and costs of firm entry at the block level.

The MLA–BDO interactions occur through both informal channels and formal institutional venues. The Panchayat Samiti, an intermediate tier of local governance at the block level, serves as the primary formal venue for these interactions. MLAs whose constituencies overlap with a block sit as ex officio members of that block’s Panchayat Samiti, while the BDO serves as the committee’s executive officer. Panchayat Samiti meetings thus create a structured setting in which MLAs regularly observe BDO performance and, when multiple MLAs are present, can also observe one another’s behavior.

When a single MLA’s constituency encompasses an entire block, that MLA serves as the sole political representative in the Panchayat Samiti. In this setting, the MLA and BDO can coordinate over rent-sharing arrangement, without facing scrutiny from rival political principals. The absence of competing political principals within the block, combined with the fact that much of this behavior operates through low-visibility bureaucratic channels, limits both external exposure and electoral discipline. As a result, patterns of rent extraction can persist even when they impose substantial aggregate costs.

When a block spans multiple constituencies, multiple MLAs sit as ex officio members of the same Panchayat Samiti. This institutional arrangement creates conditions under which each MLA's actions and demands become observable to rival politicians through regular committee meetings. MLAs with diverging political interests face electoral incentives to monitor each other and publicize behavior that appears self-serving. For the BDO, the presence of several elected principals with potentially divergent interests makes it harder to sustain an exclusive rent-sharing arrangement with any single MLA. The heightened monitoring and coordination difficulties might make MLAs reluctant to pursue extractive arrangements that risk exposure by political rivals. Consequently, blocks with multiple MLAs are more likely to experience reduced rent extraction and lower barriers to firm entry than blocks represented by a single MLA

\section{Data}
\label{sec_data}

We assemble several administrative datasets that measure economic and political activity at fine spatial resolution across India. This section provides a brief overview of the most important datasets and describes their use in subsequent analyses. We provide further details in Appendix \ref{app:data_sources}.

\subsection{Administrative and Political Boundaries} 
\subsubsection{Identifying Split Blocks} \label{block_definition}

We obtain geospatial data on administrative blocks from MLInfomap and on electoral constituency boundaries from DataMeet for both the 1977 and 2008 delimitations. Administrative blocks represent the unit of local governance in India, with each block overseen by a Block Development Officer (BDO) who manages regulatory approvals and public program implementation. Assembly constituencies define the electoral boundaries for state legislative elections, with each constituency electing one Member of Legislative Assembly (MLA).

We overlay these shapefiles to identify which blocks fall under the jurisdiction of multiple politicians. Following the institutional structure that a block wholly or partially located within an assembly constituency falls under that constituency's elected representative, we define a block as \textit{Split} if it is part of more than one assembly constituency. To account for discrepancies in boundary alignment across data sources, we require each constituency partition to occupy at least 10\% of the block's geographic area when classifying a block as \textit{Split}.

Appendix Figures \ref{app_fig_boundary} and \ref{app_fig_overlap} illustrate this procedure for the state of Karnataka. Appendix Figure \ref{app_fig_boundary_a} shows boundaries for the administrative block, while Appendix Figure \ref{app_fig_boundary_b} shows boundaries for electoral constituency within Karnataka. Appendix Figure \ref{app_fig_overlap_a} displays the overlay of these two boundary systems, and Appendix Figure \ref{app_fig_overlap_b} identifies the resulting split and unsplit blocks. Figure \ref{fig_identification} illustrates a case of adjacent split and unsplit blocks in our sample.

\textit{Split} blocks constitute 39.43\% of our sample and are not geographically concentrated (Appendix Figure \ref{map_split_a}). These blocks account for 50\% of the 2001 population and 40\% of total geographic area. Among \textit{Split} blocks, 83.08\% are governed by two politicians, 14.83\% by three politicians, and 2.09\% by more than three politicians (Appendix Figure \ref{app_fig_dist_num}).

We use villages as the primary unit of observation for our analysis. We obtain village boundaries from the 2001 Census provided by \citet{sedac2018}. We map villages to their corresponding blocks and assembly constituencies by overlaying the shapefiles. Using village centroids, we calculate each village's distance to the nearest boundary separating a \textit{Split} from an \textit{Unsplit} block, which serves as the running variable in our regression discontinuity design. The village-level resolution allows us to estimate local polynomial within narrow bandwidths around block boundaries, comparing outcomes in villages that are geographically proximate but experience differential political oversight.

\subsubsection{Exogenous Block Boundaries}
\label{sec_boundary_exogeneity}

We further refine our identification strategy by restricting analysis to boundaries with characteristics that preclude strategic placement based on economic conditions. We employ two complementary approaches: (1) identifying boundaries defined by geographic features - in this case, river segments, and (2) identifying straight-line-like boundaries. River boundaries constitute natural barriers that predate modern economic development and are unlikely to reflect strategic placement to separate economically distinct areas. Straight-line boundaries reflect historical surveying conventions that imposed geometric regularity without regard to local topographic or economic features. These sample restrictions provide robustness checks on our baseline estimates.

First, we identify boundaries determined by river courses using high-resolution geospatial data on river segments from HydroATLAS \citep{linke2019hydroatlas}. We classify a boundary segment as determined by river segments if it aligns with a river for at least 80\% of its length and the river has average discharge greater than or equal to 200 $m^3/s$ (cumecs). The discharge threshold ensures the river is large enough to serve as a natural territorial delimiter. This definition yields 3.02\% of boundaries and 5.74\% of blocks from our sample. Figure \ref{fig_boundary_river} presents representative examples of block boundaries determined by river courses.

Second, we identify straight-line boundaries. We calculate the Sinuosity Index (SI) for each boundary segment—the ratio of actual boundary length to the Euclidean distance between its endpoints. Boundaries with SI values less than 1.05 exhibit near-straight characteristics. We include boundaries with SI less than 1.05, yielding 6.24\% of boundaries and 10.68\% of blocks. Appendix Figures \ref{straight_line_ex} and \ref{wiggly_line_ex} illustrate examples of straight-line and irregular boundary segments. 

\subsection{Universe of Firm Registration}

We utilize data from the Ministry of Corporate Affairs (MCA), Government of India, which administers the statutory registry for companies incorporated under the Companies Act.\footnote{The MCA electronic filing system maintains comprehensive records for all registered companies, including private limited companies, public limited companies, and one-person companies.} Our dataset covers approximately 850,000 firms established between 2003 and 2016, providing a near-census of the formal private corporate sector in India.\footnote{The dataset excludes entities not subject to the Companies Act, such as sole proprietorships, ordinary partnerships, and cooperatives, which constitute a substantial share of India's non-corporate enterprises.} This registry serves as the foundation for India's national accounts estimates of the private corporate sector.

The Companies Act requires all registered companies to file incorporation documents and ongoing returns with the Registrar of Companies, including registration dates, operational addresses, financial statements, and information on directors and key managerial personnel. The MCA records include each firm's precise operational address and date of incorporation, enabling us to construct a spatio-temporal dataset of firm entry across India. The data also contain detailed information on company directors, including the names of these directors, which we leverage to infer the social identities of key personnel in charge of operating these firms.

\subsubsection{Geolocating Firm Entry}

The MCA registry provides information on the address of the registered office for each firm, which includes street addresses and six-digit postal codes (pincodes). This allows us to locate these firms in space and measure private sector activity at highly granular spatial resolution, an improvement over existing datasets on private sector activity in India, which typically provide aggregates at the state or district level.

We start constructing village-level measures of firm entry by first geolocating firm addresses into spatial coordinates. \footnote{The MCA records provide addresses for each firm's registered office but do not include latitude and longitude coordinates.} We approximate the location of a firm by that of the post office corresponding to its pincode. Geocoordinates for post offices are obtained using the Google Maps API. Appendix \ref{sec_app_firm_data} provides detailed discussion of the geolocation procedure, potential measurement errors associated with pincode-based assignment, and validation exercises indicating these errors are unlikely to significantly affect our results.

With these firm-level coordinates, we then map firms to villages by performing spatial joins with village boundary shapefiles. This procedure enables us to aggregate firm counts at the village level, creating a spatio-temporal dataset of firm entry by location and year from 2003 to 2016. Appendix Figure \ref{map_entry_a} and Appendix Table \ref{app_tab_summary_firms} present the spatial and temporal distribution of firm entry in our sample.

\subsubsection{Firm Directors and Social Identity}

The Companies Act requires firms to file information on their directors and key managerial personnel with the Registrar of Companies, which includes the names of directors. Surnames in India often carry identity markers that signal caste affiliation, as well as broader religious identities.\footnote{In India, \textit{jati} — endogamous subcastes that constitute narrowly defined groups — represents a salient identity around which social and political networks form \citep{munshi2019caste}.}  We leverage this feature of the data to infer the social identities of firm directors and construct measures of patronage connections between firms and local politicians.

We rely on recent advances in AI/ML methods and combine rich ethnographic documentation as well as large-scale datasets on self-reported identity affiliations to link names to social identities. This enables us to classify firms based on whether their directors share social identity markers with the local politician in power. We further discuss this in Section \ref{sec:patronage_analysis}. Appendix \ref{sec:data_patronage} provides a detailed discussion of the methodology followed to infer social identity.

\subsection{Auxiliary Data}

\subsubsection{Data on Economic Activity} 

\noindent \textit{NREGS Application Data:} The National Rural Employment Guarantee Scheme (NREGS) is the world's largest employment guarantee program, under which the government serves as employer of last resort. We collect data on individual-level applications filed for unemployment relief covering the period 2016 to 2020. The number of days for which applicants seek work through NREGS provides a measure of demand for unemployment assistance. To our knowledge, we are the first to collect and use application-level NREGS data.\newline

\noindent\textit{Harmonized Nightlights.} We use global nighttime light data from \cite{li2020harmonized}, which harmonizes data from the Defense Meteorological Satellite Program (DMSP) and the Visible Infrared Imaging Radiometer Suite (VIIRS).\footnote{DMSP provides coverage until 2013, while VIIRS coverage starts from 2012. The transition between these satellite systems occurs within our study period (2003–2016), necessitating harmonization to ensure consistent measurement of nightlight intensity over time.} We construct village-level measures of average nightlight intensity from 2003 to 2016, which serve as a proxy for economic activity \citep{donaldson2016view}.\newline

\noindent\textit{CMIE CapEx Data.} We obtain data on investment projects from the Centre for Monitoring Indian Economy (CMIE) Capital Expenditure database, covering projects announced from 2003 to 2016 that involve capital expenditure of Rs. 10 million or more. For each project, the database records announcement dates, regulatory approval dates, and implementation milestones. We use the time between project announcement and initial regulatory approval as a measure of bureaucratic processing delays.\newline

\noindent\textit{Economic Census (2013).} We use data from the 2013 Economic Census of India \citep{ecindia}
, which provides data on non-farm employment at the village level.\footnote{The 2013 Economic Census is the first comprehensive economic census conducted after the 2008 delimitation.} These data were accessed through the Socioeconomic High-resolution Rural-Urban Geographic Platform for India (SHRUG) \citep{almn2021}.

\subsubsection{Data on Legislative Elections}

\noindent\textit{Electoral Data:} We use data on politicians and election outcomes from the Trivedi Center for Political Data \citep{jensenius2017}, accessed through SHRUG \citep{almn2021}. This dataset provides election results, party affiliations, and names of candidates for state legislative assembly elections. We use this information to construct measures of political alignment, defined as the degree to which multiple politicians governing split blocks belong to different political parties. We also use candidate names to infer their social identities and measure connections to entrepreneurs entering their jurisdictions.\newline

\noindent\textit{Candidate Affidavits:} We obtain data on candidate characteristics from sworn affidavits filed with the Election Commission of India, which were digitized by the Association for Democratic Reforms (ADR).\footnote{Electoral law in India requires candidates to file sworn asset affidavits (Form 26) at the time of nomination (Section 33A, Representation of the People Act, 1951, inserted following the Supreme Court's 2002 ruling in \textit{Union of India v. Association for Democratic Reforms}). The ADR dataset is available under the Open Database License at \url{https://github.com/Vonter/india-election-affidavits}.} These affidavits include self-reported assets and liabilities, allowing us to measure changes in private wealth over a politician's term in office by comparing declarations across elections.

\subsubsection{Data on Public Good Provision}

\noindent\textit{PMGSY Road Construction:} We obtain administrative data from the Pradhan Mantri Gram Sadak Yojana (PMGSY) rural road construction program, accessed through SHRUG \citep{asher2020, almn2021}. The data cover road projects from 2001 to 2014 and include estimated costs, actual costs, and completion timing. We use cost and time overruns as measures of inefficiency in road construction, testing whether split blocks experience more efficient public goods provision.\newline

\noindent\textit{Population Census:} We use village-level data from the 2001 and 2011 Population Census of India \citep{pcindia}, accessed through SHRUG \citep{almn2021}. The Population Census provides village-level data on demographics and public infrastructure.

\section{Impact of Multiple Politicians}
\label{sec_baseline}
The primary objective of this paper is to identify the effect of multiple politicians on local economic growth, in general, and the entry of firms, in particular. The haphazard overlap of administrative block boundaries and the electoral constituency boundaries allows us to identify blocks under the supervision of one politician (unsplit block) and blocks under the supervision of multiple politicians (split blocks). Blocks entirely subsumed within an electoral constituency are classified as unsplit blocks. By contrast, blocks that span more than one electoral constituency are classified as split blocks - split between multiple politicians.

\subsection{Empirical Strategy}
\label{hypempr}
The empirical strategy hinges on comparing two administrative regions that are similar in all attributes but differ in the number of politicians at the helm of their administrative affairs. To do so, we follow two identification strategies. First, we employ a cross-sectional geographic regression discontinuity design by examining the differences in the outcome variable on either side of the boundary separating a split block and an unsplit block. Second, we employ a differences-in-discontinuity design by exploiting the 2008 delimitation of electoral constituencies. The redrawing of electoral constituency boundaries resulted in converting some unsplit blocks into split blocks and vice versa. The natural experiment of delimitation allows us to examine changes in differences across the border between two blocks in cases where delimitation changes the number of politicians in a block while keeping them fixed in the contiguous block.

\subsubsection{Spatial Regression Discontinuity}
\begin{figure}[htbp]
	\centering
	\caption{Illustration of Empirical Strategy}
	\includegraphics[width=0.75\textwidth]{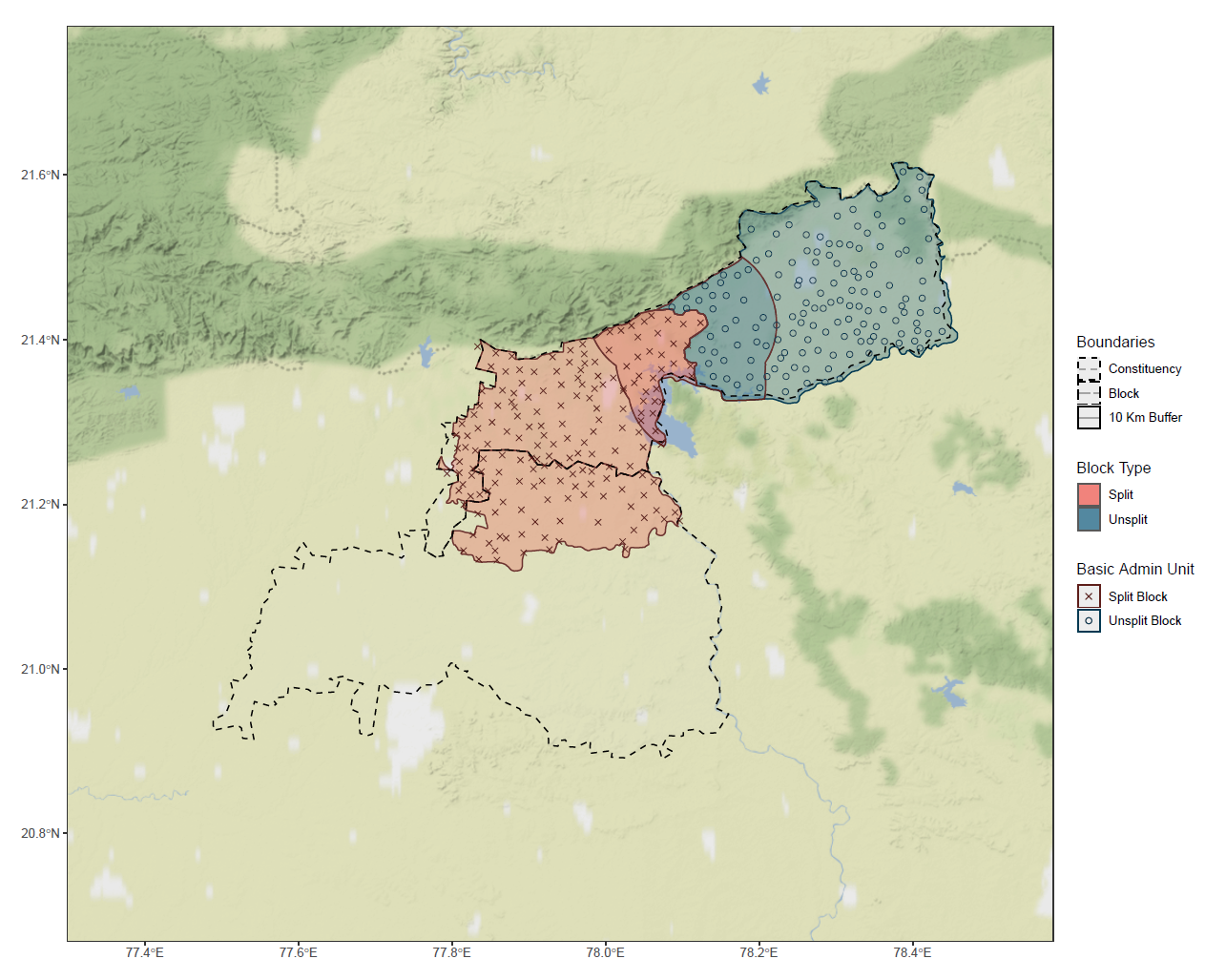}
	\label{fig_identification}
	\begin{minipage}{0.75\textwidth}
{\footnotesize This figure illustrates a case of adjacent split and unsplit blocks in our sample. The geographic area shaded in red, is the administrative block of Morshi, in the district of Amaravati, Maharashtra. This block is split between the two assembly constituencies of Morshi and Teosa. The geographic area shaded in blue is the administrative block of Warud, in the district of Amaravati, Maharashtra, and is an unsplit block.   
\par}
	\end{minipage}
\end{figure}

First, we estimate the causal effect of multiple politicians on the outcomes of interest using a regression discontinuity design. We compare outcomes of interest in a village located just inside a split block relative to villages on the other side of the boundary, within an unsplit block. We pool the outcome variables in a village across years, since 2008, and estimate the average causal effect of multiple politicians using the following regression discontinuity specification:
\begin{align}
Y_{v(v \in b(B)}  = \gamma\cdot Split_b  + f(distance_v) + \beta \cdot X_v + \phi_B + \varepsilon_{v(v \in b)}\,\, for\,\, v\in bw \label{rd}
\end{align}
where $Y_{v(v \in b(B)}$ is the outcome of interest in a village $v$ located within administrative block $b$ belonging to a pair of split and unsplit blocks that share a common boundary $B$. $Split_b$ is a binary variable that takes a value of 1 if a block spans multiple electoral constituencies and consequently has multiple politicians in charge. $X_v$ refers to a vector of village-level covariates such as population, area, distance to district headquarters, and compactness; $\phi_B$ is boundary fixed effect; and $f(location_v)$ is the RD polynomial, which controls for smooth functions of the geographic location for village $v$. Our baseline specification is a local linear polynomial in the distance of the village to the boundary $B$ estimated separately on each side of the boundary, as suggested by \cite{calonico2014robust}, \cite{cattaneo2019practical}, and \cite{gelman2019high}. Following \cite{dell2018nation}, we use a triangular weighting kernel. We estimate our specification for different bandwidths $bw$, taking values of 5, 10, 20, and 50 km on either side of the boundary. Figure \ref{fig_identification} illustrates this geographic RD design.\footnote{Figure \ref{fig_identification} illustrates a case of adjacent split and unsplit blocks in our sample. The geographic area shaded in red, is the administrative block of Morshi, in the district of Amaravati, Maharashtra. This block is split between the two assembly constituencies of Morshi and Teosa. The geographic area shaded in blue is the administrative block of Warud, an unsplit block, in the district of Amaravati, Maharashtra.} We check robustness to using various other forms of the RD polynomial, kernel, and bandwidths in section \ref{sec_result_robustness}.

Our coefficient of interest is $\gamma$, the effect of being just inside a split block, and governed by multiple politicians, on our outcome of interest. The interpretation of $\gamma$ as a causal effect of multiple politicians requires two identifying assumptions. First, all relevant factors before the 2008 delimitation varied smoothly at the block boundaries. This assumption implies the villages located close to the boundary are similar in terms of observables and unobservables before being sorted into split and unsplit blocks in 2008. Second, no selective sorting occurs across the RD threshold; that is villages cannot select in or out of a split block. This assumption ensures a village, within a narrow bandwidth of the boundary, is arbitrarily allocated into a split or an unsplit block. The two assumptions, taken together, ensure the villages, located in unsplit blocks, just outside the boundary of a split block are a valid counterfactual to villages situated just inside the split block.

The two identifying assumptions are likely to be true for our empirical design. As discussed in section \ref{sec_insti}, a block is defined as either split or unsplit based on the 2008 delimitation, which was largely politically neutral, devoid of gerrymandering, and focused primarily on equalizing populations across electoral constituencies \citep{iyer2013redrawing}. These electoral constituencies are large relative to the villages in the narrow bandwidth. Hence, villages along the boundaries did not have the ability to negotiate whether their blocks would be governed by single or multiple politicians.

Moreover, Table \ref{tab_summary} provides empirical support for the smoothness requirement by presenting a balance test across multiple socio-economic dimensions, demonstrating the similarity of villages on both sides of the boundaries. We address concerns of selective sorting by focusing our analysis on a subset of (1) boundaries that are close to a straight line because squiggly boundaries can indicate selective sorting \citep{polsby1991third, alesina2011artificial}, and (2) boundaries defined by salient geographic characteristics, in this case, rivers and river basins.

\subsubsection{Time Series Difference in Spatial Discontinuity}
Next, we examine the effect of time-series variation in the number of politicians governing a block by exploiting the exogenous delimitation of electoral constituencies in 2008. The natural experiment allows us to control for time-invariant unobservables across villages by including village fixed effects and time-varying heterogeneity across adjacent block pairs through boundary $\times$ year fixed effects. Specifically, the natural experiment allows us to investigate the variations that occur across the border between two blocks when the delimitation process alters the number of politicians governing a block while keeping it unchanged in the neighboring block. We estimate the following differences-in-discontinuity regression specifications:
\begingroup\makeatletter\def\f@size{9.5}\check@mathfonts
\def\maketag@@@#1{\hbox{\m@th\large\normalfont#1}}%
\begin{align}
\footnotesize
Y_{v(v \in b(B),t)}  & = \gamma\cdot (Unsplit\rightarrow Split)_{b}\times Post_t  +f(distance_v) + \beta_v + \phi_{Bt} +\varepsilon_{v(v \in b,t)}\,\, \forall \text{ } v\in bw  \label{did_1} \\
Y_{v(v \in b(B),t)}  & = \gamma\cdot (Split\rightarrow Unsplit)_{b}\times Post_t  +f(distance_v) + \beta_v + \phi_{Bt} +\varepsilon_{v(v \in b,t)}\,\, \forall \text{ } v\in bw \label{did_2}
\end{align} \endgroup
In regression specification \ref{did_1}, $(Unsplit\rightarrow Split)_{b}$ is a binary variable that takes a value of 1 for blocks that transitioned from being unsplit to split between multiple politicians following the 2008 delimitation. We compare these unsplit to split blocks with contiguous blocks that remained unsplit throughout. In regression specification \ref{did_2}, $(Split\rightarrow Unsplit)_{b}$ is a binary variable that takes a value of 1 for blocks that transitioned from being split to unsplit following the 2008 delimitation. We compare these split to unsplit blocks with contiguous blocks that remained split throughout. $Post_{t}$ is a dummy variable taking the value of 1 for the years following the 2008 delimitation. $\phi_{Bt}$ controls for the time-varying boundary fixed effect that allows $\gamma$ to be estimated from the differences between split and unsplit blocks sharing the same boundary. $\beta_v$ denotes the village fixed effect which allows the estimation to come from the time-series variation of the same village and consequently the administrative block switching its split status. Finally, the inclusion of the RD polynomial $f(distance_v)$ allows $\gamma$ to capture the difference in geographic discontinuity over time.

\subsection{Spatial Discontinuity Results}
\label{did}
\begin{figure}[htbp]
    \centering
    \caption{RD Plots for Firm Entry and Nightlights}
    \begin{subfigure}{.49\textwidth}
    \centering
    \includegraphics[width = \textwidth]{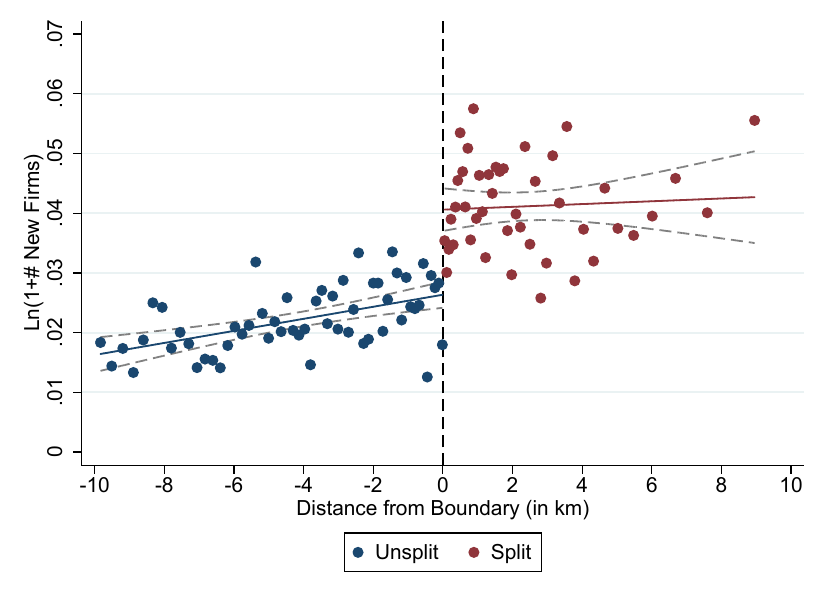}
    \caption{Firm Entry}
    \label{fig_entry_bin}
    \end{subfigure} %
    \begin{subfigure}{.49\textwidth}
    \centering
    \includegraphics[width = \textwidth]{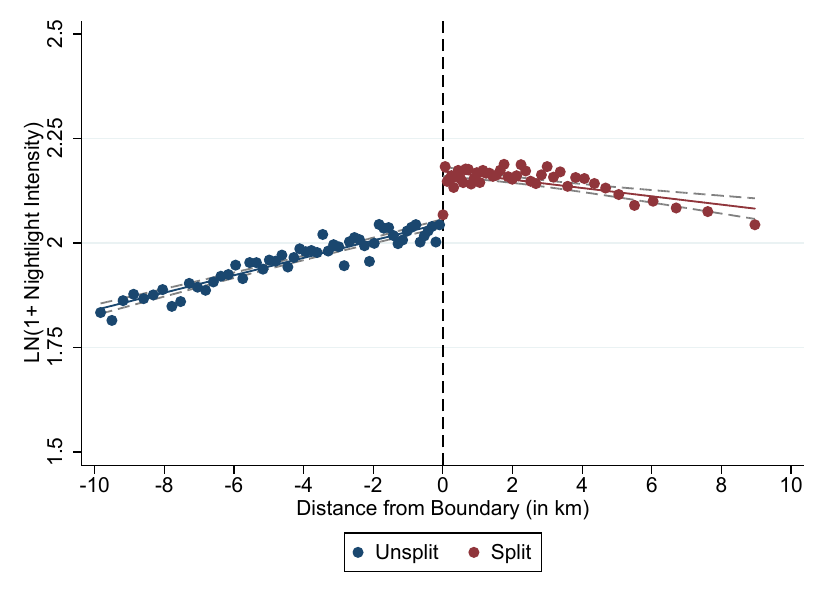}
    \caption{Nightlights}
    \label{fig_nl_bin}
    \end{subfigure} %
    \label{fig_rd_bin}
    \begin{minipage}{1.00\textwidth}
	\begin{center}
		\end{center}
		{\footnotesize 	The figure presents RD plots for our main outcomes and the mean value of each outcome variable at different bins along the running variable (distance to boundary) as well as with a local linear trend estimated separately, along with 95\% confidence intervals, on each side of the discontinuity. The unit of observation is a village. The red dots indicate villages in the split blocks and the blue dots indicate villages in the unsplit blocks. Figure \ref{fig_entry_bin} presents the RD plots for firm entry. Firm Entry is defined as the total number of firms that have entered a village from 2008 to 2016. Figure \ref{fig_nl_bin} presents the RD plots for nightlights. Nightlights is defined as the average nightlight intensity in a village from 2008 to 2016.  \par}
\end{minipage}
\end{figure}

This section examines the spatial difference in the propensity of firms to enter and local economic activity between split and unsplit blocks. To do this, we aggregate the primary variables -- number of new firms and night light intensity -- at the village level for the years post delimitation, i.e. 2008-2016. Figure \ref{fig_entry_bin} and \ref{fig_nl_bin} present the RD plots for firm entry and nightlight intensity, respectively, with distance to the boundary as the running variable. We observe a discontinuous increase in firm entry and nightlight intensity at the boundary when moving from an unsplit block to a split block.

Table \ref{sptd_firm_entry} reports estimates for specification \ref{rd} using the natural logarithm of 0.001 plus firm entry and nightlight intensity as the dependent variables in Panel A and B, respectively. We use different bandwidths of 5, 10, 20, and 50 km on either side of the boundary, separating a split block from an unsplit block in columns (1), (2), (3), and (4), respectively. Across all specifications, the coefficient of interest is positive and statistically significant. Our estimates indicate that villages just inside the split block exhibit 3\% higher firm entry and 7\% higher nightlight intensity relative to villages in unsplit blocks just outside the boundary. The estimate in Panel B is higher as it may represent the aggregate effect of multiple politicians on the formal as well as the informal economy. Appendix Table \ref{sptd_firm_entry_multi} reports cross-sectional heterogeneity based on the number of politicians in a split block. The magnitude of the coefficients increases in the number of politicians.

Using census data, we find 6\% higher employment in villages just inside a split block than in villages in unsplit blocks just outside the boundary (see Panel A of Table \ref{app_tab_employment}). Lastly, we document 5\% lower applications for unemployment benefits filed under National Rural Employment Guarantee Scheme (NREGA), suggesting lower demand for unemployment benefits (see Panel B of Table \ref{app_tab_employment}). This lower demand for NREGA is consistent with the \cite{gulzar2017politicians}, who show that multiple politicians are associated with lower unemployment benefits. 

\subsubsection{Exploiting Exogenous Block Boundaries}
Although the electoral constituency borders are drawn by a non-partisan committee to equate population across electoral constituencies, there might be a concern that the boundary separating two blocks could be drawn to place certain villages in split or unsplit blocks. We address this concern in two ways, using a subsample of boundaries defined by rivers and a subsample of straight-line-like boundaries.

\begin{figure}[H]
    \centering
    \caption{Block Boundaries Delimited by Natural Features}
    \begin{subfigure}{.45\textwidth}
    \centering
    \includegraphics[width = \textwidth]{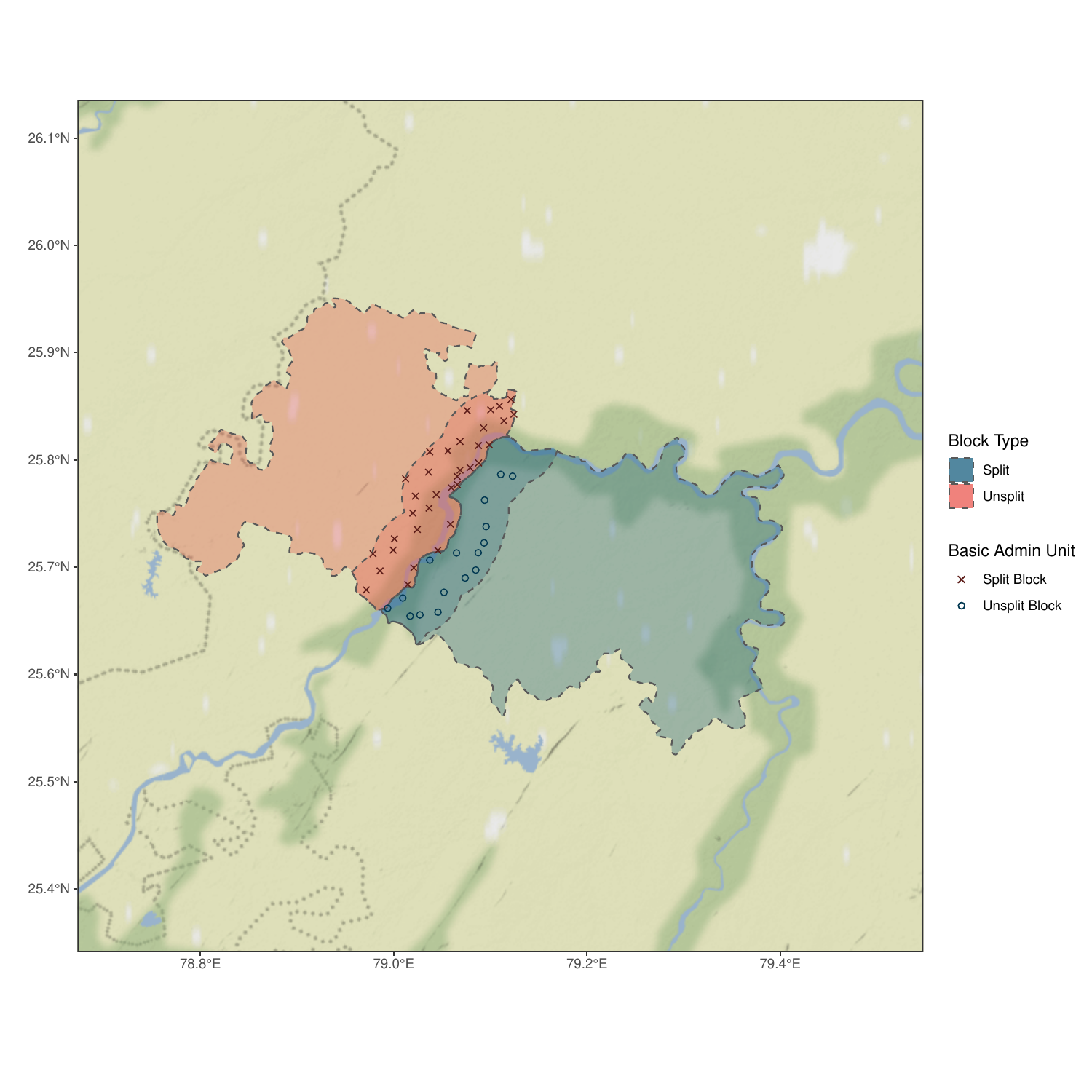}
    \caption{Example 1}
    \end{subfigure} %
    \begin{subfigure}{.45\textwidth}
    \centering
    \includegraphics[width = \textwidth]{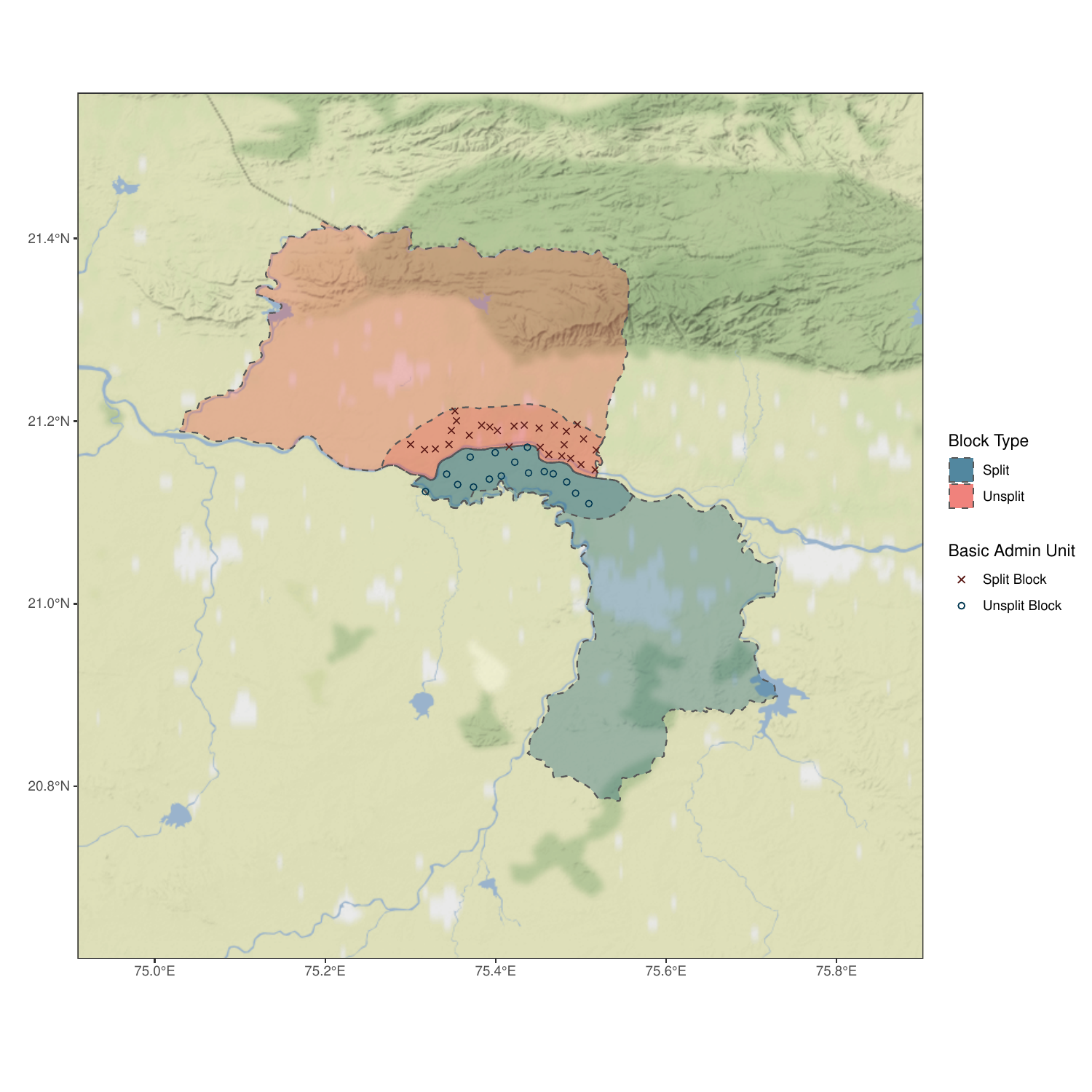}
    \caption{Example 2}
    \end{subfigure} %
     \begin{subfigure}{.45\textwidth}
    \centering
    \includegraphics[width = \textwidth]{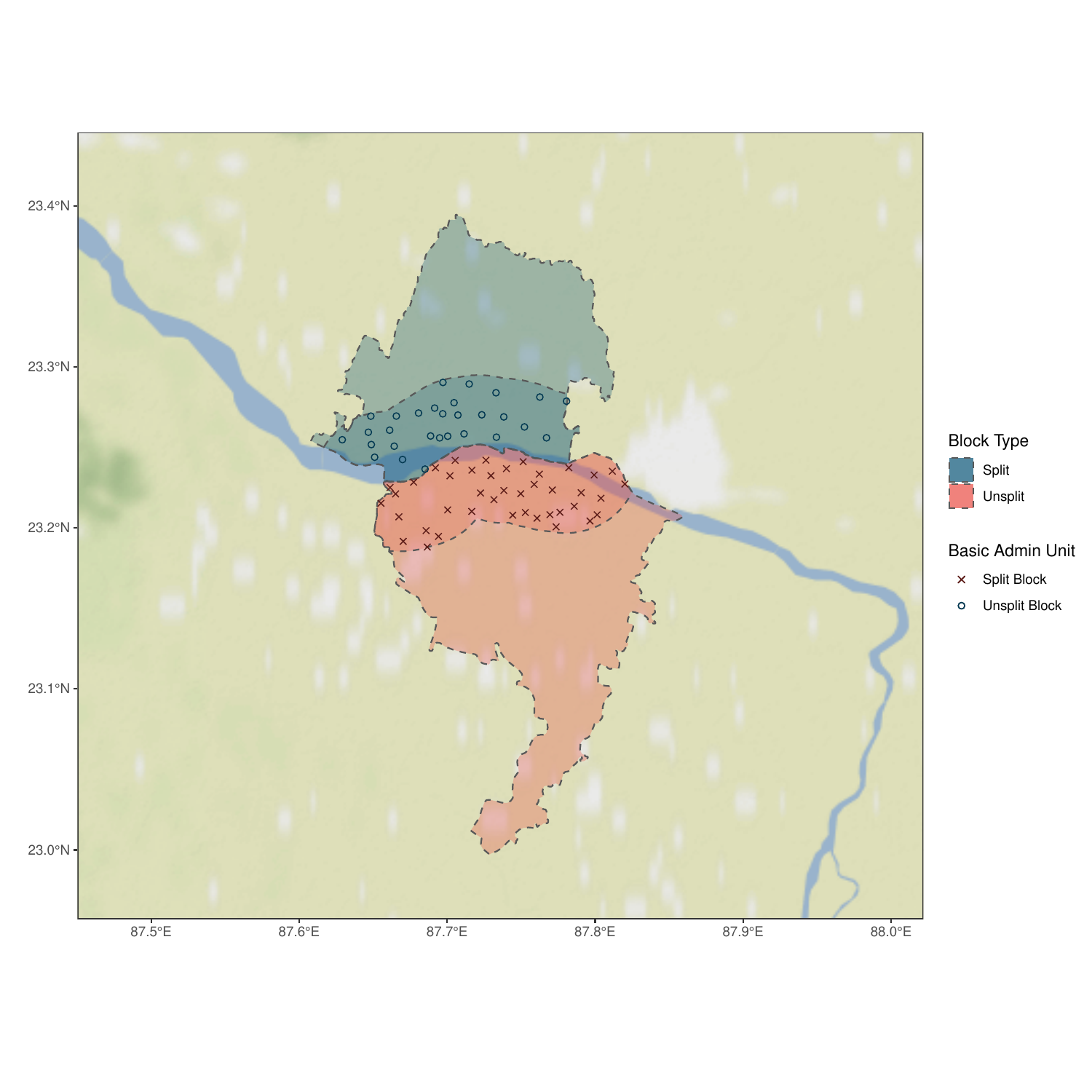}
    \caption{Example 3}
    \end{subfigure} %
    \begin{subfigure}{.45\textwidth}
    \centering
    \includegraphics[width = \textwidth]{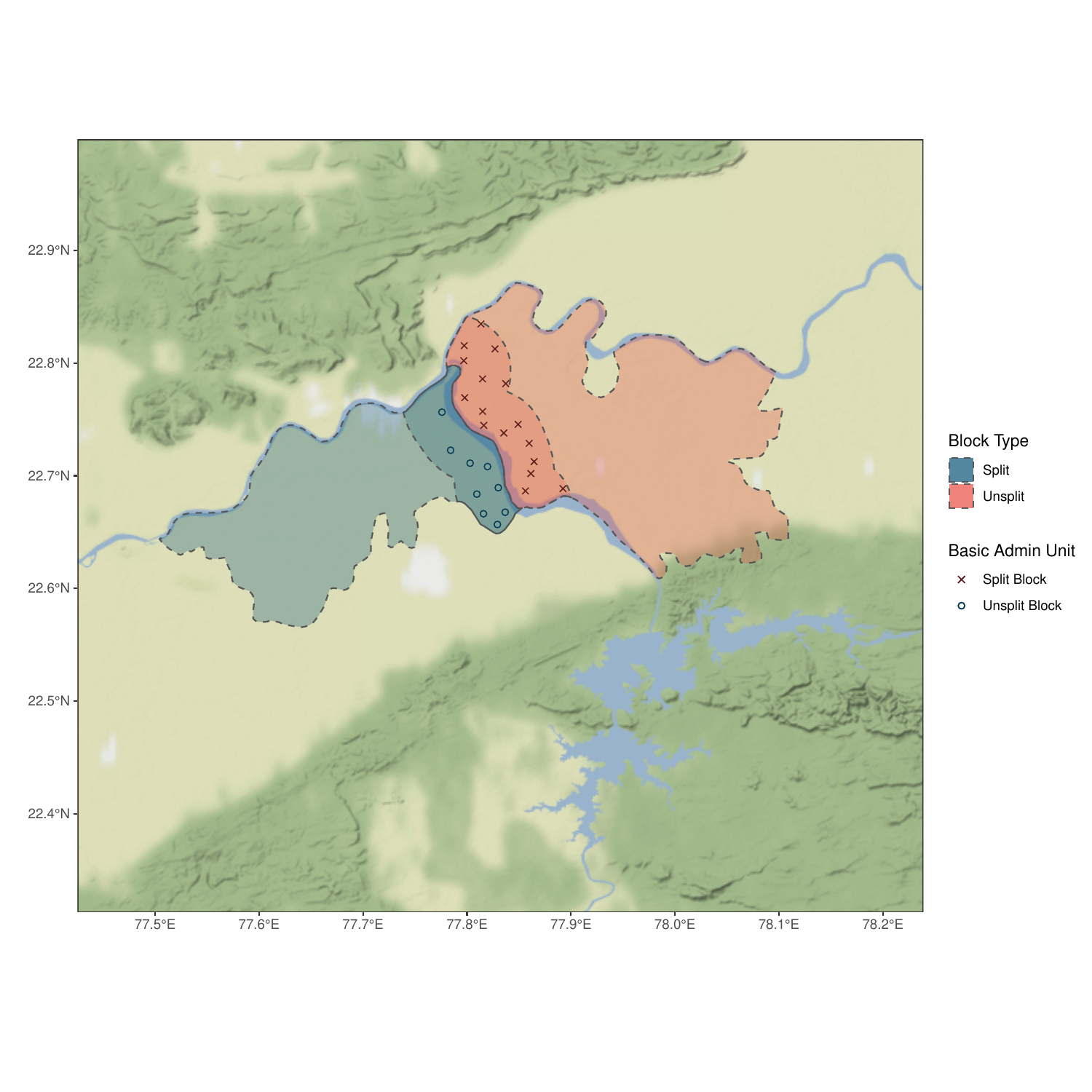}
    \caption{Example 4}
    \end{subfigure} %
    \label{fig_boundary_river}
    \begin{minipage}{1.00\textwidth}
	\begin{center}
		\end{center}
		{\footnotesize This figure presents four representative examples from the sample of boundaries that are delimited by geographic features, in this case, river segment. Example 1 illustrates River Betwa delimiting the boundary between the blocks, Bamaur and Moth, in the district of Jhansi, Uttar Pradesh. Example 2 illustrates River Tapi  delimiting the boundary between the blocks, Jalgaon and Chopda, in the district of Jalgaon, Maharashtra. Example 3 illustrates River Damodar  delimiting the boundary between the blocks, Galsi II and Khandaghosh, in the district of Purba Bardhaman, West Bengal;. Example 4 illustrates River Tawa delimiting the boundary between the blocks, Babai and Hoshangabad, in the district of Hoshangabad, Madhya Pradesh.    \par}
\end{minipage}
\end{figure}

First, we restrict the sample to boundaries between split and unsplit blocks defined by geographic features, in this case, rivers, and river basins. The intuition of this estimation is that boundaries defined by geographic features are likely to be devoid of selective sorting. Figure \ref{fig_boundary_river} presents representative examples illustrating block boundaries determined by a river segment. Panel A of Table \ref{tab_exo_boundary} reports the results using a sample of units partitioned by boundaries defined by rivers and river basins. The estimate is positive and statistically significant indicating that our baseline results are unlikely to be plagued by selective sorting of villages into split and unsplit blocks.

Second, we restrict estimation to a sample of split and unsplit blocks partitioned by straight-line-like borders. Appendix figures \ref{straight_line_ex} and \ref{wiggly_line_ex} present a representative example of a straight-line-like and a squiggly boundary segment, respectively. The intuition behind this estimation is that straight-line-like boundaries are more likely to be arbitrary than squiggly ones as discussed in \cite{polsby1991third} and \cite{alesina2011artificial}. We define a boundary as a straight line if the Sinuosity Index (SI) measure for the boundary is less than 1.05. Panel B of Table \ref{tab_exo_boundary} reports the results restricting sample to units partitioned by boundaries that follow a straight line. We find qualitatively similar results using the sub-sample of straight-line-like boundaries.\footnote{Note that the magnitude of the estimate using both the river-defined and straight-line-like boundaries is greater than the magnitude reported in Panel A of Table \ref{sptd_firm_entry} indicating the bias is likely to be positive. This suggests that selective sorting may be under-biasing the true effect.} 

\subsubsection{Robustness}
\label{sec_result_robustness}
We conduct various robustness tests to ensure our results are not driven by: a particular econometric specification, specific sample or transformation of the dependent variable, differences across villages along the boundary and covariates, firm exit, spatial auto-correlation, and spurious correlation.

First, we verify our results are robust to alternative econometric specifications. Appendix Table \ref{appendix_tab_poly2} shows the results are robust to using a local quadratic polynomial instead of a local linear polynomial. The results are robust to using a uniform kernel as shown in appendix Table \ref{appendix_tab_uni_kernel}. Appendix Table \ref{tab:appendix_tab_spat} shows that the results are robust to modification of $f(.)$ in specification \ref{rd} to be a bivariate function of the latitude and the longitude of the village. Appendix Figure \ref{fig_cont_bw} shows the results are robust to choosing any bandwidth between 2 km and 50 km. Appendix Table \ref{appendix_tab_coneley} reports that the results are robust to inference based on \cite{conley1999gmm} standard errors, which allows for spatial dependence of an unknown form, with cutoff distances of 2 km, 5 km, 10 km, and 20 km. Moreover, our inference is robust to alternative clustering at village, district, boundary, and state level as shown in Appendix Table \ref{appendix_tab_alt_cluster}.

Second, we verify that the results are not an artifact of the sample and transformation of the key dependent variable. Appendix Table \ref{appendix_tab_alt_sample_ihs} shows the results are robust to dropping small and large blocks, indicating the outliers are unlikely to drive our results.\footnote{A block is defined as a small block if the total number of firms entering the block during our sample period is less than 0, 5, 10, and 20. A block is defined as a large block if the total number of firms entering the block during our sample period is greater than 1000.} Panel A of appendix Table \ref{appendix_tab_alt_sample_ihs_ppml} shows results are robust to re-estimating our baseline specification using Poisson regression.\footnote{\cite{chen2022log} argue that manipulations of the variables such as adding a number before log transformation may affect the empirical results and suggest performing Poisson regression.}. Panel B of appendix Table \ref{appendix_tab_alt_sample_ihs_ppml} shows results are robust to using inverse hyperbolic sine (IHS) transformation of the key dependent variable, in alternative samples, instead of $LN(0.001+x)$ transformation.\footnote{The IHS transformation has been used to overcome problems in regression analyses with right-skewed censored dependent variables like ours \citep{carboni2012empirical} and to address the issue of log transformation with zeros \citep{burbidge1988alternative}.} In appendix Table \ref{tab:appendix_tab_nonpool}, we verify that the results are not driven by aggregating observations across time at the village level, by running a non-pooled regression using village-year as the unit of observation. Lastly, we verify that the results are not driven by using the village as the primary unit of analysis. Villages are a predetermined unit of aggregation and could potentially influence our results. For robustness, we create arbitrary grids of cell size - $1 \times 1$, $2 \times 2$, and $5 \times 5$ and aggregate data on firm entry at the cell level. Appendix Table \ref{tab:appendix_tab_grid} shows the results are robust to using these arbitrarily defined cells.

Third, we address the concern that the villages located close to the border on either side of the boundary can be different in characteristics, and thereby, the observed impact of having multiple politicians would just be an artifact of such difference in characteristics. Our first defense to this concern is in Table \ref{tab_summary} where we show various pre-existing village specific characteristics such as population, education, health, infrastructure, and geography are largely similar across villages along the border. Moreover, we document that firm entry and nightlights vary smoothly across these boundaries before the 2008 delimitation (see appendix Table \ref{tab:appendix_tab_balance_firm_nl}). Next, we include several covariates in the regression specification \ref{rd} to account for several other differences across villages and blocks. These covariates include the 2001 Census population, geographic area of the village, the geographic distance of the village from the district headquarters, and block compactness measured as in \cite{harari2020}.\footnote{\cite{bardhan2002} and \cite{michalopoulos2014national} highlight the role of a region's distance to capital as a determinant of growth. \cite{harari2020} highlights the impact of a city's shape on growth - the more compact the shape of the city, the greater is the growth potential.} Appendix Table \ref{appendix_tab_controls} shows the results are robust to inclusion of these covariates. Another potential concern with our analysis is that the bordering split and unsplit blocks may systematically differ by the quality of the politician. We address this issue by augmenting our baseline specification with constituency fixed effects. Appendix Table \ref{tab:appendix_ac_fe} reports these results and finds similar results, indicating differences in politician quality are unlikely to drive our results.

Fourth, we verify that the firms that enter split blocks are not more likely to exit. This test addresses the concern that additional firms entering split blocks are of poor quality and hence less likely to survive. We examine the probability of a firm exiting within six months, one year, and two years of its entry as a function of the block being split or unsplit. Appendix Table \ref{tab:appendix_prob_firm_exit} presents these results. The estimate of interest is negative and mostly statistically insignificant. This indicates that firms entering in split blocks are not more likely to exit than firms entering in unsplit blocks. Hence, we can rule out the possibility that firms entering unsplit blocks are of systematically lower quality and less likely to survive.

Fifth, we show our main regression is capturing an effect that only appears as we cross the actual boundary between split and unsplit blocks. As a falsification test, we rerun the analysis using arbitrary borders. In particular, we draw fake borders inside each block in our sample such that these borders divide a block into three concentric zones of equal area, as shown in appendix Figure \ref{app_fig_schematic}. Appendix Table \ref{tab_app_falsification} presents the results of comparing villages within a block that are only separated by fake borders and have the same number of politicians. The coefficients are insignificant in every case, both economically and statistically. As a placebo test, we randomly define a block as split or unsplit and rerun our analysis using the fake split status assignment. We repeat this exercise and estimate the RD coefficient 10,000 times. Figure \ref{fig_placebo} reports these results and finds insignificant results. Hence, the null effects in the falsification and the placebo test indicate our baseline estimation captures an effect that only appears as we cross the actual boundary between split and unsplit block and is unlikely to be spurious or driven by spatial autocorrelation.

\subsection{Differences-in-Discontinuity Results}\label{results_timedid}
Next, we examine the effect of the exogenous change in the number of politicians governing a block, following the 2008 delimitation by estimating specifications  \ref{did_1} and \ref{did_2}. The specification includes village and boundary $\times$ year fixed effects. The estimate is identified by comparing villages in the treatment group before and after delimitation with the contiguous villages in the control group, while controlling for boundary-specific shocks to investment opportunities. 

Panel A of Table \ref{sptd_firm_entry_switcher} presents estimates from specification \ref{did_1}, focusing on the treatment group of blocks that transitioned from unsplit (single politician) to split (multiple politicians) after the 2008 delimitation. The control group consists of contiguous blocks that remained unsplit throughout. The results show an increase in firm entry by 1.1\% - 1.6\% when a block transitions from being governed by a single politician to multiple politicians. Panel B of Table \ref{sptd_firm_entry_switcher} reports estimates from specification \ref{did_2} focusing on the treatment group of blocks that transitioned from split to unsplit after the 2008 delimitation. The control group consists of contiguous blocks that remained split throughout. In contrast to the estimate presented in Panel A, the estimate in Panel B is negative indicating a decline in firm entry by 0.7\% - 1.0\% when a block switches from multiple politicians to a single politician.

\begin{figure}[htbp]
    \centering
    \caption{Differences-in-Discontinuity Design}
    \begin{subfigure}{.49\textwidth}
    \centering
    \includegraphics[width = \textwidth]{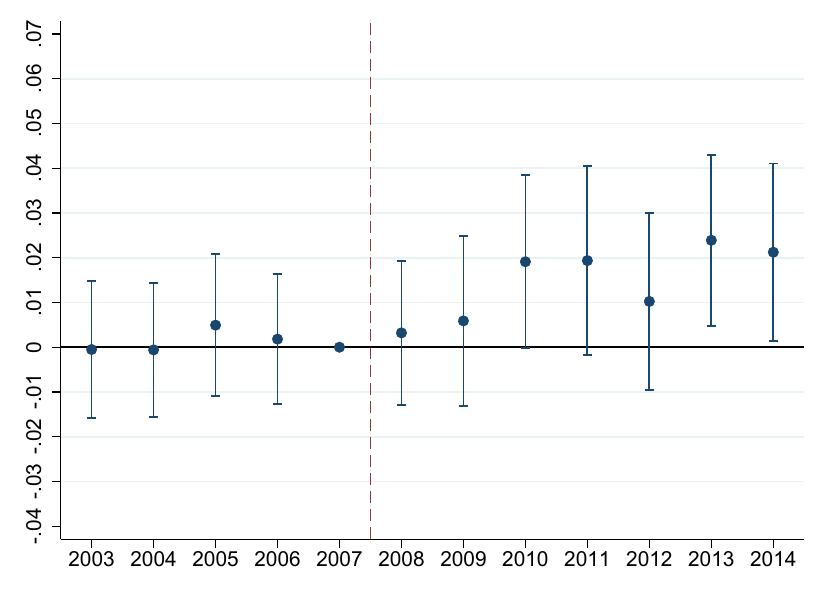}
    \caption{Unsplit $\rightarrow$ Split}
    \label{figure_switcher_1}
    \end{subfigure} %
    \begin{subfigure}{.49\textwidth}
    \centering
    \includegraphics[width = \textwidth]{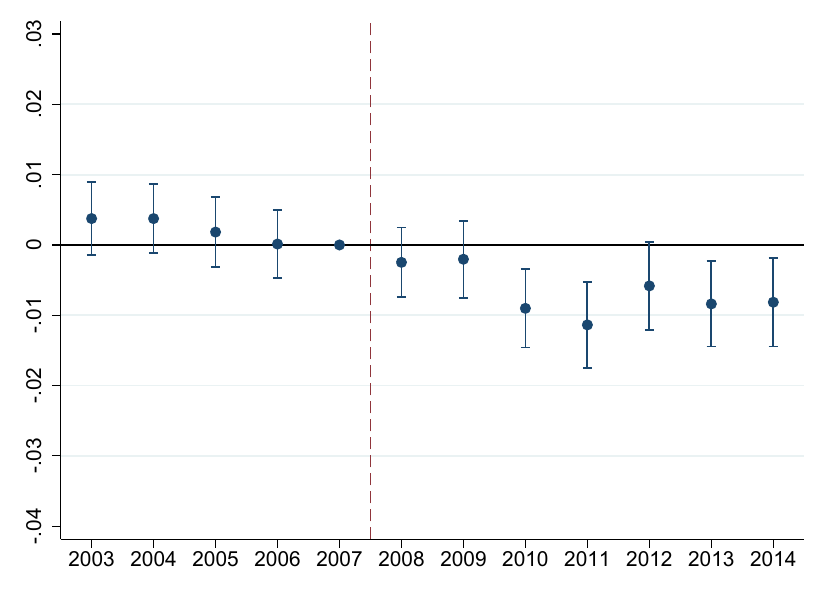}
    \caption{Split $\rightarrow$ Unsplit}
    \label{figure_switcher_2}
    \end{subfigure} %
    \label{figure_switcher}
    \begin{minipage}{1.0\textwidth}
	\begin{center}
		\end{center}
		{\footnotesize 	The figure presents the yearly RD estimates between split and unsplit blocks for each year before and after the 2008 delimitation. Panel (a) plots the yearly RD estimates from specification \ref{did_1} between the treated blocks that switched from being unsplit to split following the 2008 delimitation and the control group that remained unsplit throughout. Panel (b) plots the yearly RD estimates from specification \ref{did_2} between the treated blocks that switched from being split to unsplit following the 2008 delimitation and the control group that remained split throughout. The unit of observation is a village-year. The dependent variable is the natural logarithm of 0.001 plus the number of new firms in a village during the year. The estimates are plotted with the 95\% confidence interval based on clustering the standard errors at the block level. The specification is estimated for the bandwidth of 10 km. All regressions include boundary $\times$ year and village fixed effects, a local linear specification estimated separately on each side of the boundary, and use a triangular kernel.  \par}
\end{minipage}
\end{figure}

Figure \ref{figure_switcher}, Panels A and B, present the estimates of the dynamic specification associated with equations \ref{did_1} and \ref{did_2}, respectively. The dynamic plots offer two key takeaways. First, we do not find discontinuity between the treatment and the control group before 2008, indicating an absence of pre-trends. Second, we find evidence of discontinuity between the treatment and the control group after 2008, gradually developing over time.

We note that although the delimitation announcement took place in 2008, the actual changes in assembly constituency boundaries occurred in the subsequent state election. However, employing an event study design using the actual election year after 2008 would violate the no-anticipation assumption. We argue that a part of the impact of the boundary changes is internalized at the time of the announcement, as firms are forward-looking agents. Therefore, an empirical design exploiting the announcement of new electoral boundaries is more appropriate. We verify this by augmenting the previously estimated specification from Table \ref{sptd_firm_entry_switcher} with a Treatment $\times$ Post-Election variable. The results presented in Appendix Table \ref{delim_election} provide support to our argument.


\section{Mechanism}
\label{mech}

This section investigates the mechanism through which multiple politicians foster firm entry. Firm entry involves engaging with local governance apparatus for regulatory compliance and resource access. Politicians hold discretion over how such policies are interpreted and implemented, creating opportunities for rent extraction, affecting firm entry. When multiple politicians share governance over the same area, they can observe each other's actions and impose mutual checks and balances, limiting individual rent extraction. We first establish that checks and balances are operative, by leveraging institutional features that induce variation in mutual oversight intensity. We show that firm entry and politician rent extraction respond to this variation. We then provide further evidence that checks and balances potentially enhance firm entry by examining domains where unchecked political discretion deters firm creation.

\subsection{Checks and Balances}

The institutional structure facilitates checks and balances in split blocks, where multiple electoral constituencies overlap across the same administrative block. The MLAs governing these constituencies operate alongside the same BDO and are required to participate jointly in block-level Panchayat Samiti meetings. Such an arrangement places politicians in a common information environment where each can observe the others' interactions with bureaucrats, creating mutual observability of administrative decisions.

We begin by providing suggestive evidence that checks and balances operate by leveraging institutional features that generate variation in the intensity of checks and balances. First, we rely on partisan composition of politicians governing split blocks. Politicians from different parties have stronger incentives to monitor one another, as they benefit electorally from exposing rivals' wrongdoings and face greater barriers to collusive arrangements. Consistent with this, we find that split blocks governed by non-aligned politicians — those from different parties — experience significantly higher firm entry compared to those governed by aligned politicians from the same party. Second, we leverage variation in the share of split blocks across constituencies. Politicians governing constituencies with a higher share of split blocks face greater oversight, as more of their administrative decisions occur in blocks that overlap with other constituencies where rival politicians can observe their actions. Consistent with this, we find that politicians governing such constituencies exhibit lower gains private wealth.

\subsubsection{Non-Alignment Across Politicians}

We first examine variation in the intensity of checks and balances generated by partisan composition of politicians governing split blocks. Politicians from different parties have stronger incentives to monitor each other, as they gain electoral advantage by exposing rivals' corrupt behavior. Moreover, ideological differences and political animosity between parties create barriers to collusive rent extraction, making coordinated corruption more difficult to sustain \citep{spenkuch2021ideology, kempf2021partisan}. Conversely, politicians from the same party can more easily coordinate rent-seeking arrangements because they share electoral interests and face fewer coordination barriers. Partisan composition thus generates variation in the intensity of checks and balances across split blocks. We examine whether this variation corresponds with patterns of firm entry. 

The variation in political alignment operates both across split blocks at any given time and within split blocks over time, as party composition of governing MLAs changes around elections. We estimate specification \ref{rd} augmented with the interaction between split blocks and the fraction of non-aligned politicians. The specification includes boundary × year and village fixed effects, relying on within-village variation after 2008. The coefficient of interest is estimated from changes in the fraction of non-aligned politicians over time while controlling for boundary-specific investment opportunities. Table \ref{alignment} reports estimates showing that split blocks with non-aligned politicians experience significantly greater firm entry than those with aligned politicians. These estimates are consistent with checks and balances operating to increase firm entry

\subsubsection{Private Returns of Politicians}

We provide further evidence that checks and balances operate by examining whether politicians subject to stronger oversight accumulate less private wealth. Politicians can extract rents through their discretionary authority over administrative processes, and such rent extraction often manifests in personal wealth accumulation. Electoral law in India requires candidates to file sworn asset affidavits at the time of nomination, allowing us to measure changes in private wealth over a politician's term in office. Following \cite{bhavnani2012using} and \cite{fisman2014private}, we proxy for rent extraction through the winner's premium, defined as the differential asset growth of election winners relative to runners-up over a term. By comparing similar candidates who differ only in electoral success, this measure isolates wealth gains from holding office. 

A politician's constituency contains multiple blocks. Haphazard overlap of boundaries implies that constituencies vary in the extent to which they share split blocks with the adjacent constituencies. Politicians governing constituencies with more split blocks face greater oversight from politicians in the overlapping constituencies. We test whether the winner's premium varies across constituencies with varying intensity of such mutual oversight. A key challenge in estimating the winner's premium is accounting for unobserved skills or resources that politicians possess independent of electoral success. We address this by restricting our sample to elections with a margin of victory below 5 percent. This allows us to compare politicians with plausibly similar characteristics who differ primarily in whether they won office.

Table \ref{private_return_table} presents estimates examining how rent-seeking (winner's premium) is affected by the intensity of checks and balances. We construct \textit{Splitness}, which captures the intensity of checks and balances that a constituency-level politician faces. Splitness is measured two ways: (i) a continuous measure—the normalized share of the constituency’s population living in split blocks—and (ii) a binary indicator equal to one if this share exceeds 50\% and zero otherwise. Columns 1 and 2 report estimates separately for constituencies with differing intensity of checks and balances. In constituencies with low intensity of checks and balances (Splitness = 0), we observe a 16\% increase in the assets of the winner relative to the runner-up. We find no statistically significant winner's premium in constituencies with high intensity of checks and balances (Splitness = 1). Columns 3 and 4 utilize the complete sample and include a specification with an interaction between the Winner and the Splitness variable to test whether the winner's premium varies with the intensity of checks and balances. The negative estimate associated with the interaction term indicates that the winner's premium decreases in constituencies with a higher share of split blocks. These estimates suggest that greater intensity of mutual oversight reduces politicians' private wealth accumulation, consistent with checks and balances operating to constrain rent extraction.

\subsection{How Do Checks and Balances Increase Firm Entry?}

We provide further evidence for checks and balances in operation by examining domains where unchecked political discretion impedes firm entry. First, we document reduced preferential treatment in market access, showing increased entry of firms outside politicians' co-ethnic networks and lower entry in industries prone to cronyism. Second, we provide evidence of reduced regulatory barriers, finding shorter approval delays and higher entry in industries with greater regulatory intensity where unchecked power typically creates hold-up. Third, we demonstrate improved efficiency in public good provision, which serves as a productive input for private firms while representing a domain where political rent extraction is prevalent.

\subsubsection{Reduced Preferential Treatment}
\label{subsec:preferential_treatment}

Firm entry typically requires obtaining regulatory permits across different levels of governance, including at the block level where local politicians direct implementation through bureaucrats. When political institutions provide insufficient constraints on such authority, politicians gain discretionary power that often leads to preferential allocation based on personal connections. Such preferential treatment can manifest through patronage networks that favor co-ethnic individuals and cronyistic arrangements that benefit politically valuable industries. Checks and balances among politicians might therefore foster firm entry by constraining preferential allocation arising from such discretionary authority.

India's political landscape is characterized by ethnic networks that influence resource allocation \citep{chandra2007ethnic}, while well-documented quid pro quo arrangements exist between politicians and rent-seeking industries \citep{kapur2013quid, crabtree2018billionaire}. Against this backdrop, we examine firm entry across contexts in which preferential treatment would create differential advantages to connected agents. First, we study relative entry of firms with directors sharing caste identity with local politicians, where co-ethnic connections might facilitate preferential access. Second, we exploit heterogeneity across industries prone to cronyism, where rent-seeking sectors typically receive political favoritism.

\paragraph{Lower Entry of Patron Firms}
\label{sec:patronage_analysis}

Political patronage often operates through shared identities that facilitate preferential access to resources. In India, \textit{jati} — endogamous subcastes that constitute narrowly defined groups — represents a salient identity around which social and political networks form \citep{munshi2019caste}. We proxy for patronage by identifying firms led by entrepreneurs that share the same jati with local politicians, classifying such firms as potentially connected through patronage. 

Surnames in India often carry identity markers that signal jati affiliation. We use the names of firm directors and politicians to deduce whether they share common identity markers. We rely on the People of India Project \citep{singh1996communities}, a comprehensive anthropological study that systematically documented 2,205 jatis with their associated surnames at the state level. The salience of jati as an identity is particularly pronounced for the Hindu community compared to other religious groups, and surname-caste associations are more systematic in North Indian states compared to South Indian states.\footnote{South Indian naming conventions reduce the informativeness of surnames for caste inference. In many southern states, people commonly use patronymics and initials (e.g., father's given name as an initial) rather than fixed, hereditary surnames; caste suffixes that were historically used have been legally discouraged and socially abandoned; and names may include neutral given names or village identifiers. By contrast, North Indian names more often retain inherited family surnames that correlate with jati, making surname-based inference more reliable.} This limits our sample to blocks in North Indian states governed by Hindu politicians, where we can reliably establish caste connections through surname analysis. Processing Indian names presents substantial complexity, hence we leverage recent advances in language processing tools to identify potential patronage connections between politicians and firm directors.\footnote{Challenges include varying surname conventions across states, non-standardized naming formats, transliteration differences creating multiple spelling variants, and complex name structures.} A detailed description of our methodology is provided in Appendix \ref{sec:data_patronage}.

We test whether patterns of firm entry are consistent with reduced patronage by estimating specification \ref{rd} augmented with the interaction term Split × Connected and include boundary × connected and village fixed effects. Connected is an indicator variable equal to one for firms with directors sharing caste identity with the local politician and zero otherwise. Estimates in Table \ref{tab:patronage_test} show that patronage-connected firms experience 0.9 to 1.3 percentage points lower entry relative to non-connected firms in split blocks compared to unsplit blocks. This pattern suggests that the relative advantages of patronage-connected firms diminish in split blocks, consistent with checks and balances constraining preferential treatment when multiple politicians potentially provide mutual oversight.

\paragraph{Lower Entry in Industries Prone to Cronyism}

Cronyistic arrangements, where industries offer rents in exchange for preferential treatment, are symptomatic of political favoritism. Industries vary in their vulnerability to cronyism, with sectors characterized by high rents, extensive state interaction, and opaque regulations being structurally more susceptible. We rely on such heterogeneity to examine relative entry patterns across industries that vary in their propensity for cronyism. We capture this by estimating specification \ref{rd} augmented with interactions between Split and Crony. We also include village and boundary × industry fixed effects. Crony is an indicator for industries vulnerable to cronyism.\footnote{We use the industry-level cronyism index created by \emph{The Economist} using \emph{Transparency International} methodology. See \href{https://www.economist.com/international/2014/03/15/planet-plutocrat}{Planet Plutocrat}. Appendix Table \ref{tab_app_crony_classification} lists industries with higher vulnerability to cronyism in our sample.} Estimates in Table \ref{tab:crony_industries} show that the differential entry between industries vulnerable to cronyism and those less vulnerable is 6 percentage points lower in split blocks compared to unsplit blocks. This pattern suggests that multiple politicians foster firm entry by constraining (rather than enhancing) cronyistic arrangements, consistent with checks and balances operating to limit political favoritism towards connected sectors.

\subsubsection{Reduced Hold-Up in Regulatory Implementation} \label{cross section}

Political discretion in the issuance of regulatory permits enables hold-up opportunities by inducing artificial delays. Such regulatory implementation is a documented source of high entry costs in developing economies \citep{djankov2002regulation, djankov2009regulation, campos2010corruption}. Checks and balances among politicians might therefore foster firm entry by reducing costs arising from such discretionary implementation of regulations.

We examine firm entry across contexts in which regulatory discretion would create distinct cost burdens. First, we study approvals for regulatory permits, where discretionary delays impose direct costs. Second, we exploit heterogeneity across industries in regulatory intensity, where industries requiring more approvals face higher entry barriers. Third, we examine size-based heterogeneity, where smaller firms' entry is disproportionately affected by fixed regulatory costs because of capital constraints.

\paragraph{Faster Regulatory Approvals: } Approval times for regulatory permits can reflect costs arising from discretionary implementation. We examine whether the presence of multiple politicians reduces such discretionary delays by measuring the time required for private firms to receive regulatory approvals for their projects. Using the CapEx database, we compute approval time as the difference between project announcement and initial approval dates. Table \ref{tab_delay_time} reports estimates examining blocks that transition between single and multiple politician governance compared to blocks maintaining constant governance arrangements, following equations \ref{did_1} and \ref{did_2}. We find that approval times decrease by 80\% when blocks switch from single to multiple politician governance, and increase by 57\% in the reverse transition. These patterns suggest that checks and balances may reduce discretionary delays in regulatory implementation.

\paragraph{Higher Entry in Regulated Industries}
Regulatory complexity varies across industries. Those with more complex regulatory requirements are potentially more vulnerable to administrative delays that create barriers to entry. We rely on this heterogeneity to examine relative entry patterns across industries with different regulatory intensities. To capture this, we estimate specification \ref{rd} augmented with interactions between Split and Regulated. We also include village and boundary × industry fixed effects. Regulated is an indicator for industries subject to extensive regulatory requirements.\footnote{We define high regulatory cost industries broadly following \cite{pittman1977market} methodology augmented with India-specific factors from \cite{awasthi2019cl}. Appendix Table \ref{tab_app_regulated_classification} lists classified industries.} Estimates in Table \ref{tab:regulated_industries} show that the differential entry between highly regulated and less regulated industries is 3.5 percentage points higher in split blocks relative to unsplit blocks. This pattern suggests that checks and balances may promote entry of firms in industries where regulatory hold-up is more likely to create entry barriers.

\begin{figure}[H]
	\centering
	\caption{RD Estimates by Size Deciles}
	\includegraphics[width=0.5\textwidth]{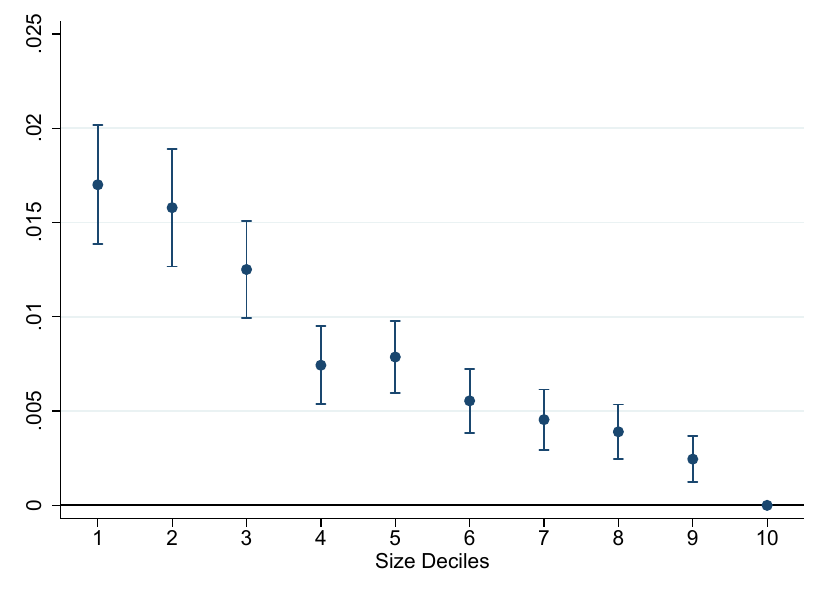}
	\label{fig_size}
	\begin{minipage}{1.0\textwidth}
{\footnotesize This figure presents the estimates for specification \ref{rd} using the natural logarithm of 0.001 plus firm entry as the dependent variable. The estimates are obtained by running the baseline specification \ref{rd} augmented to include the interaction terms of size-decile and split binary variable. All estimates are reported relative to the estimate for the firms in 10$^{th}$ decile. We measure firm size based on the book value of assets reported by the firm in the year of entry. We group firms into size deciles with 1$^{st}$ decile referring to firms in the smallest size group, and  10$^{th}$ decile referring to the firms in the largest size group. The unit of observation is a village-size decile. The dependent variable is the natural logarithm of 0.001 plus the number of new firms in a village within a  size-decile. The estimates are plotted with the 95\% confidence interval based on clustering the standard errors at the block level. The specification is estimated for the bandwidth of 10 km. All regressions include village and boundary $\times$ size decile fixed effects, a local linear specification estimated separately on each side of the boundary, and use a triangular kernel. \par}
	\end{minipage}
\end{figure}

\paragraph{Higher Entry for Small Firms} Regulatory compliance often involves a fixed cost component that impose disproportionate barriers on smaller firms due to their capital constraints. When regulatory costs decline through enhanced checks and balances, smaller firms ought to experience relatively higher firm entry. We group firms into size deciles, with the first (tenth) decile referring to firms in the smallest (largest) size group. We augment regression specification \ref{rd} with interaction terms between Split and each size decile. We also include village and boundary × size decile fixed effects. Figure \ref{fig_size} shows that differential entry is higher for smaller firms, with the effect decreasing monotonically with firm size. This pattern suggests that checks and balances operate to reduce regulatory barriers with smaller firms experiencing disproportionately larger improvements.

\subsubsection{Improved Efficiency in Public Goods Provision} \label{efficiency}

Public goods serve as productive inputs for private firms \citep{munnell1992policy}. Provision of public goods in developing countries is frequently characterized by inefficient delivery, cost overruns, and resource misallocation due to rent extraction by politicians and bureaucrats \citep{olken2007monitoring}. We examine two domains — provision of roads and electricity — that represent key inputs to production while also being sectors prone to political rent extraction \citep{lehne2018building, mahadevan2024price}. We provide evidence that patterns of public goods provision are consistent with improved efficiency when multiple politicians govern the same jurisdiction, suggesting that mutual checks and balances may enhance the delivery of such essential productive inputs.

The \textit{Pradhan Mantri Gram Sadak Yojna} (PMGSY) is a large nationwide road construction program. While provisioned by the Central Government, implementation is overseen by local authorities, creating opportunities for rent extraction during the construction process. The program has faced significant corruption allegations, with the Comptroller and Auditor General reporting construction delays, cost overruns relative to initial bids, and award of contracts to dubious contractors.\footnote{Standing Committee on Rural Development and Panchayati Raj, "Pradhan Mantri Gram Sadak Yojana (PMGSY)," Sixteenth Lok Sabha, March 20, 2017.} We examine whether the presence of multiple politicians in a jurisdiction is associated with patterns consistent with improved efficiency in road construction within the program. We measure cost overruns — our proxy for implementation efficiency — as the ratio of actual implementation costs to estimated costs at the time of tender.

We rely on the 2008 Delimitation and use a differences-in-discontinuity design discussed previously in Section \ref{results_timedid}. Panels A and B of Table \ref{efficiency_infra} report estimates from specifications (\ref{did_1}) and (\ref{did_2}), respectively, augmented with boundary $\times$ year and block fixed effects.\footnote{We cannot use village fixed effects as only one road is constructed per village under the PMGSY scheme during our sample period spanning from 2001 until 2014.} Estimates in Panel A across all bandwidths are negative and statistically significant. These estimates indicate that the transition of a block from being governed by a single politician to multiple politicians is associated with a 16\% decline in cost overruns on a conservative note. Estimates reported in Panel B are positive and indicate an increase in cost overruns of 3-5\% when a block switches from being governed by multiple politicians to a single politician. Taken together, estimates from Table \ref{efficiency_infra} suggest that governance by multiple politicians is associated with improved efficiency in public goods provision, patterns consistent with enhanced mutual oversight constraining rent extraction during implementation processes.

We supplement this analysis by examining electricity provision for commercial use. Commercial power supply serves as another essential productive input that is state-controlled, creating scope for rent extraction through inefficient delivery. Appendix Table \ref{tab_app_efficiency_power} reports estimates using reliable commercial power supply as the dependent variable.\footnote{We define a village as having reliable power supply if it receives commercial electricity more than 70\% of the time.} The estimates are positive and statistically significant, indicating that villages in split blocks are 0.8 - 1\% more likely to receive reliable commercial power relative to villages in unsplit blocks. These complementary findings across road construction and electricity provision indicate that public goods delivery patterns are consistent with improved efficiency when multiple politicians govern the same jurisdiction. The evidence suggests that mutual oversight constrains rent extraction across different infrastructure domains that directly affect firm entry.

\section{Conclusion} \label{conc}
Political institutions play a vital role in shaping the economy. Hence, understanding what type of political institutions are relatively better at fostering economic growth is of utmost importance. In this paper, we examine a particular feature of political-institutional design -- multiple politicians governing an area.

Our empirical investigation is motivated by the theoretical ambiguity surrounding the potential effect of multiple politicians. On one hand, multiple politicians can hurt the local economy due to issues of coordination, free-rider problem, common agency problems, and too many grabbing hands. On the other hand, multiple politicians can improve the local economy by reducing the overall concentration of power, imposing checks and balances on each other, bringing different skills to the table, and the division of labor.

Using a geographic regression discontinuity design across a boundary separating a unit governed by a single politician and multiple politicians, we show that multiple politicians improve the local economy, evidenced by greater firm entry and economic growth. Additionally, we use a differences-in-discontinuity design using delimitation of electoral constituency in 2008 to show firm entry increases when an area transitions from being governed by a single politician to multiple politicians. Furthermore, we find the results are driven by an increase in checks and balances in the presence of multiple politicians, as manifested by a higher impact when they belong to different political parties and castes. We argue that checks and balances imposed by multiple politicians can serve as a safeguard against the abuse of power. This ensures that no single entity becomes overly dominant and that the government operates in a manner that ensures accountability and transparency. Consequently, this can lead to lower regulatory obstacles, less cronyism, and improved provision of public infrastructure, creating an economically favorable environment for the entry of new firms.

A few caveats of our paper are in order. First, our results do not suggest that multiple politicians solely influence economic growth through the checks and balances channel. There may be other concurrent channels at play, such as the division of labor and the presence of diverse skills. Our study provides evidence for a specific channel, but future research could delve into exploring these alternative channels. Second, our findings do not imply that increasing the number of politicians indefinitely will always lead to a positive effect on economic growth. Our results are bounded by our empirical setting, which allows us to infer the effect of increasing the number of politicians up to a certain point. Beyond this observed range, there might be a possibility that increasing the number of politicians further could yield no additional benefits or even have a negative impact on economic growth.

The results expand our understanding of a specific feature of the political-institutional design that is particularly relevant in understanding the effect of horizontal decentralization, common across several decentralized governance systems. Our results strengthen the faith in the conjecture that imposing checks and balances on agents with authority can result in better governance. Moreover, our results expand our understanding of the costs and benefits of multiple principals and a potential channel through which multiple principals can have a positive impact. The results may be useful in understanding the role of multiple managers that arise in a variety of situations, such as a firm or startup being managed by multiple managers or co-founders, the same entity being regulated by multiple regulatory authorities, doctoral students being advised by multiple chairs, courses being co-taught by multiple instructors, among others.


\clearpage
\newpage
\singlespacing
\bibliographystyle{ecma}
\bibliography{localbibliography}



\newpage

\clearpage
\newpage

\clearpage
\newpage

\clearpage
\newpage

\clearpage
\newpage
\begin{table}[ht!]
  \centering
  \caption{Balance on Pre-existing Characteristics}
  \label{tab_summary}
  \resizebox{\textwidth}{!}{ 
  \begin{threeparttable}
\begin{tabular}{lcccccc}
    \toprule
    \toprule
                         & \multicolumn{2}{c}{Full Sample} & \multicolumn{2}{c}{\begin{tabular}[c]{@{}c@{}}Split Block \\ (Pre Delim)\end{tabular}} & \multicolumn{2}{c}{\begin{tabular}[c]{@{}c@{}}Unsplit Block\\ (Pre Delim)\end{tabular}} \\ \cmidrule(lr){2-3} \cmidrule(lr){4-5} \cmidrule(lr){6-7}
                         & (1) & (2) &(3) & (4) & (5) & (6) \\ \cmidrule(lr){2-3} \cmidrule(lr){4-5} \cmidrule(lr){6-7}
          & \multicolumn{1}{c}{Estimate} & \multicolumn{1}{c}{SE} & \multicolumn{1}{c}{Estimate} & \multicolumn{1}{c}{SE} & \multicolumn{1}{c}{Estimate} & \multicolumn{1}{c}{SE} \\
    \midrule
\textbf{Population}     &         &          &         &          &         &          \\
Ln(0.001 + Population)                   & 0.0153  & (0.0251) & 0.0278  & (0.0301) & 0.0833  & (0.0840) \\
\textbf{Education}      &         &          &         &          &         &          \\
Ln(1 + Schools)                          & -0.0037 & (0.0036) & -0.0027 & (0.0044) & 0.0178  & (0.0139) \\
Ln(1 + Colleges)                         & -0.0002 & (0.0003) & -0.0003 & (0.0004) & -0.0014 & (0.0011) \\
\textbf{Demographics}   &         &          &         &          &         &          \\
Ln(0.001 + Female)                       & 0.0123  & (0.0245) & 0.0257  & (0.0295) & 0.0825  & (0.0822) \\
Ln(0.001 + Lower Caste)                  & 0.0172  & (0.0325) & 0.0237  & (0.0396) & 0.1620  & (0.1172) \\
\textbf{Health}         &         &          &         &          &         &          \\
Ln(1 + Hospital)                         & -0.0001 & (0.0005) & 0.0000  & (0.0006) & -0.0017 & (0.0016) \\
Ln(1 + Dispensaries)                     & 0.0014  & (0.0012) & 0.0009  & (0.0015) & -0.0005 & (0.0027) \\
\textbf{Infrastructure} &         &          &         &          &         &          \\
Ln(1 + Banks)                            & 0.0005  & (0.0012) & -0.0007 & (0.0013) & 0.0038  & (0.0070) \\
Ln(1 + Cinemas)                          & 0.0008  & (0.0011) & -0.0012 & (0.0011) & 0.0047  & (0.0068) \\
Ln(1 + Auditorium)                       & 0.0039  & (0.0021) & 0.0020  & (0.0031) & 0.0096  & (0.0084) \\
\textbf{Geography}      &         &          &         &          &         &          \\
Ln(Area)                                 & -0.0090 & (0.0086) & -0.0039 & (0.0103) & 0.0238  & (0.0350) \\
Ln(Forest Land)                          & -0.0173 & (0.0267) & 0.0021  & (0.0327) & -0.0263 & (0.1101) \\
    \bottomrule
    \bottomrule
    \end{tabular}
    \begin{tablenotes}
\footnotesize	
\item This table presents the test for balance on pre-existing characteristics for split and unsplit blocks. Columns (1) and (2) report the RD estimate and the associated standard errors for the full sample. Column (3) and (4) report the RD coefficient on the split variable -- defined using the 2008 delimitation -- and the standard errors for the sample of blocks that were split before the 2008 delimitation. Column (5) and (6) report the RD coefficient on the split variable -- defined using the 2008 delimitation -- and the standard errors for the sample of blocks that were unsplit before the 2008 delimitation. The unit of observation is a village that lies within the 10 km bandwidth of the boundary separating a split and an unsplit block. A block is defined as a split block based on the haphazard overlap of block boundaries with electoral boundaries as per the 2008 delimitation. The data on the key variables used in the table comes from the 2001 Indian Census. Village-level characteristics are divided into population, education, demographics, health infrastructure, other infrastructure, and geography. Population includes the natural logarithm of the number of people. Education includes the natural logarithm of the number of schools and colleges. Demographics includes the natural logarithm of female, and lower caste population. The latter includes scheduled castes (SC) and scheduled tribes (ST). Health infrastructure includes the natural logarithm of the number of hospitals and dispensaries. Other infrastructure includes the natural logarithm of the number of banks, cinema halls, and auditoriums. Geography includes the natural logarithm of geographic area and area under forest land.  \\
\end{tablenotes}
\end{threeparttable}
}
\end{table}

\clearpage
\newpage
\begin{table}[ht!]
  \centering
    \caption{RD Estimate: Firm Entry, Nightlights, and Split Block}
    \label{sptd_firm_entry}%
    \resizebox{\textwidth}{!}{ 
    \begin{threeparttable}
    \begin{tabular}{lcccc}
    \toprule
    \toprule
    \multicolumn{5}{c}{\textit{Panel A: Number of Firms}}\\
    \hline
    Dep Var: LN(0.001+\# New Firms) & (1)   & (2)   & (3)   & (4) \\
    \midrule  \\
    Split(=1) & 0.0331*** & 0.0326*** & 0.0312*** & 0.0309*** \\
          & (0.0071) & (0.0060) & (0.0056) & (0.0055) \\ \\
    \midrule
    \#Obs & {263,307 } & {356,260 } & {416,694 } & {436,980 } \\
    $R^2$ & {0.1537} & {0.1293} & {0.1147} & {0.1062} \\
    Bandwidth & 5 KM  & 10 KM & 20 KM & 50 KM \\
    Boundary FE & Yes & Yes & Yes & Yes \\
    \bottomrule 
    \multicolumn{5}{c}{\textit{Panel B: Nightlights}}\\
    \hline
    Dep Var: LN(0.001+Nightlight) & (1)   & (2)   & (3)   & (4) \\
    \midrule \\
    Split(=1) & 0.0703*** & 0.0782*** & 0.0773*** & 0.0661*** \\
          & (0.0085) & (0.0100) & (0.0114) & (0.0127) \\ \\
    \midrule
    \#Obs & {257,787} & {348,539} & {407,616} & {427,659} \\
    $R^2$ & {0.6054} & {0.5736} & {0.5453} & {0.5239} \\
    Bandwidth & 5 KM  & 10 KM & 20 KM & 50 KM \\
    Boundary FE & Yes & Yes & Yes & Yes \\
    \bottomrule
    \bottomrule
    \end{tabular}%
      \begin{tablenotes}
\footnotesize	
\item This table presents estimates for specification (\ref{rd}) using the natural logarithm of 0.001 plus firm entry and nightlight intensity as the dependent variables in Panels A and B, respectively. We use different bandwidths of 5, 10, 20, and 50 km on either side of the boundary, separating a split block from an unsplit block in columns (1), (2), (3), and (4), respectively.  The unit of observation is a village that lies within the narrow bandwidth of the boundary separating a split and an unsplit block. A block is defined as a split block based on the haphazard overlap of block boundaries with electoral boundaries as per the 2008 delimitation. Firm entry is defined as the total number of firms that entered a village from 2008 to 2016. Nightlight is defined as the average nightlight intensity in a village from 2008 to 2016.  All regressions include boundary fixed effects, a local linear specification estimated separately on each side of the boundary, and use a triangular kernel. Standard errors reported in parentheses are clustered at the block level. \sym{*} \(p<0.1\), \sym{**} \(p<0.05\), \sym{***} \(p<0.01\). \\
\end{tablenotes}
\end{threeparttable}
}
\end{table}%

\clearpage
\newpage
\begin{table}[ht!]
  \centering
    \caption{RD Estimate: Employment \& Applications for Unemployment Assistance}
    \label{app_tab_employment}%
    \resizebox{\textwidth}{!}{ 
    \begin{threeparttable}
    \begin{tabular}{lcccc}
    \toprule
    \toprule
    \multicolumn{5}{c}{\textit{Panel A: Employment}}\\
    \hline
    Dep Var: LN(0.001+ Employment) & (1)   & (2)   & (3)   & (4) \\
    \midrule  \\
    Split (=1) & 0.0595*** & 0.0597*** & 0.0599*** & 0.0587*** \\
          & (0.0197) & (0.0180) & (0.0179) & (0.0183) \\ \\
    \midrule
    \# Obs & 226,010 & 304,539 & 355,420 & 372,431 \\
    $R^2$  & 0.4420 & 0.4286 & 0.4188 & 0.4107 \\
    Bandwidth & 5 KM  & 10 KM & 20 KM & 50 KM \\
    Boundary FE & Yes & Yes & Yes & Yes \\
    \bottomrule 
    \multicolumn{5}{c}{\textit{Panel B: Applications for Employment Benefits under NREGA}}\\
    \hline
    Dep Var: LN(Days Applied) & (1)   & (2)   & (3)   & (4) \\
    \midrule \\
    Split (=1) & -0.0622** & -0.0543** & -0.0543** & -0.0561** \\
          & (0.0284) & (0.0256) & (0.0255) & (0.0255) \\ \\
    \midrule 
    \# Obs & 131,620 & 189,722 & 228,799 & 241,962 \\
    $R^2$  & 0.3454 & 0.3378 & 0.3327 & 0.3325 \\
    Bandwidth & 5 KM  & 10 KM & 20 KM & 50 KM \\
    Boundary $\times$ Year FE & Yes & Yes & Yes & Yes \\
    \bottomrule
    \bottomrule
    \end{tabular}%
      \begin{tablenotes}
\footnotesize	
\item This table presents estimates for specification (\ref{rd}) using the natural logarithm of 0.001 plus the total number of people employed and the natural logarithm of the total number of days for which unemployment relief is applied for under the National Rural Employment Guarantee Scheme (NREGA) as the dependent variables in Panels A and B, respectively. We use different bandwidths of 5, 10, 20, and 50 km on either side of the boundary, separating a split block from an unsplit block in columns (1), (2), (3), and (4), respectively.  The unit of observation is a village that lies within the narrow bandwidth of the boundary separating a split and an unsplit block. A block is defined as a split block based on the haphazard overlap of block boundaries with electoral boundaries as per the 2008 delimitation. The data on employment comes from the 2013 Economic Survey of India and includes the number of people employed in all non-farm activities. The data on applications for unemployment relief applied under the National Rural Employment Guarantee Scheme (NREGA)  comes from the Ministry of Rural Development and spans from 2016 to 2020.  All regressions include boundary fixed effects, a local linear specification estimated separately on each side of the boundary, and use a triangular kernel. Standard errors reported in parentheses are clustered at the block level. \sym{*} \(p<0.1\), \sym{**} \(p<0.05\), \sym{***} \(p<0.01\). \\
\end{tablenotes}
\end{threeparttable}
}
\end{table}%

\clearpage
\newpage

\begin{table}[ht!]
  \centering
    \caption{RD Estimate: Using Exogenous Block Boundaries}
    \label{tab_exo_boundary}%
    \resizebox{\textwidth}{!}{ 
    \begin{threeparttable}
    \begin{tabular}{lcccc}
    \toprule
    \toprule 
    \multicolumn{5}{c}{\textit{Panel A: Boundaries defined by Geographic Features (Rivers \& River Basins)}}\\
    \hline
    Dep Var: LN(0.001+\# New Firms) & (1)   & (2)   & (3)   & (4) \\
    \midrule \\
    Split (=1) & 0.0965** & 0.0913*** & 0.0819*** & 0.0777*** \\
          & (0.0428) & (0.0343) & (0.0309) & (0.0295) \\ \\
    \midrule
    \# Obs & 7,137  & 9,804  & 11,372 & 11,792 \\
    $R^2$  & 0.0823 & 0.0711 & 0.0618 & 0.0562 \\
    Bandwidth & 5 KM  & 10 KM & 20 KM & 50 KM \\
    Boundary FE & Yes & Yes & Yes & Yes \\
    \bottomrule
    \multicolumn{5}{c}{\textit{Panel B: Boundaries defined by Straight Lines}}\\
    \hline
    Dep Var: LN(0.001+\# New Firms) & (1)   & (2)   & (3)   & (4) \\
    \midrule \\
    Split(=1) & 0.1763*** & 0.1360*** & 0.1106*** & 0.1076*** \\
          & (0.0606) & (0.0485) & (0.0406) & (0.0379) \\ \\
    \midrule
    \#Obs & {7,018} & {10,394} & {13,426} & {14,328} \\
    $R^2$ & {0.2012} & {0.1698} & {0.1471} & {0.1331} \\
    Bandwidth & 5 KM  & 10 KM & 20 KM & 50 KM \\
    Boundary FE & Yes & Yes & Yes & Yes \\
    \bottomrule
    \bottomrule
    \end{tabular}%
      \begin{tablenotes}
\footnotesize	
\item This table presents estimates for specification \ref{rd} using the natural logarithm of 0.001 plus firm entry as the dependent variables. Panel A restricts the sample to boundaries between split and unsplit blocks that are straight lines. We define a boundary as a straight line if the Sinuosity Index (SI) is less than 1.05. Sinuosity Index is defined as the ratio of the length of the curve and the Euclidean distance between the endpoints of the curve. Panel B restricts the sample to boundaries between split and unsplit blocks defined by geographic features -- rivers and river basins. We only consider rivers and river basins with an average discharge greater than or equal to 200 $m^3/s$ or cumecs. We describe a boundary to be defined by a river or river basin if the boundary follows a river for at least 80\% of its length. Appendix figure \ref{fig_boundary_river} provides four representative examples of block boundaries separating a split and unsplit block that are defined by rivers or river basins. We use different bandwidths of 5, 10, 20, and 50 km on either side of the boundary, separating a split block from an unsplit block in columns (1), (2), (3), and (4), respectively.  The unit of observation is a village that lies within the narrow bandwidth of the boundary separating a split and an unsplit block. A block is defined as a split block based on the haphazard overlap of block boundaries with electoral boundaries as per the 2008 delimitation. Firm Entry is defined as the total number of firms that have entered a village from 2008 to 2016. All regressions include boundary fixed effect, a local linear specification estimated separately on each side of the boundary, and use a triangular kernel. Standard errors reported in parentheses are clustered at the block level. \sym{*} \(p<0.1\), \sym{**} \(p<0.05\), \sym{***} \(p<0.01\). \\
\end{tablenotes}
\end{threeparttable}
}
\end{table}%

\clearpage
\newpage
\begin{table}[htbp]
  \centering
  \caption{Firm Entry, Change in the Number of Politicians, and 2008 Delimitation}
  \label{sptd_firm_entry_switcher}%
  \resizebox{0.84\textwidth}{!}{ 
  \begin{threeparttable} 
    \begin{tabular}{lcccc}
    \toprule
    \toprule 
    \multicolumn{5}{c}{\textit{Panel A : Unsplit to Split}} \\
    \midrule
    Dep Var:  LN(0.001+\# New Firms) & (1)   & (2)   & (3)   & (4) \\
    \midrule 
    Treat x Post & 0.0157** & 0.0136** & 0.0123** & 0.0113** \\
          & (0.0074) & (0.0062) & (0.0055) & (0.0051) \\ 
    \midrule
    \# Obs & 687,792 & 975,312 & 1,185,816 & 1,263,456 \\
    $R^2$  & 0.6489 & 0.6329 & 0.6203 & 0.6129 \\
    Bandwidth & 5 KM  & 10 KM & 20 KM & 50 KM \\
    Boundary $\times$ Year FE & Yes   & Yes   & Yes   & Yes \\
    Village FE & Yes   & Yes   & Yes   & Yes \\
    \midrule
    \multicolumn{5}{c}{\textit{Panel B : Split to Unsplit}} \\
    \midrule
    Dep Var:  LN(0.001+\# New Firms) & (1)   & (2)   & (3)   & (4) \\
    \midrule 
    Treat x Post & -0.0104*** & -0.0086*** & -0.0074*** & -0.0073*** \\
          & (0.0023) & (0.0020) & (0.0018) & (0.0018) \\ 
    \midrule
    \# Obs & 2,469,408 & 3,297,648 & 3,812,364 & 3,978,144 \\
    $R^2$  & 0.6888 & 0.6861 & 0.6817 & 0.6769 \\
    Bandwidth & 5 KM  & 10 KM & 20 KM & 50 KM \\
    Boundary $\times$ Year FE & Yes   & Yes   & Yes   & Yes \\
    Village FE & Yes   & Yes   & Yes   & Yes \\
    \bottomrule
    \bottomrule
    \end{tabular}%
 \begin{tablenotes}
\footnotesize	
\item This table presents estimates for specification (\ref{did_1}) and (\ref{did_2}) using the natural logarithm of 0.001 plus firm entry as the dependent variable in Panels A and B, respectively. In Panel A the treated group is the set of blocks that switched from being unsplit to split following the 2008 delimitation, whereas the control group comprises of blocks that were always unsplit both before and after the delimitation and border the treated group. In Panel B, the treated group is the set of blocks that switched from being split to unsplit following the 2008 delimitation, whereas the control group comprises blocks that are always split among multiple politicians both before and after the 2008 delimitation and border the treated group. The variable $Post$ takes a value of 1 for all years since 2008. We use different bandwidths of 5, 10, 20, and 50 km on either side of the boundary, separating the treatment and the control groups in columns (1), (2), (3), and (4), respectively.  The unit of observation is a village-year that lies within the narrow bandwidth of the boundary separating the treatment and the control group. Firm entry is defined as the total number of firms that have entered a village during the year. All regressions include village and boundary $\times$ year fixed effects, a local linear specification estimated separately on each side of the boundary, and use a triangular kernel. Standard errors reported in parentheses are clustered at the block level. \sym{*} \(p<0.1\), \sym{**} \(p<0.05\), \sym{***} \(p<0.01\). \\
\end{tablenotes}
\end{threeparttable}   
}
\end{table}%

\newpage

\begin{table}[!htbp]
  \centering
\caption{Political Alignment and Firm Entry}
  \label{alignment}
  \scalebox{1}{
\begin{tabular}{lcccc}
    \toprule
    \toprule 
    Dep Var: LN(0.001+\# New Firms) & (1)   & (2)   & (3)   & (4) \\ 
    \midrule 
    Frac. Non Aligned x Split & 0.0053 & 0.0094** & 0.0114*** & 0.0117*** \\
          & (0.0056) & (0.0045) & (0.0041) & (0.0039) \\ 
    \midrule
    \# Obs & 1,813,190 & 2,497,757 & 2,958,673 & 3,120,412 \\
    \(R^2\)  & 0.7637 & 0.7581 & 0.7521 & 0.7466 \\
    Within \(R^2\) & 0.0001 & 0.0002 & 0.0002 & 0.0002 \\
    Bandwidth & 5 KM  & 10 KM & 20 KM & 50 KM \\
    Boundary \(\times\) Year FE  & Yes & Yes & Yes & Yes \\
    Village FE & Yes & Yes & Yes & Yes \\
    \bottomrule
    \bottomrule
    \end{tabular}}
\parbox{\textwidth}{
\vspace{1ex}      
\footnotesize   
\noindent This table presents estimates for specification (\ref{rd}) augmented with the interaction term between split blocks and the fraction of non-aligned politicians, using the natural logarithm of 0.001 plus firm entry as the dependent variable. We use bandwidths of 5, 10, 20, and 50 km on either side of the boundary—separating split blocks from unsplit blocks—in columns (1) through (4), respectively. The unit of observation is a village-year within the narrow bandwidth around the boundary. A block is classified as split based on the random overlap of block boundaries with electoral boundaries following the 2008 delimitation. Firm entry is defined as the total number of new firms entering a village in a given year, for the period from 2008 to 2016. Results are reported for an estimation where non-alignment is measured solely based on political party affiliation. The fraction of non-aligned politicians is a continuous variable that takes a value between 0 and 1: it equals 0 when all politicians belong to the same party and 1 when each is affiliated with a different party. In unsplit blocks, this measure is always 0. All regressions include boundary \(\times\) year fixed effects, village fixed effects, a local linear specification estimated separately on each side of the boundary, and use a triangular kernel. Standard errors reported in parentheses are clustered at the block level. \(\sym{*}\,p<0.1\), \(\sym{**}\,p<0.05\), \(\sym{***}\,p<0.01\).
}
\end{table}%

\newpage

\begin{table}[!htbp]
\centering
\caption{Winner's Premium and Multiple Politicians} 
\label{private_return_table}
\scalebox{0.8}{
\begin{tabular}{lcccc}
\toprule \toprule
Dep Var: Ln(Final Asset)     & (1)            & (2)          & (3)       & (4)       \\ 
\midrule 
Winner                         & 0.1605**       & -0.0800      & 0.1589*   & 0.0579    \\
                               & (0.0809)       & (0.0921)     & (0.0811)  & (0.0607)  \\
Winner X Splitness (Binary)    &                &              & -0.2348*  &           \\
                               &                &              & (0.1226)  &           \\
Winner X Splitness (Continous) &                &              &           & -0.1413** \\
                               &                &              &           & (0.0615)  \\
Ln(Initial Asset)              & 0.6112***      & 0.7224***    & 0.6580*** & 0.6583*** \\
                               & (0.0630)       & (0.0764)     & (0.0485)  & (0.0482)  \\  
\midrule
Constituency FE                & Yes            & Yes          & Yes       & Yes       \\
\#Obs                         & 330            & 252          & 582       & 582       \\
$R^2$                          & 0.8195         & 0.824        & 0.8233    & 0.8243    \\
Within $R^2$ & 0.0001 & 0.0002 & 0.0002 & 0.0002 \\
Sample                         & Splitness (Binary) = 0 & Splitness (Binary) = 1 & All       & All      \\ 
\bottomrule \bottomrule
\end{tabular}}
\parbox{\textwidth}{
\vspace{1ex}      
\footnotesize	
\noindent This table relates the difference in asset accumulated over an election cycle by a winner vis-a-vis runner up to the degree of splitness of an electoral constituency. Splitness (Binary) is defined as a binary variable taking a value of one if more than 50\% population in the constituency resides in split block and zero otherwise. We also measure Splitness as a continuous variable: defined as the (normalized) proportion of the assembly constituency population that resides in split blocks. In column (1) and (2) we report the results from the following regression specification performed separately for electoral constituency where splitness is zero and one respectively. In columns (3) and (4) we run the regression specification using all electoral constituencies. In column (3) splitness is defined as a binary variable taking a value of one if more than 50\% of the constituency population resides in split block. In column (4) splitness is continuous variable, defined as the normalized proportion of constituency population that resides in split blocks. Standard errors reported in parentheses are robust. \sym{*} \(p<0.1\), \sym{**} \(p<0.05\), \sym{***} \(p<0.01\).
}
\end{table}

\clearpage

\begin{table}[!htbp]
  \centering
\caption{Firm Entry by Political Patronage}
  \label{tab:patronage_test}
  \scalebox{1}{
\begin{tabular}{lcccc}
      \toprule
      \toprule
      Dep Var: LN(0.001+\# Firms) & (1)   & (2)   & (3)   & (4) \\
      \midrule
      Split $\times$ Connected & -0.0104* & -0.0104** & -0.0108** & -0.0108*** \\
            & (0.0056) & (0.0047) & (0.0042) & (0.0039) \\
      \midrule
      \#Obs & 350,704 & 487,846 & 584,470 & 619,864 \\
      $R^2$ & 0.6832 & 0.6810 & 0.6780 & 0.6745 \\
      Within $R^2$ & 0.0002 & 0.0002 & 0.0002 & 0.0002 \\
      Bandwidth & 5 KM  & 10 KM & 20 KM & 50 KM \\
      Boundary $\times$ Connected FE  & Yes & Yes & Yes & Yes \\
      Village FE & Yes & Yes & Yes & Yes \\
      \bottomrule
      \bottomrule
      \end{tabular}}
\parbox{\textwidth}{
\vspace{1ex}      
\footnotesize	
\noindent This table presents estimates for specification (1) augmented for the interaction term of split with connected. The dependent variable is the natural logarithm of 0.001 plus firm entry. Firm entry is defined as the total number of firms of each type entering a village in the first political cycle after the delimitation. A block is defined as a split block based on the haphazard overlap of block boundaries with electoral boundaries as per the 2008 delimitation. Firms are classified into two types: (1) Connected firms if they have at least one director sharing the same caste identity as the local politician in power, determined through ethnographic mapping using the Persons of India database, and (2) Non-connected firms if they do not share caste identity with the local politician. Connected is an indicator variable that equals one for connected firms and zero for non-connected firms. The interaction term captures the differential effect of block splitting on connected versus non-connected firms. We use different bandwidths of 5, 10, 20, and 50 km on either side of the boundary, separating a split block from an unsplit block in columns (1), (2), (3), and (4), respectively. The unit of observation is the village-firm type that lies within the narrow bandwidth of the boundary separating a split and an unsplit block. The data is structured in long format where each village appears twice: once for connected firms and once for non-connected firms. All regressions include boundary $\times$ connected fixed effects, village fixed effects, a local linear specification estimated separately on each side of the boundary, and use a triangular kernel. Standard errors reported in parentheses are clustered at the block level. \sym{*} \(p<0.1\), \sym{**} \(p<0.05\), \sym{***} \(p<0.01\).
}
\end{table}
\clearpage

\begin{table}[!htbp]
  \centering
  \caption{Firm Entry across Industries Vulnerable to Cronyism}
  \label{tab:crony_industries}
  \scalebox{1}{
    \begin{tabular}{lcccc}
    \toprule
    \toprule
    Dep Var:  LN(0.001+\# New Firms) & (1)   & (2)   & (3)   & (4) \\
    \midrule 
    Crony x Split(=1) & -0.0494** & -0.0607*** & -0.0599*** & -0.0578*** \\
          & (0.0210) & (0.0170) & (0.0155) & (0.0150) \\ 
    \midrule
    \#Obs & 328,104 & 419,364 & 465,966 & 478,656 \\
    $R^2$  & 0.3623 & 0.3574 & 0.3538 & 0.3524 \\
    Within $R^2$ & 0.0001 & 0.0002 & 0.0002 & 0.0002 \\
    Bandwidth & 5 KM  & 10 KM & 20 KM & 50 KM \\
    Boundary $\times$ Industry FE  & Yes & Yes & Yes & Yes \\
    Village FE & Yes & Yes & Yes & Yes \\
    \bottomrule
    \bottomrule
    \end{tabular}}
\parbox{\textwidth}{
\vspace{1ex}      
\footnotesize	
\noindent This table presents estimates for specification (1) augmented with the interaction term between Split and Crony. The dependent variable is the natural logarithm of 0.001 plus firm entry. We use different bandwidths of 5, 10, 20, and 50 km on either side of the boundary separating split and unsplit blocks in columns (1), (2), (3), and (4), respectively. The unit of observation is the village-industry within the narrow bandwidth of the boundary. A block is defined as split based on the haphazard overlap of block boundaries with electoral boundaries following the 2008 delimitation. Firm entry is defined as the total number of firms in an industry entering a village from 2008 to 2016. All regressions include boundary $\times$ industry fixed effects, village fixed effects, a local linear specification estimated separately on each side of the boundary, and use a triangular kernel. Crony is an indicator for industries vulnerable to cronyism, based on the industry-level cronyism index created by \emph{The Economist} using \emph{Transparency International} methodology. The detailed description can be found in \href{https://www.economist.com/international/2014/03/15/planet-plutocrat}{Planet Plutocrat}. Standard errors reported in parentheses are clustered at the block level. \sym{*} \(p<0.1\), \sym{**} \(p<0.05\), \sym{***} \(p<0.01\).
}
\end{table}

\begin{table}[!htbp]
  \centering
  \caption{Multiple Politicians and Time for Regulatory Approvals}
  \label{tab_delay_time}
  \scalebox{1}{
    \begin{tabular}{lcc}
    \toprule
    \toprule
      $LN(\text{Days to obtain approval})$    & (1)   & (2) \\
    \midrule 
    Treat x Post & -0.8084*** & 0.5706* \\
          & (0.2122) & (0.3461) \\
    \midrule
    Sample & Unsplit to Split & Split to Unsplit \\
    \# Obs & 2284  & 9091 \\
    $R^2$    & 0.2449 & 0.2745 \\
    Within $R^2$ & 0.0001 & 0.0002 \\
    Bandwidth & 10 KM & 10 KM \\
    Boundary x Year FE & Yes   & Yes \\
    Block FE & Yes   & Yes \\
    \bottomrule
    \bottomrule
    \end{tabular}}
\parbox{\textwidth}{
\vspace{1ex}      
\footnotesize	
\noindent This table presents estimates for specification (1) and (2) using the natural logarithm of "Days taken to obtain regulatory approvals" as the dependent variable. In column (1) the treated group is the set of blocks that switched from being unsplit to split following the 2008 delimitation, whereas the control group comprises blocks that were always unsplit both before and after the delimitation and border the treated group. In column (2), the treated group is the set of blocks that switched from being split to unsplit following the 2008 delimitation, whereas the control group comprises of blocks that are always split among multiple politicians both before and after the 2008 delimitation and border the treated group. The variable $Post$ takes a value of 1 for all years since 2008. We use a bandwidths of 10 km on either side of the boundary, separating the treatment and the control groups. The unit of observation is a project, announced between 2003 and 2016 that lies within the narrow bandwidth of the boundary separating the treatment and the control group. All regressions include block and boundary $\times$ year fixed effects, a local linear specification estimated separately on each side of the boundary, and use a triangular kernel. Standard errors reported in parentheses are clustered at the block level. \sym{*} \(p<0.1\), \sym{**} \(p<0.05\), \sym{***} \(p<0.01\).
}
\end{table}

\newpage

\begin{table}[!htbp]
  \centering
  \caption{Firm Entry across Industries with Varying Regulatory Intensity}
  \label{tab:regulated_industries}
  \scalebox{1}{
    \begin{tabular}{lcccc}
    \toprule
    \toprule
    Dep Var:  LN(0.001+\# New Firms) & (1)   & (2)   & (3)   & (4) \\
    \midrule 
    Regulated x Split(=1) & 0.0349* & 0.0346** & 0.0359** & 0.0343** \\
          & (0.0208) & (0.0172) & (0.0156) & (0.0150) \\ 
    \midrule
    \#Obs & 328,104 & 419,364 & 465,966 & 478,656 \\
    $R^2$ & 0.3623 & 0.3574 & 0.3538 & 0.3524 \\
    Within $R^2$ & 0.0001 & 0.0002 & 0.0002 & 0.0002 \\
    Bandwidth & 5 KM  & 10 KM & 20 KM & 50 KM \\
    Boundary $\times$ Industry FE  & Yes & Yes & Yes & Yes \\
    Village FE & Yes & Yes & Yes & Yes \\
    \bottomrule
    \bottomrule
    \end{tabular}}
\parbox{\textwidth}{
\vspace{1ex}      
\footnotesize	
\noindent This table presents estimates for specification (1) augmented for the interaction term of split with regulated industries. The dependent variable is the natural logarithm of 0.001 plus firm entry. We use different bandwidths of 5, 10, 20, and 50 km on either side of the boundary, separating a split block from an unsplit block in columns (1), (2), (3), and (4), respectively. The unit of observation is the village-industry that lies within the narrow bandwidth of the boundary separating a split and an unsplit block. A block is defined as a split block based on the haphazard overlap of block boundaries with electoral boundaries as per the 2008 delimitation. Firm entry is defined as the total number of firms in an industry entering a village from 2008 to 2016. All regressions include boundary $\times$ industry fixed effect, village fixed effects, a local linear specification estimated separately on each side of the boundary, and use a triangular kernel. We define industries with high regulatory costs following the methodology for the classification of regulated industries in the United States based on Pittman (1977) augmented with India specific regulations and factors as discussed in Awasthi et al. (2019). Standard errors reported in parentheses are clustered at the block level. \sym{*} \(p<0.1\), \sym{**} \(p<0.05\), \sym{***} \(p<0.01\).
}
\end{table}

\newpage

\begin{table}[!htbp]
  \centering
  \caption{Multiple Politicians and Cost Overruns in Road Construction}
  \label{efficiency_infra}
  \scalebox{1}{
    \begin{tabular}{lcccc}
    \toprule
    \toprule 
    \multicolumn{5}{c}{\textit{Panel A : Unsplit to Split}} \\
    \midrule
    Dep Var:  $LN(\frac{Actual Cost}{Estimated Cost})$ & (1)   & (2)   & (3)   & (4) \\
    \midrule 
    Treat $\times$ Post & -0.2586***   & -0.1947**    & -0.1658**    & -0.1647**    \\
                    & (0.0980)     & (0.0925)     & (0.0821)     & (0.0770)   \\ 
    \midrule
    \# Obs & 5,819 & 8,754 & 11,124 & 11,897 \\
    $R^2$  & 0.7154 & 0.6898 & 0.6646 & 0.6492 \\
    Within $R^2$ & 0.0001 & 0.0002 & 0.0002 & 0.0002 \\
    Bandwidth & 5 KM  & 10 KM & 20 KM & 50 KM \\
    Boundary $\times$ Year FE & Yes   & Yes   & Yes   & Yes \\
    Block FE & Yes   & Yes   & Yes   & Yes \\
    \midrule 
    \multicolumn{5}{c}{\textit{Panel B : Split to Unsplit}} \\
    \midrule
    Dep Var:  $LN(\frac{Actual Cost}{Estimated Cost})$ & (1)   & (2)   & (3)   & (4) \\
    \midrule 
    Treat $\times$ Post & 0.0313       & 0.0514*      & 0.0532**     & 0.0538**     \\
                    & (0.0316)     & (0.0269)     & (0.0230)     & (0.0227)   \\ 
    \midrule
    \# Obs & 22,796 & 32,363 & 38,527 & 40,638 \\
    $R^2$  & 0.6713 & 0.6462 & 0.6319 & 0.6192 \\
    Within $R^2$ & 0.0001 & 0.0002 & 0.0002 & 0.0002 \\
    Bandwidth & 5 KM  & 10 KM & 20 KM & 50 KM \\
    Boundary $\times$ Year FE & Yes   & Yes   & Yes   & Yes \\
    Block FE & Yes   & Yes   & Yes   & Yes \\
    \bottomrule
    \bottomrule
    \end{tabular}}
\parbox{\textwidth}{
\vspace{1ex}      
\footnotesize	
\noindent This table presents estimates for specification (1) and (2) using the natural logarithm of actual cost minus the natural logarithm of estimated cost as the dependent variable in panels A and B, respectively. In Panel A the treated group is the set of blocks that switched from being unsplit to split following the 2008 delimitation, whereas the control group comprises of blocks that were always unsplit both before and after the delimitation and border the treated group. In Panel B, the treated group is the set of blocks that switched from being split to unsplit following the 2008 delimitation, whereas the control group comprises of blocks that are always split among multiple politicians both before and after the 2008 delimitation and border the treated group. The variable $Post$ takes a value of 1 for all years since 2008. We use different bandwidths of 5, 10, 20, and 50 km on either side of the boundary, separating the treatment and the control groups in columns (1), (2), (3), and (4), respectively. The unit of observation is a village-year spanning from 2001 to 2014 that lies within the narrow bandwidth of the boundary separating the treatment and the control group. A village appears only once in the dataset because only one road could be constructed under the PMGSY per village. All regressions include block and boundary $\times$ year fixed effects, a local linear specification estimated separately on each side of the boundary, and use a triangular kernel. Standard errors reported in parentheses are clustered at the block level. \sym{*} \(p<0.1\), \sym{**} \(p<0.05\), \sym{***} \(p<0.01\).
}
\end{table}%

\clearpage
\newpage
\pagenumbering{arabic}
\renewcommand*{\thepage}{A\arabic{page}}
\appendix
\begin{appendices}
\numberwithin{equation}{section}
\counterwithin{figure}{section}
\numberwithin{table}{section}

\begin{center}
    \textbf{Online Appendix for} \\
    \vspace{0.25in}
    \title{Political Power-Sharing, Firm Entry, and Economic Growth: Evidence from Multiple Elected Representatives}
\end{center}

\section{Background \& Institutional Details}
\label{app_insti_details}
\begin{figure}[!ht]
	\centering
	\caption{The Six Tiers of Government in India}
	\includegraphics[width=0.7\textwidth]{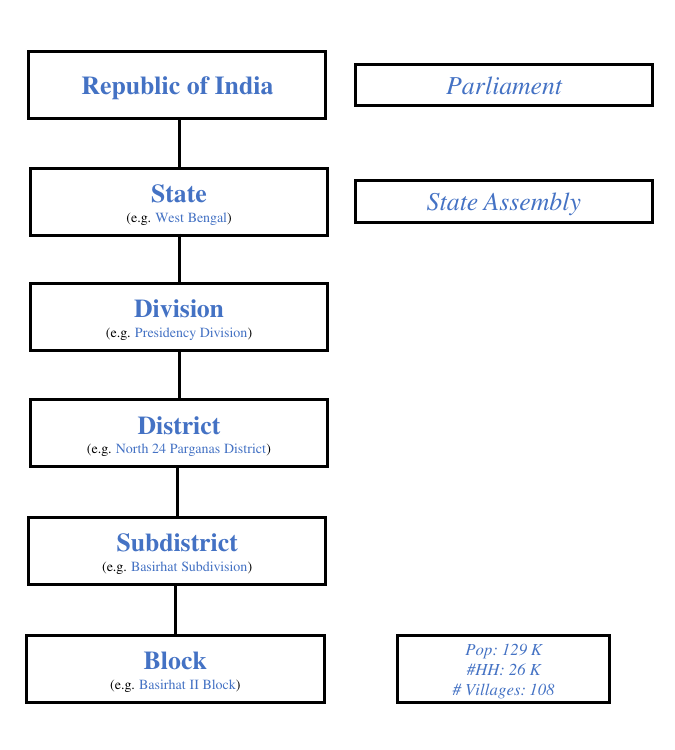}
	\label{app_fig_six_tiers}
	\begin{minipage}{0.8\textwidth}
{\footnotesize The figure illustrates the six tiers of government in India. This hierarchy includes six tiers of government, starting from the federal government of the Republic of India, followed by states, division, district, sub-district, and finally, the local governing bodies. We refer to the local governing bodies in urban and rural areas as blocks which include towns, villages, or clusters of towns and villages. India has a parliamentary system as defined by its constitution, with power distributed between the central government (Parliament) and the states (State Assembly).
\par}
	\end{minipage}
\end{figure}

\clearpage

\subsection{Roles and Responsibilities of the Block Development Officer}
\label{app:bdo_powers}

This appendix examines the statutory powers of Block Development Officers (BDOs) using Karnataka's governance structure as a case study.\footnote{For terminological consistency, we refer to the Executive Officer of the Taluk Panchayat as the Block Development Officer (BDO) and the Taluk Panchayat as the block throughout this paper.} Karnataka implements a three-tier local self-governance system through the Karnataka Panchayat Raj Act, 1993: Zilla Panchayat (district level), Taluk Panchayat (block), and Grama Panchayat (village level). The State Appointed Executive Officer (BDO) of the Taluk Panchayat serves as the administrative head with powers that establish a hierarchical relationship affecting governance processes at the village level. The discussion that follows details the BDO's legal mandate, specific oversight mechanisms for Grama Panchayats, and how these powers enable administrative control over local economic and governance activities.

\subsubsection{The Block (Taluka) and its BDO: Definitions, Mandate, and General Powers}
\label{app:bdo_mandate}

\paragraph{The Block (Taluk Panchayat) as an Administrative Unit}

The Taluk Panchayat (block) serves as the intermediate tier in Karnataka's rural local self-government system, functioning as a pivotal unit for area-based planning, implementation of development programs, and administrative coordination for Grama Panchayats (villages) within its boundaries. Its mandate includes consolidating village development plans, executing inter-village or block-wide works, and providing administrative and technical support to village-level bodies. The block thus plays a crucial role in translating district-level policies into actionable programs at the block level.

\begin{itemize}
    \item \textbf{Legal Basis:} The establishment and jurisdiction of the Taluk Panchayat are detailed in Section 119 of the Act. Its functions are enumerated in Section 145 and further elaborated in Schedule II of the Act.
\end{itemize}

\paragraph{The BDO: Appointment and General Statutory Authority within the Block}

The BDO serves as the chief executive of the Taluk Panchayat (block), appointed by the state government as a career civil servant. The BDO's general statutory authority encompasses the management of the block's day-to-day administration, its financial resources, and the implementation of various developmental schemes and public works undertaken by the Taluk Panchayat.

\begin{itemize}
    \item \textbf{Legal Basis for Appointment \& Status:} Section 155(1) of the Act mandates that ``The Government shall appoint a Group-A Officer of the State Civil Services or equal to the rank of the Assistant Commissioner to be the Executive Officer of the Taluk Panchayat.''
    \item \textbf{Legal Basis for General Powers:} Section 156(1) of the Act outlines the primary functions of the BDO (Executive Officer), which include:
    \begin{itemize}
        \item ``exercise all the powers specifically imposed or conferred upon him by or under this Act, or under any other law for the time being in force'' (Section 156(1)(a));
        \item ``control the officers and officials of, or holding office under the Taluk Panchayat subject to the general superintendence and control of the Adhyaksha and such rules as may be prescribed'' (Section 156(1)(b));
        \item ``supervise and control the execution of all works of the Taluk Panchayat'' (Section 156(1)(c));
        \item ``take necessary measures for the speedy execution of all works and developmental schemes of the Taluk Panchayat'' (Section 156(1)(d)). The term ``necessary measures'' inherently implies a degree of discretion in operational decisions.
        \item ``have custody of all papers and documents connected with the proceedings of the meetings of the Taluk Panchayat and of its Committees'' (Section 156(1)(e));
        \item ``draw and disburse monies out of the Taluk Panchayat fund'' (Section 156(1)(f)).
    \end{itemize}
\end{itemize}

\subsubsection{The BDO's Authority and Influence over Grama Panchayats}
\label{app:bdo_village_oversight}

The Taluk Panchayat's (block) operational mandate includes oversight of all Grama Panchayats (village institutions) within its jurisdiction, making this a core BDO responsibility. As the Taluk Panchayat's executive arm, the BDO connects block and village governance through statutory powers that provide discretionary authority over the functioning and resource management of Grama Panchayats. These powers enable the BDO to shape governance outcomes at the village level, with mechanisms to ensure village compliance with block-level policy preferences. The statutory provisions conferring these jurisdictional powers are examined below.

\subparagraph{Financial and Budgetary Control}

The BDO exercises control over Grama Panchayat finances through budget approval requirements and audit authority, creating dependence of village institutions on block-level authorization.

\begin{itemize}
    \item \textbf{Legal Basis (Budget):} Section 241(2-5) mandates Grama Panchayat budget submission to the Taluk Panchayat, which holds modification rights to ``secure compliance.'' If a village fails to pass its budget, the Taluk Panchayat ``shall approve it with or without modification,'' with the BDO certifying the final document. This creates a direct approval requirement whereby village-level spending plans cannot proceed without BDO certification.
    \item \textbf{Legal Basis (Audit):} Section 246 grants the BDO post-expenditure control, including powers to ``direct that the defects or irregularities... shall be removed or remedied'' (Sec 246(7)) and to ``disallow any item of expenditure'' and ``surcharge the amount thereof on the person making or authorising the illegal payment'' (Sec 246(8)). These provisions establish personal financial liability for Grama Panchayat officials who make expenditures later rejected by the BDO—creating incentives for compliance with BDO preferences.
\end{itemize}

\subparagraph{Mandatory Compliance and BDO Inspection Directives}

The BDO holds authority to conduct inspections of Grama Panchayat operations and issue binding directives that village institutions must implement.

\begin{itemize}
    \item \textbf{Legal Basis:} Section 232(a) explicitly states the BDO may ``inspect the offices or premises or works taken up by the... Grama Panchayat'' and that ``the... Grama Panchayat concerned shall comply with the instructions issued after such inspections.'' This mandatory compliance requirement transforms routine inspections into a directive mechanism through which the BDO can direct operational changes at the village level without requiring further authorization.
\end{itemize}

\subparagraph{Non-Compliance Response and Takeover Authority}

The Act establishes intervention mechanisms for addressing deficiencies in Grama Panchayat performance, with the BDO serving as the implementing authority for Taluk Panchayat directives.

\begin{itemize}
    \item \textbf{Legal Basis:} Section 235(1) authorizes the Taluk Panchayat to determine when a Grama Panchayat has ``made default in performing any duty'' and to ``fix a period for the performance of that duty.'' Upon continued non-compliance, the Taluk Panchayat may ``appoint a person to perform it'' and ``direct that the expense... shall be forthwith paid by the Grama Panchayat.'' As the executive officer responsible for executing Taluk Panchayat resolutions under Section 156(1), the BDO operationalizes these interventions, effectively serving as the administrative channel through which higher-level directives supplant village-level authority.
\end{itemize}

\subparagraph{Information Access and Document Requisition Powers}

The Act grants the BDO powers to requisition, seize, and investigate Grama Panchayat records and finances.

\begin{itemize}
    \item \textbf{Legal Basis:} Section 157(1-3) empowers the BDO to requisition ``moneys or... accounts, records or other property'' from Grama Panchayats, with authority to ``issue a search warrant and exercise all such powers... as may lawfully be exercised by a Magistrate.'' Section 157(3) creates a legal obligation for any person with knowledge of concealed village resources to report this to the BDO. These powers extend the BDO's authority beyond administrative oversight into investigative domains.
    \item \textbf{Legal Basis (Routine Information):} Section 232(b) further authorizes the BDO to ``call for any return, statement, account or report'' from Grama Panchayats, establishing documentation requirements that facilitate monitoring.
\end{itemize}

\subparagraph{Personnel Control Through Appellate Authority}

The BDO's authority extends to personnel matters, establishing administrative judgment over disciplinary actions against Grama Panchayat employees.
\begin{itemize}
    \item \textbf{Legal Basis:} Section 113(4) establishes that appeals against disciplinary orders issued by the Panchayat Development Officer of a Grama Panchayat ``shall lie... to the Executive Officer [BDO].'' This appellate authority creates a supervisory relationship between the BDO and village-level staff, affecting the authority of elected Grama Panchayat officials in personnel matters.
\end{itemize}

\clearpage

\subsection{Business Permits and Licensing Authority by the Local Administration}
\label{app:business_permits}

This appendix examines the business permit and licensing framework at the local level, using Karnataka's governance structure to illustrate how local regulatory authority operates in practice. We focus on the comprehensive licensing powers formally vested in Grama Panchayats under the Karnataka Gram Swaraj and Panchayat Raj Act, 1993, which serve as foundational requirements for business establishment and operation. The Act grants Grama Panchayats substantial discretionary authority in permit decisions, including power to impose conditions specified in such permission'' and to refuse to grant the permission... for reasons to be recorded in writing.''\footnote{Sections 66, 67, 68, 69 for conditions specification; Section 70(2) for refusal authority, Karnataka Panchayat Raj Act, 1993.} This discretionary space in legal interpretation allows local authorities to influence the cost and feasibility of business operations. Grama Panchayats also possess enforcement powers to penalize non-compliance, issue violation notices, and determine penalty levels.\footnote{Sections 64(3), 70(2F), and 292 provide enforcement authority and discretionary compliance standards, Karnataka Panchayat Raj Act, 1993.} These local permits frequently serve as prerequisites for state-level clearances, creating cascading dependencies where denial or delay at the village level can obstruct multiple higher-level approvals required for business operation. While these decisions occur under Grama Panchayat authority, they operate within a broader administrative framework where Block Development Officers exercise supervisory oversight through multiple mechanisms, including appellate authority, budget approval, and compliance monitoring.\footnote{As detailed in the previous appendix, BDOs possess inspection powers (Section 232), audit authority (Section 246), and serve as final appellate authority for permit decisions (Sections 64(5) and 70(3), Karnataka Panchayat Raj Act, 1993).} The statutory provisions detailed below demonstrate the foundational role of local permits in business formation and the mechanisms through which administrative oversight can influence firm entry outcomes.

\subsubsection*{Mandatory Business Permits Issued by Grama Panchayats}

\paragraph{Trade Licenses (Section 69)} 
\begin{itemize}
    \item \textbf{Legal Mandate}: Section 69 mandates that ``No place within the jurisdiction of a Grama Panchayat shall be used as a shop whether permanently or temporarily... except under a licence granted or renewed by the Grama Panchayat.''
    \item \textbf{Coverage}: All commercial establishments, retail shops, service centers, and trading activities within GP jurisdiction.
    \item \textbf{Downstream Requirements}: Bank Accounts, KSPCB Hazardous Waste Authorization, Municipal Solid Waste Authorization.
\end{itemize}

\paragraph{Factory and Industrial Establishment Permissions (Section 66)}
\begin{itemize}
    \item \textbf{Legal Mandate}: Section 66 requires that ``No person shall, without the permission of the Grama Panchayat... construct or establish any factory, workshop or workplace... to employ steam power, water power or other mechanical power or electrical power, or... install... any machinery or manufacturing plant driven by any power.''
    \item \textbf{Coverage}: All mechanized production facilities, requiring permission before factory construction or machinery installation.
    \item \textbf{Downstream Requirements}: Factory Plan Approval, Factory License.
\end{itemize}

\paragraph{Building Construction Permits (Section 64)}
\begin{itemize}
    \item \textbf{Legal Mandate}: Section 64(1) states ``no person shall erect any building or alter or add to any existing building... or reconstruct any building without the written permission of the Grama Panchayat.''
    \item \textbf{Coverage}: All new construction, alterations, additions, and reconstructions within GP limits, including all business premises, factories, and commercial buildings.
    \item \textbf{Downstream Requirements}: Occupancy Certificate, Commencement Certificate.
\end{itemize}

\paragraph{Specialized Business Activity Licenses (Section 67)}
\begin{itemize}
    \item \textbf{Legal Mandate}: Section 67 mandates that ``No place within the jurisdiction of Grama Panchayat shall be used for the purpose of any trade, business or industry which the Government may, by notification declare to be offensive or dangerous, except under a licence granted or renewed by the Grama Panchayat.''
    \item \textbf{Coverage}: Government-notified hazardous, offensive, or dangerous trades and industries.
    \item \textbf{Downstream Requirements}: Environmental clearances, KSPCB authorizations.
\end{itemize}

\paragraph{General Business NOCs (Section 60)}
\begin{itemize}
    \item \textbf{Legal Mandate}: Section 60 grants that ``Grama Panchayat shall have powers to do all acts necessary for or incidental to the carrying out of the functions entrusted, assigned or delegated to it.''
    \item \textbf{Coverage}: No Objection Certificates and other supporting documentation required by state-level regulatory authorities as prerequisites for higher-level licenses.
    \item \textbf{Downstream Requirements}: Bank Accounts, Land conversion Approvals, Factory Plan Approval, Electricity Connection.
\end{itemize}

\clearpage

\clearpage

\newpage
\begin{figure}[!htbp]
    \centering
    \caption{Administrative and Electoral Boundaries}
    \begin{subfigure}{.40\textwidth}
    \centering
    \includegraphics[width = \textwidth]{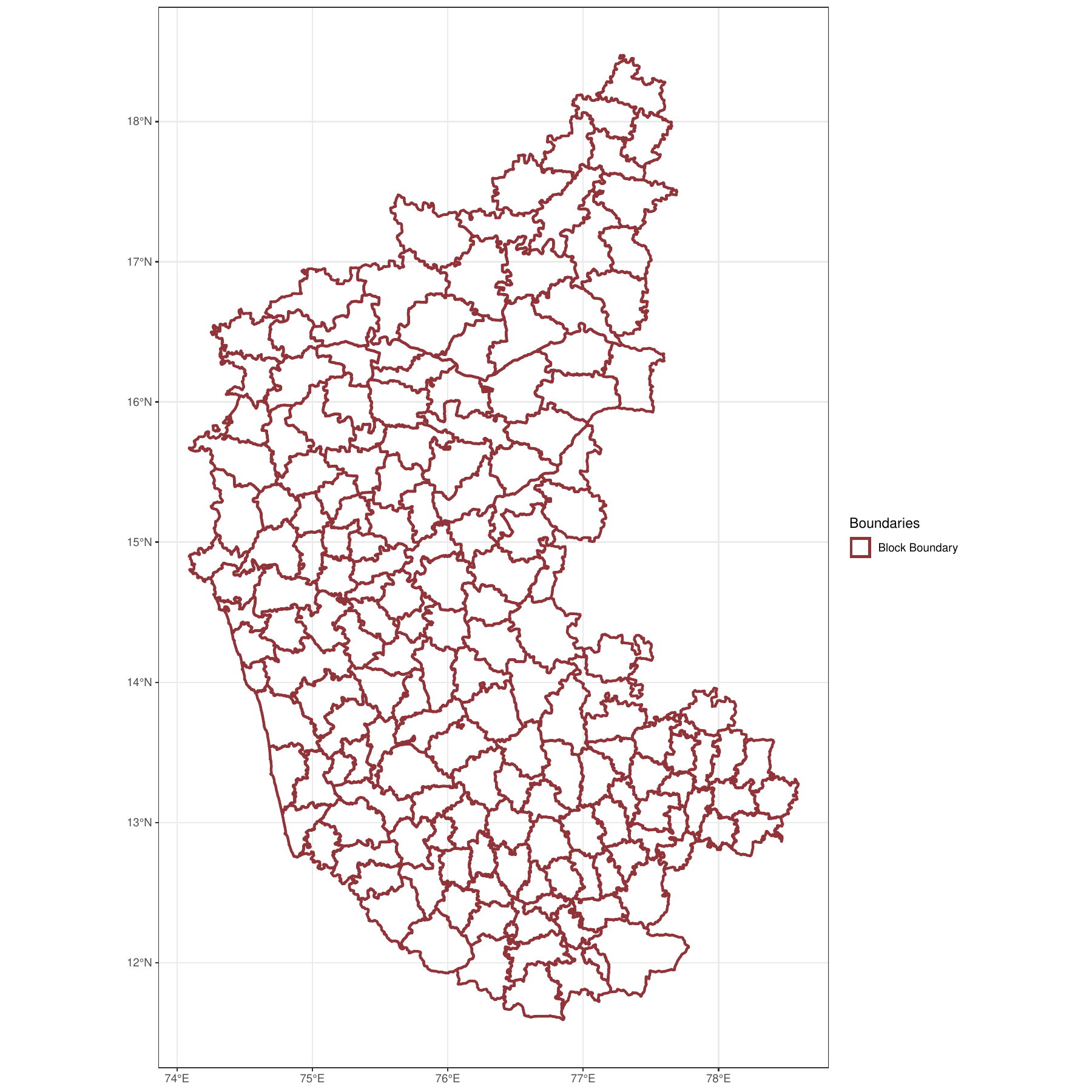}
    \caption{Administrative Boundaries (Blocks)}
    \label{app_fig_boundary_a}
    \end{subfigure} %
    \begin{subfigure}{.40\textwidth}
    \centering
    \includegraphics[width = \textwidth]{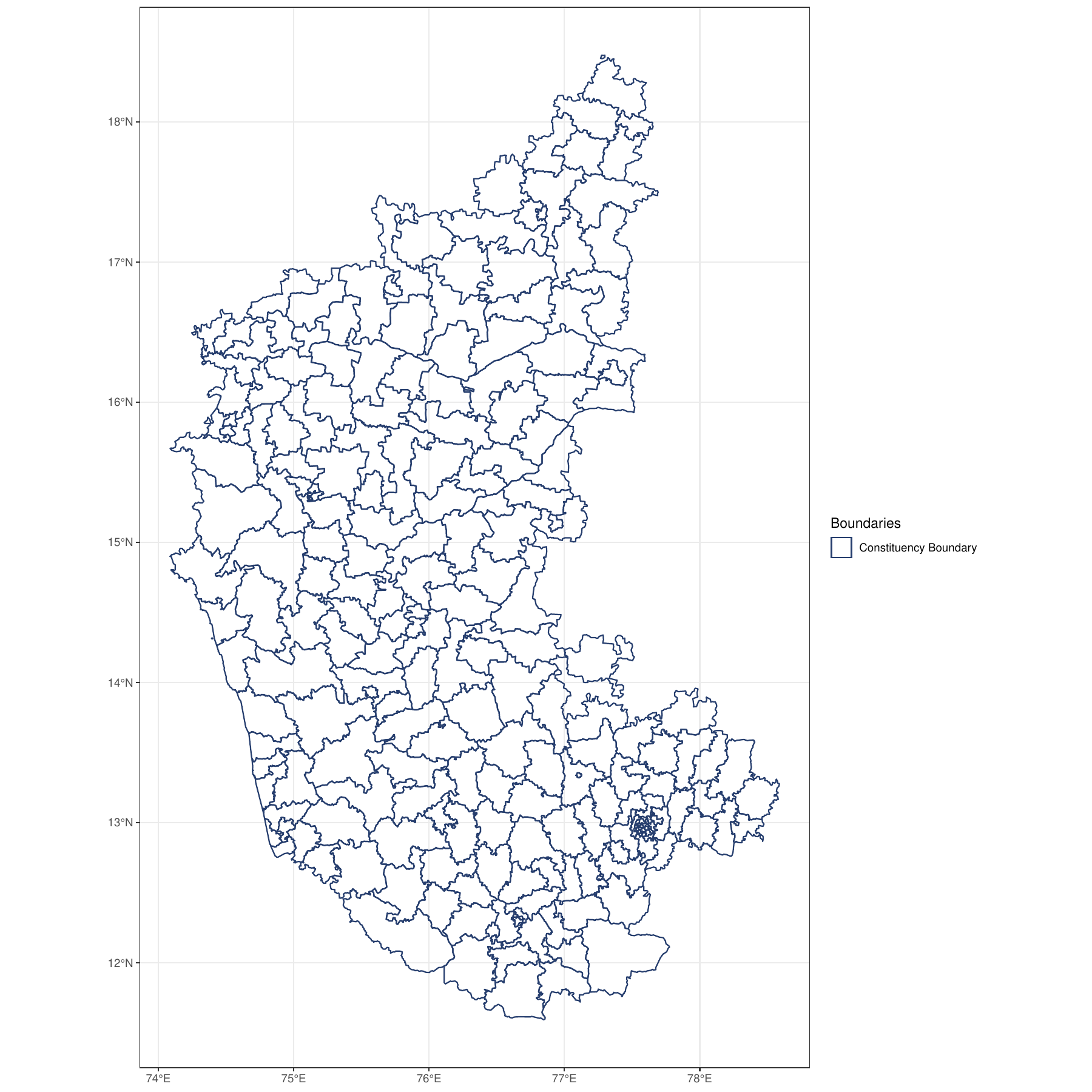}
    \caption{Electoral Boundaries (Assembly Constituency)}
    \label{app_fig_boundary_b}
    \end{subfigure} %
    \label{app_fig_boundary}
    \begin{minipage}{1.00\textwidth}
	\begin{center}
		\end{center}
		{\footnotesize 	The figure presents the administrative and electoral boundaries, as an example, from the state of Karnataka. The administrative boundaries shown in Figure \ref{app_fig_boundary_a} in red are based on the basic administrative unit - block. The electoral boundaries shown in Figure \ref{app_fig_boundary_b} in blue are based on the 2008 delimitation. \par}
\end{minipage}
\end{figure}

\begin{figure}[!htbp]
    \centering
    \caption{Overlap of Administrative and Electoral Boundaries}
    \begin{subfigure}{.40\textwidth}
    \centering
    \includegraphics[width = \textwidth]{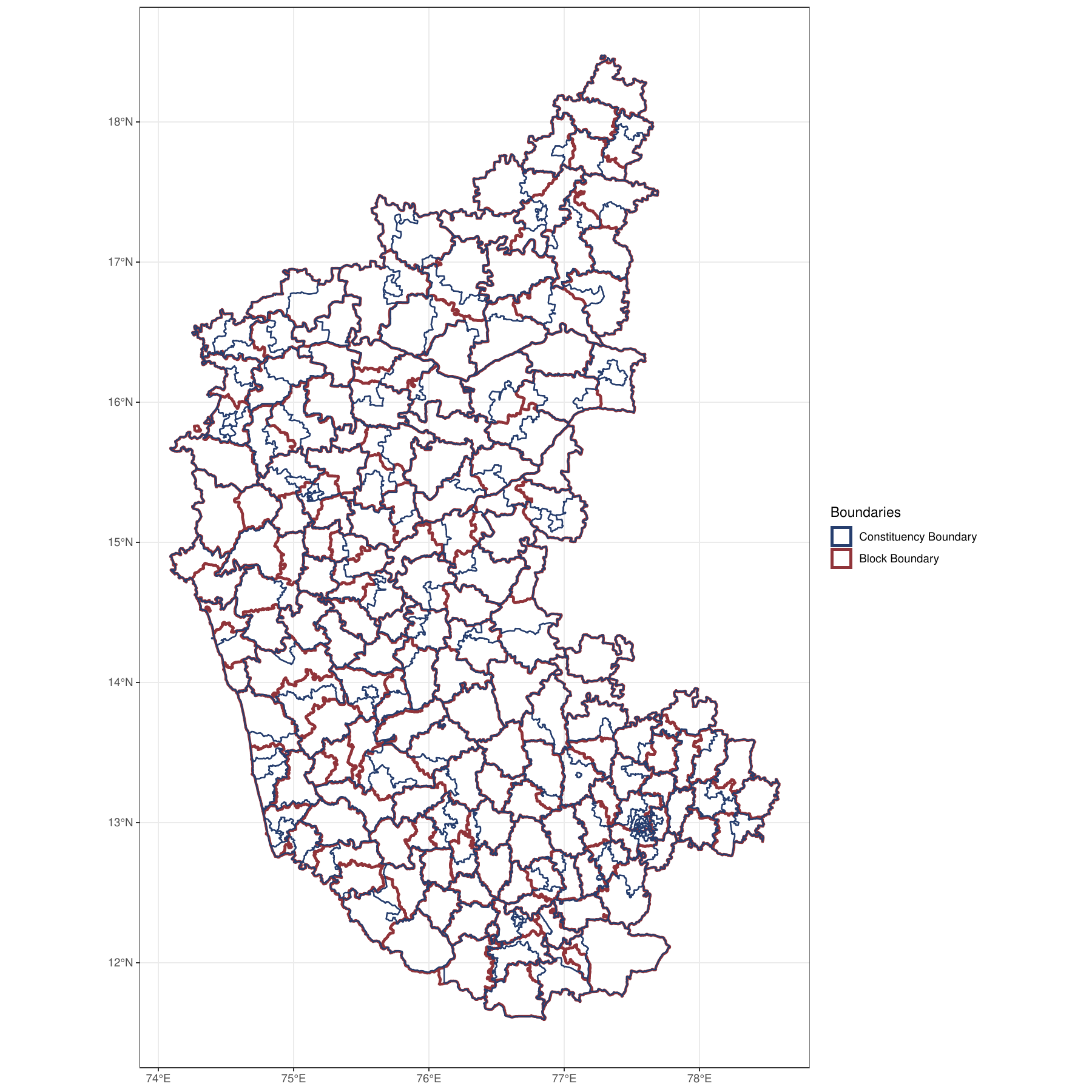}
    \caption{Overlap of Boundaries}
    \label{app_fig_overlap_a}
    \end{subfigure} %
    \begin{subfigure}{.40\textwidth}
    \centering
    \includegraphics[width = \textwidth]{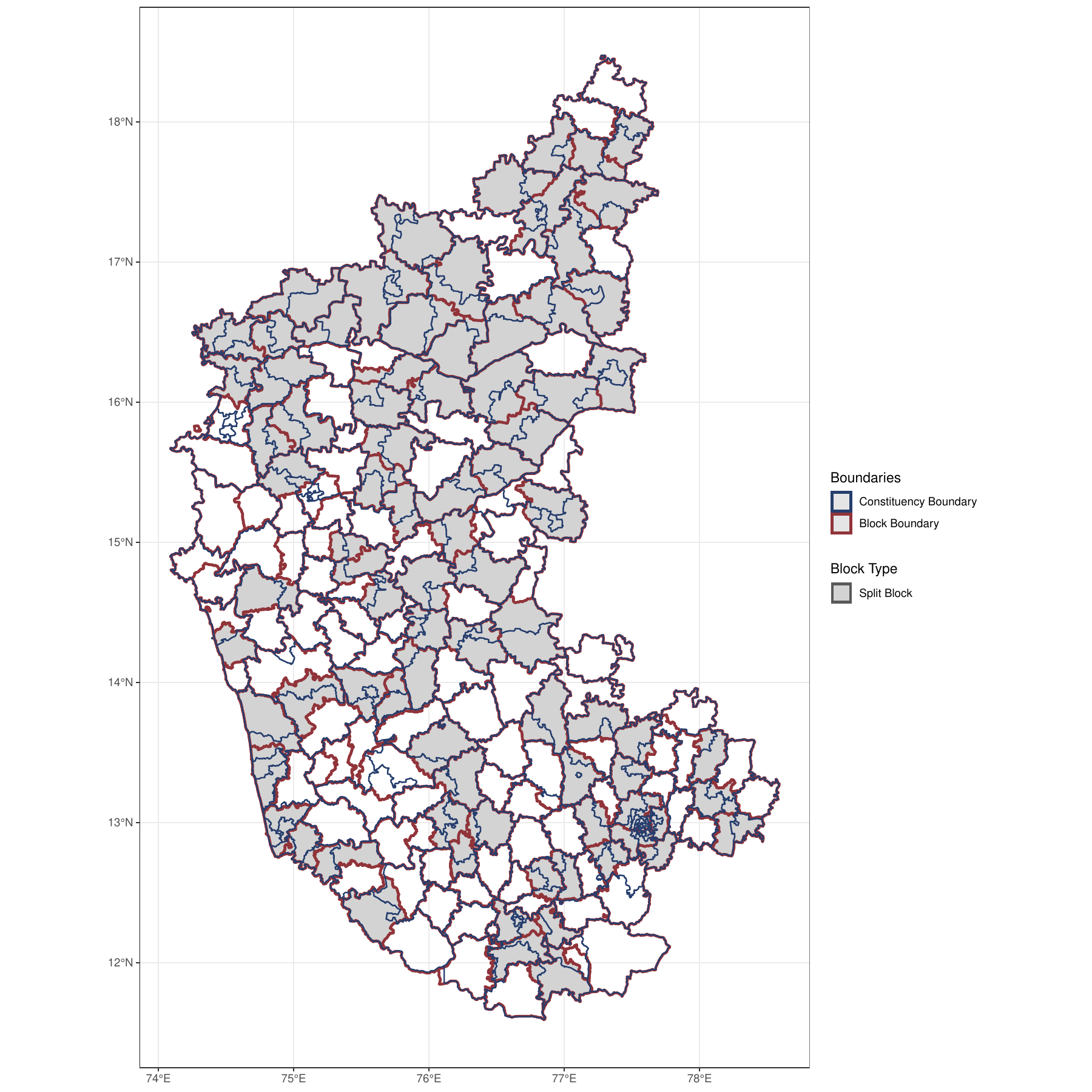}
    \caption{Split \& Unsplit Blocks}
    \label{app_fig_overlap_b}
    \end{subfigure} %
    \label{app_fig_overlap}
    \begin{minipage}{1.00\textwidth}
	\begin{center}
		\end{center}
		{\footnotesize 	The figure presents the overlap of the administrative and electoral boundaries, as an example, from the state of Karnataka and the formation of split and unsplit blocks. The administrative boundaries in red are based on the basic administrative unit - block. The electoral boundaries shown in blue are based on the 2008 delimitation. Figure \ref{app_fig_overlap_a} shows the overlap of the electoral and administrative boundaries. Figure \ref{app_fig_overlap_b} presents a graphic depiction of split blocks in grey and unsplit blocks in white. \par}
\end{minipage}
\end{figure}

\clearpage
\newpage
\begin{figure}[!htbp]
	\centering
	\caption{Geography of Blocks Administered by Multiple Politicians}
	\includegraphics[width=0.55\textwidth]{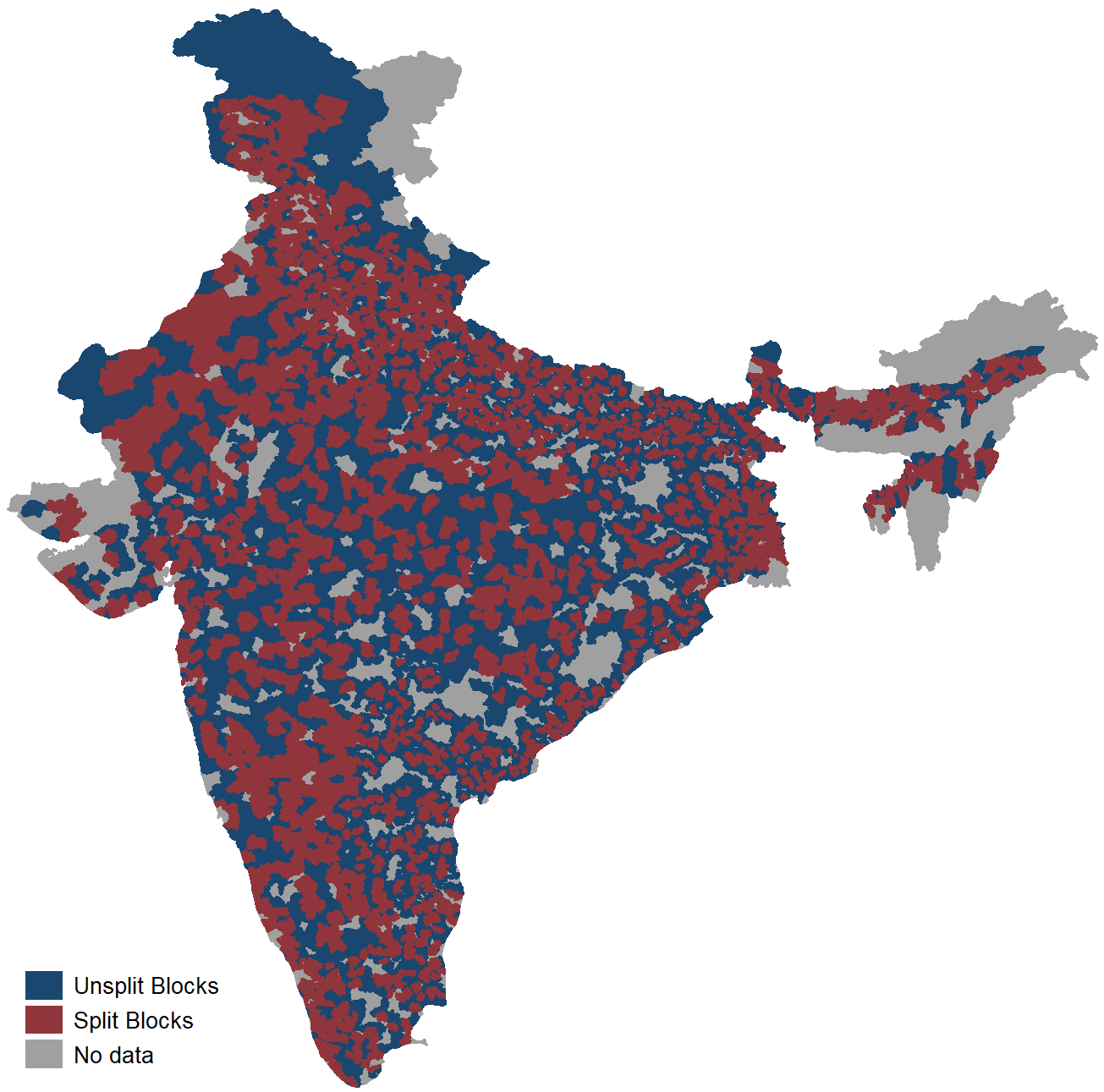}
	\label{map_split_a}
	\begin{minipage}{0.8\textwidth}
{\footnotesize This figure plots the geographic distribution of split and non-split blocks across all blocks in India. A block is defined as a split block if it is split across two or more state legislative assembly constituencies, with at least two constituencies, each accounting for at least 10\% of the geographic area of the block based on 2008 assembly constituency boundaries. 
\par}
	\end{minipage}
\end{figure}

\begin{figure}[!htbp]
	\centering
	\caption{Distribution of Blocks by Number of Governing Politicians}
	\includegraphics[width=0.5\textwidth]{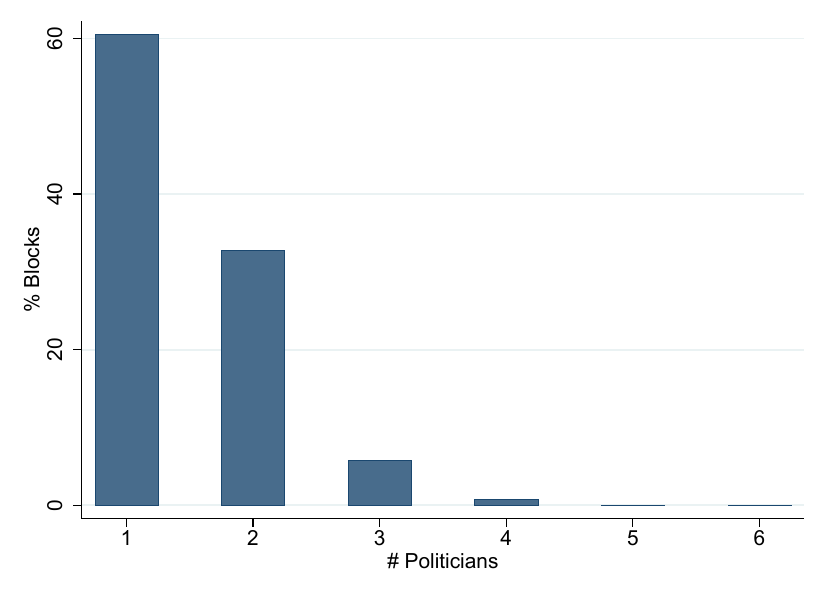}
	\label{app_fig_dist_num}
	\begin{minipage}{0.6\textwidth}
{\footnotesize The figure illustrates the distribution of blocks by the number of governing politicians. A block is defined as a split block based on the haphazard overlap of block boundaries with electoral boundaries as per the 2008 delimitation.
\par}
	\end{minipage}
\end{figure}

\clearpage
\newpage

\section{Data}

\subsection{Firm Geolocation Methodology}
\label{sec_app_firm_data}
This subsection outlines the methodology used to approximate the locations of companies registered with the Ministry of Corporate Affairs (MCA) in India. We also discuss potential biases associated with the approximation algorithm and provide robustness checks to validate our baseline estimates against these concerns.

\subsubsection{Geolocating Companies Registered with the MCA}
We follow a three-step process to geolocate companies registered with the Ministry of Corporate Affairs (MCA):

\begin{itemize}
\item \textbf{Step 1:} Extract the pincode from the company's registered address as provided in the MCA dataset.\footnote{A pincode is a six-digit code in the Indian postal code system used by India Post to simplify the manual sorting and delivery of mail by eliminating confusion over incorrect addresses. It is similar to the concept of zipcode in the United States.}

\item \textbf{Step 2:} Estimate the approximate geographic coordinates of each pincode by using the addresses of corresponding post offices.

\item \textbf{Step 3:} Determine the approximate geographic location of companies using the pincode extracted in Step 1 and the coordinates derived in Step 2.

\end{itemize}

\subsubsection*{Step 1: Extract the pincode from the company's registered address}
Section 12 of the Companies Act of 2013 mandates that all businesses, including Limited Liability Partnerships (LLPs), must have a registered office either at the time of incorporation or within 30 days of their registration. This registered office serves as the official address of the company for all communications and legal purposes. The promoters of the company are responsible for designating a specific location as the registered office and must ensure that this address is formally registered with the Ministry of Corporate Affairs (MCA). \\

We use a regular expression (regex) pattern to extract the pincode from the company's registered address. In India, a valid pincode is always six digits long, does not start with zero, has a first digit ranging from 1 to 9, and the remaining five digits can be any number from 0 to 9. The regex pattern ensures that the extracted pincode adheres to this standard format. Table \ref{app_data_firm_address} presents a few illustrative examples of the address and the extracted pincodes, highlighted in green.

\begin{table}[H]
\centering
\caption{Illustrative Examples of Registered Addresses with Extracted Pincodes}
\label{app_data_firm_address}
\begin{threeparttable}
\begin{tabular}{lp{10cm}}
   \toprule \toprule
\textbf{Company Identification Number} & \textbf{Registered Address} \\
\midrule
U01110KA2016PTC097248 & Salughatte Farmers Producer Company Limited, \newline Sira TQ 138, Hunasekatte, Sira, Karnataka \colorbox{lime}{572115} \\\\
U15323CT2014PTC001592 & Jai Hanuman Agroprocessing Pvt. Limited, \newline M.G. Road, beside Makhan Vihar, H.No. 05,\newline Ambikapur, Chhattisgarh \colorbox{lime}{497001} \\\\
U25114OR2006PTC008532 & Utkal Retread Pvt. Ltd., Keonjhar, \newline Plot No. 2577/3298, AT/PO: Raisuar, Odisha \colorbox{lime}{758013} \\
\bottomrule
\bottomrule
\end{tabular}
\begin{tablenotes}
\footnotesize	
\item 
\end{tablenotes}
\end{threeparttable}
\end{table}

\subsubsection*{Step 2: Approximate geographic coordinates of a pincode}
The country is divided into multiple pin zones, each corresponding to a specific geographic region. A single pincode can be served by more than one post office. To approximate the location of firms within a pincode, we use the location of one of the post offices in that pincode. The following steps outline how we estimate the geographic coordinates for each pincode.

\begin{enumerate}
\item Collection of Post Office Data
\begin{itemize}
\item We obtain a dataset of post offices, including their names, addresses, and the pincodes they serve from India Post
\item This dataset provides attributes for 154,797 post offices across the country serving 19,100 unique pincodes
\end{itemize}
\item Approximating Pincodes to Post Offices
\begin{itemize}
\item We approximate the location of each pincode by first sorting the associated post offices alphabetically by name\footnote{Our estimates are robust to this choice, and we further test the robustness of this selection method in subsection \ref{subsec:postoffice_robustness_checks}}
\item  We then select the first post office in the list to represent the pincode's geographic location, ensuring a consistent and singular geographic point for each pincode
\end{itemize}
\item Geocoding Post Office Addresses
\begin{itemize}
\item We submit the address of the post office chosen in the previous step to Google Maps API for geocoding
\item  The API returns the latitude and longitude of the address, representing the geographic coordinates of the post office
\end{itemize}
\end{enumerate}

\noindent\textbf{Illustrative Example}\\

To illustrate the process of approximating pincode coordinates, consider pincode \textit{\textbf{225303}}, which is served by three post offices: \textit{Basauli B.O}, \textit{Mohammad Pur Khaley S.O}, and \textit{Ram Mandai B.O}, all located within the Taluk of Fatehpur in the Barabanki district of Uttar Pradesh (see Table \ref{app_data_firm_pincode_po}). 

\begin{table}[H]
\centering
\caption{Post Offices serving the Pincode \textit{225303}}
\label{app_data_firm_pincode_po}
\begin{tabular}{clclll}
\toprule
\textbf{Sno} & \textbf{Post Office Name} & \textbf{Pincode} & \textbf{Taluk} & \textbf{District} & \textbf{State} \\ \midrule
1 & Basauli B.O & 225303 & Fatehpur & Barabanki & Uttar Pradesh \\ 
2 & Mohammad Pur Khaley S.O & 225303 & Fatehpur & Barabanki & Uttar Pradesh \\ 
3 & Ram Mandai B.O & 225303 & Fatehpur & Barabanki & Uttar Pradesh \\ \bottomrule
\end{tabular}
\label{tab:post_office}
\end{table}
\noindent
We approximate the geographic coordinates of this pincode using the following steps:

\begin{enumerate}

\item \textbf{Alphabetical Sorting of Post Offices:} The post offices are sorted alphabetically by their names, resulting in the sequence: \textit{Basauli B.O}, \textit{Mohammad Pur Khaley S.O}, and \textit{Ram Mandai B.O}.

\item \textbf{Selection of First Post Office:}  From the alphabetically sorted list, \textit{Basauli B.O} is the first post office and is selected to represent the geographic location of pincode 225303. 

\item \textbf{Geocoding the Selected Post Office:} The address of the selected post office, "\textit{Basauli B.O, Fatehpur Taluk, Barabanki District, Uttar Pradesh}", is then geocoded using the Google Maps API. The API returns the latitude and longitude for the location of \textit{Basauli B.O}, which are then used as the approximate geographic coordinates for pincode 225303.
\end{enumerate}

\subsubsection*{Step 3: Approximate geographic coordinates of companies}

Using the pincode extracted from a company's registered address in \textbf{Step 1}, we approximate the company's geographic location based on the coordinates of its associated pincode, as determined in \textbf{Step 2}. We assume that a company's registered office, as recorded in the Ministry of Corporate Affairs (MCA) dataset, is generally located near the post office serving its pincode.   

\subsubsection*{Data Attrition}
The original Ministry of Corporate Affairs (MCA) dataset comprises of 908,267 firms. We are able to retain 900,396 firms after employing the regex pattern matching to extract pincodes form their registered address. The pincode extraction process results in 7,871 firms (0.86\%) being dropped due to invalid or missing pincodes in their addresses. Finally, we were able to geolocate 825,980 firms to villages, based on the 2001 census definitions, encompassing 91.7\% of the original dataset.

\subsubsection{Accuracy of Custom Geocoding Method}
We assess the accuracy of our geocoding process by comparing the coordinates generated by our methodology with those obtained from an external service, the Latlong Maps API. We perform this validation on a sample of 195,000 firms registered across four states -- Gujarat, Uttar Pradesh, Karnataka, and Tamil Nadu. We measure the reliability of our method by calculating the distance between the firm geo-coordinates based on our methodology and firm geo-coordinates provided by Latlong Maps API. We express the distance error between the two sets of coordinates in kilometers (km). Table \ref{table:validation_results} summarizes the statistical distribution of the distance error.

\begin{table}[H]
\centering
\caption{Statistical Distribution of Distance Errors (in Km)}
\label{table:validation_results}
\begin{threeparttable}
\begin{tabular}{cccccc}
\toprule
\textbf{\# Firms} & \textbf{p25} & \textbf{p50} & \textbf{p75} & \textbf{p90} & \textbf{Mean} \\ \midrule
195,000  & 0.75 & 1.58 & 3.83 & 11.27 & 6.42  \\ 
\bottomrule
\end{tabular}
\begin{tablenotes}
\footnotesize   
\item This table presents the statistical distribution of distance errors in kilometers (km) between the coordinates derived from our geocoding methodology and those obtained from the Latlong Maps API. The columns display the number of firms, along with the 25th percentile (p25), median (p50), 75th percentile (p75), 90th percentile (p90), and mean distance errors, with the mean calculated after winsorizing the right tail at 1\%
\end{tablenotes}
\end{threeparttable}
\end{table}

Table \ref{table:validation_results} shows that the median distance error between our geocoding results and those from the Latlong Maps API is 1.58 km. Given that the median block area is 313 square km, corresponding to a median block diameter of roughly 20 km, this error is relatively minor. This suggests that our geocoding method accurately places firms within the correct block.  \\

To evaluate the robustness of our baseline results to our geocoding approach, we re-estimate our baseline specification using data derived from both our custom geocoding method and the Latlong Maps API. The results, shown in Table \ref{table:geocode_comparison}, indicate that our geocoding method does not materially affect the study's conclusions. In fact, comparing the estimates in Panel A (based on our methodology) and Panel B (based on geo-coordinates from Latlong API) suggests that the measurement error in firm location due to our approximation methodology likely leads to a downward bias in the baseline estimates for this sample.

\begin{table}[H]
  \centering
    \caption{RD Estimates: Using Different Geocoding Approaches}
    \label{table:geocode_comparison}%
    \begin{threeparttable}
    \begin{tabular}{lcccc}
    \hline \hline 
    \multicolumn{5}{c}{\textit{Panel A: Data constructed using Custom Geocoding Method}}\\
    \hline
    Dep Var: LN(0.001+\# New Firms) & (1)   & (2)   & (3)   & (4) \\
    \midrule
    \textbf{Split (=1)} & 0.0663$^{***}$ & 0.0605$^{***}$ & 0.0587$^{***}$ & 0.0558$^{***}$\\
& (0.0135) & (0.0116) & (0.0112) & (0.0111)\\
\midrule
{R$^2$} & 0.11139 & 0.09334 & 0.08321 & 0.07799\\
{Observations} & 80,790 & 105,325 & 119,053 & 122,257\\
    Bandwidth & 5 KM  & 10 KM & 20 KM & 50 KM \\
    Boundary FE & Yes & Yes & Yes & Yes \\
    \bottomrule \\
    \multicolumn{5}{c}{\textit{Panel B: Data constructed using Latlong Maps API}}\\
    \hline
    Dep Var: LN(0.001+\# New Firms) & (1)   & (2)   & (3)   & (4) \\
    \midrule
    \textbf{Split (=1)} & 0.0873$^{***}$ & 0.0955$^{***}$ & 0.0866$^{***}$ & 0.0737$^{***}$\\
& (0.0199) & (0.0176) & (0.0179) & (0.0189)\\
\midrule
{R$^2$} & 0.20503 & 0.18110 & 0.16414 & 0.15450\\
{Observations} & 80,790 & 105,325 & 119,053 & 122,257\\
    Bandwidth & 5 KM  & 10 KM & 20 KM & 50 KM \\
    Boundary FE & Yes & Yes & Yes & Yes \\
    \hline \hline 
    \end{tabular}%
      \begin{tablenotes}
\footnotesize   
\item This table presents RD estimates comparing the effectiveness of two geocoding methods: our custom method and the Latlong Maps API. Panel A uses data constructed via our custom geocoding method, while Panel B uses data from the Latlong Maps API. The dependent variable is the natural logarithm of 0.001 plus the number of new firms. We use different bandwidths of 5, 10, 20, and 50 km on either side of the boundary separating a split block from an unsplit block in columns 1, 2, 3, and 4, respectively. Each regression includes boundary fixed effects, a local linear specification estimated separately on each side of the boundary, and a triangular kernel. Standard errors are clustered at the block level. \sym{*} \(p<0.1\), \sym{**} \(p<0.05\), \sym{***} \(p<0.01\). \\
\end{tablenotes}
\end{threeparttable}
\end{table}

\subsubsection{Robustness Checks on Post Office Selection Within Pincodes\label{subsec:postoffice_robustness_checks}}

While our initial validation confirms the overall accuracy of our geocoding method, we recognize that the choice of post office within each pincode could introduce bias into our results. This issue is especially relevant when a pincode spans both split and unsplit blocks. Each pincode can be served by multiple post offices. In our geocoding process, we select the first post office listed in alphabetical order to represent the pincode. This selection method may bias our estimates if these chosen post offices are disproportionately located within split blocks. To address this concern and further validate our methodology, we conduct additional analyses on the impact of post office selection within pincodes. The following section outlines our approach and presents the results of our robustness checks.\\

\subsubsection*{Distribution of Split and Unsplit Blocks within each pincode}
We start by analyzing the spatial distribution of split and unsplit blocks within each pincode. By utilizing shapefiles for pincodes and administrative blocks, we observe their overlap and measure the fraction of the area occupied by split blocks for each pincode using the following expression:

$$\text{Fraction(Split)} = \sum_{i = 1}^{N} w_i \times 1(\text{Split})$$

\noindent here, \( N \) denotes the total number of blocks within the pincode, \( i \) is the index for each block, \( w_i \) represents the fractional area of block \( i \) within the pincode, and \( 1(\text{Split}) \) is an indicator function that equals 1 if block \( i \) is a split block and 0 otherwise. Note that if the value of Fraction(Split) is close to 0, it indicates that the pincode is entirely within unsplit blocks. Conversely, a value close to 1 suggests that the pincode is fully contained within split blocks. \\

\begin{figure}[H]
    \centering
    \caption{Visualization of Pincode Overlap Across Split and Unsplit Blocks}
    \includegraphics[width=0.35\textwidth]{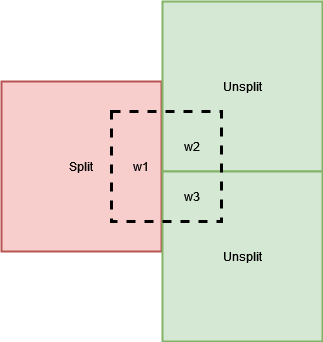}
    \label{fig:split_unsplit_distribution}
    \begin{minipage}{0.65\textwidth}
    \footnotesize This figure illustrates the overlap of a pincode with split and unsplit blocks. The shaded squares denote the different block types, with the red area representing a split block and the green areas indicating unsplit blocks. The dashed outline defines the pincode boundary. The fractional contributions of each block to the area of the pincode are labeled as \( w_1 \), \( w_2 \), and \( w_3 \).
\end{minipage}
\end{figure}

Figure \ref{fig:split_unsplit_distribution} provides a visual representation of this calculation. The figure shows how a pincode overlaps with both split and unsplit blocks. The shaded squares represent different block types: the red square indicates a split block, while the green squares represent unsplit blocks. The dashed outline marks the pincode boundary. The fractional contributions of each block to the pincode area are labeled as as \( w_1 \), \( w_2 \), and \( w_3 \).

\begin{figure}[H]
    \centering
    \caption{Histogram of proportion of area occupied by split blocks within each pincode}
    \includegraphics[width=0.85\textwidth]{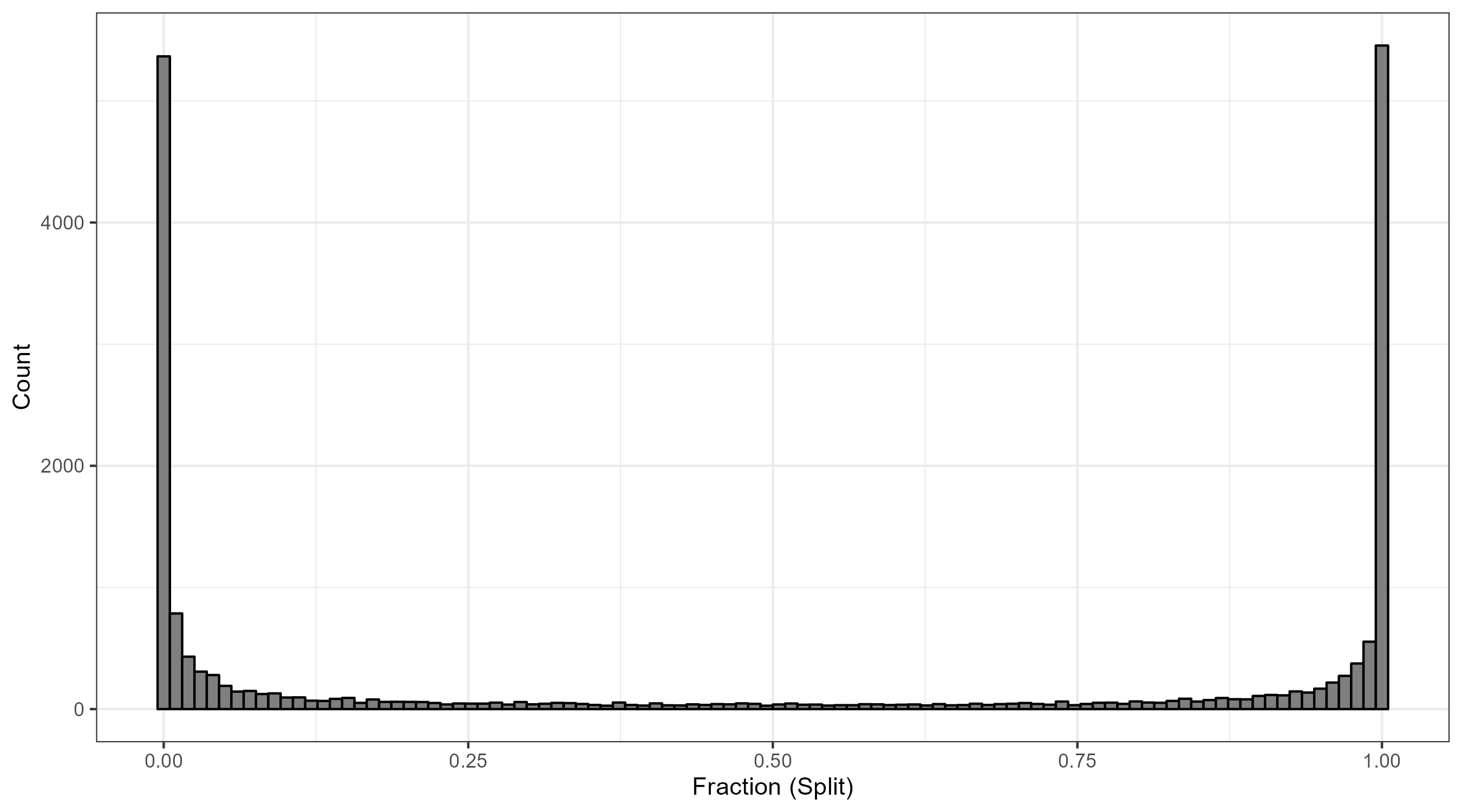}
    \label{fig:split_fraction_distribution}
    \begin{minipage}{0.85\textwidth}
    {\footnotesize This figure displays a histogram depicting the distribution of the fraction of split blocks (Fraction(Split)) across all pincodes. Each bar represents the count of pincodes falling within specified ranges of Fraction(Split), illustrating how the area of split blocks is distributed among the pincodes analyzed.}
\end{minipage}
\end{figure}

Figure \ref{fig:split_fraction_distribution} presents the distribution of the fraction of split blocks across all pincodes. The distribution is predominantly clustered around 0 and 1, indicating that most pincodes are either entirely within split blocks or entirely within unsplit blocks, as reflected by Fraction(Split) values close to 0 or 1. Specifically, 15,559 out of 19,448 pincodes ($\approx$ 80\%) have a \( \text{Fraction(Split)} \) of less than 0.1 or greater than 0.9, indicating that most pincodes are subsumed within split or unsplit blocks. \\

We estimate our baseline specification using only firms located in pincodes that are clearly either entirely within split blocks or entirely within unsplit blocks. This approach ensures that our geocoding method, which approximates firm locations based on the first post office, does not unfairly favor post offices in split blocks. By focusing on these distinctly categorized pincodes, we can confirm that our baseline results are robust and not unlikely to be influenced by selection bias. Table \ref{tab:baseline_specification_fully_subsumed} presents the results of this analysis across three panels, each with progressively stricter criteria for including pincodes.

\begin{table}[H]
\centering
\caption{Estimates for Firms in pincodes that are subsumed within Split and Unsplit Blocks}
\label{tab:baseline_specification_fully_subsumed}
\begin{threeparttable}
\begin{tabular}{lcccc}
   \toprule \toprule
   \multicolumn{5}{c}{\textit{Panel A: Pincodes with \( \text{Fraction(Split)} \) $<$ 0.1 or \( \text{Fraction(Split)}\) $>$ 0.9}}\\
   \midrule
    Dep Var: LN(0.001+\# New Firms) & (1)   & (2)   & (3)   & (4) \\
   \midrule
   \textbf{Split (=1)}     & 0.0441$^{***}$ & 0.0438$^{***}$ & 0.0415$^{***}$ & 0.0416$^{***}$\\
                           & (0.0106)       & (0.0090)       & (0.0083)       & (0.0080)\\     
   \midrule  
    R$^2$                   & 0.20149        & 0.16425        & 0.14214        & 0.12960\\  
   Observations            & 165,922        & 244,004        & 300,788        & 320,556\\ 
   Bandwidth               & 5 KM           & 10 KM          & 20 KM          & 50 KM\\
   Boundary FE             & Yes            & Yes            & Yes            & Yes\\
   \bottomrule\\
   \multicolumn{5}{c}{\textit{Panel B: Pincodes with \( \text{Fraction(Split)} \) $<$ 0.05 or \( \text{Fraction(Split)}\) $>$ 0.95}}\\
   \midrule
    Dep Var: LN(0.001+\# New Firms) & (1)   & (2)   & (3)   & (4) \\
   \midrule
   \textbf{Split (=1)}              & 0.0499$^{***}$ & 0.0489$^{***}$ & 0.0466$^{***}$ & 0.0465$^{***}$\\
                           & (0.0123)       & (0.0104)       & (0.0095)       & (0.0092)\\     
   \midrule  
    R$^2$                   & 0.22580        & 0.18416        & 0.15802        & 0.14298\\
   Observations            & 136,362        & 206,603        & 260,789        & 280,213\\ 
   Bandwidth               & 5 KM           & 10 KM          & 20 KM          & 50 KM\\
   Boundary FE             & Yes            & Yes            & Yes            & Yes\\
   \bottomrule\\
   \multicolumn{5}{c}{\textit{Panel C: Pincodes with \( \text{Fraction(Split)} \) $<$ 0.01 or \( \text{Fraction(Split)}\) $>$ 0.99}}\\
   \midrule
    Dep Var: LN(0.001+\# New Firms) & (1)   & (2)   & (3)   & (4) \\
   \midrule
   g10\_split              & 0.1054$^{***}$ & 0.1021$^{***}$ & 0.0966$^{***}$ & 0.0929$^{***}$\\
                           & (0.0240)       & (0.0197)       & (0.0178)       & (0.0169)\\     
   \midrule  
    R$^2$                   & 0.28650        & 0.23506        & 0.19889        & 0.17754\\
   Observations            & 76,044         & 122,538        & 165,610        & 183,343\\ 
   Bandwidth               & 5 KM           & 10 KM          & 20 KM          & 50 KM\\
   Boundary FE             & Yes            & Yes            & Yes            & Yes\\
   \bottomrule
   \bottomrule
\end{tabular}
\begin{tablenotes}
\footnotesize	
\item This table presents estimates for the natural logarithm of 0.001 plus the number of new firms as the dependent variable. Each panel progressively narrows the sample to include villages in pincodes that are more homogeneously within either split or unsplit blocks. Panel A includes villages in pincodes with \( \text{Fraction(Split)} \) $<$ 0.1 or \( \text{Fraction(Split)}\) $>$ 0.9, Panel B further restricts to \( \text{Fraction(Split)} \) $<$ 0.05 or \( \text{Fraction(Split)}\) $>$ 0.95, and Panel C to \( \text{Fraction(Split)} \) $<$ 0.01 or \( \text{Fraction(Split)}\) $>$ 0.99. The models use different bandwidths of 5, 10, 20, and 50 km on either side of the boundary separating split and unsplit blocks in columns 1, 2, 3, and 4, respectively. The unit of observation is a village that lies within the specified bandwidths. A split block is defined based on the overlap of block boundaries with electoral boundaries as per the 2008 delimitation. Firm entry is calculated as the total number of new firms from 2008 to 2016. All regressions include boundary fixed effects. Standard errors, reported in parentheses, are clustered at the block level. \sym{*} \(p<0.1\), \sym{**} \(p<0.05\), \sym{***} \(p<0.01\)
\end{tablenotes}
\end{threeparttable}
\end{table}

To further assess the robustness of our results against potential biases from arbitrary post office selection, we use a bootstrapping approach. We generate 1,000 samples, retaining the original post office choice for the 15,559 pincodes that are entirely within either split or unsplit blocks. For the 3,889 pincodes that overlap, we geocode all post offices and randomly select one post office per pincode. This introduces variability in firm location approximations. We then infer firm coordinates from these selected post offices and map the firms to villages based on these coordinates. \\ 

\begin{figure}[H]
    \centering       
    \caption{Bootstrap Estimates}
    \begin{subfigure}{0.49\textwidth}
        \centering
        \includegraphics[width=\textwidth]{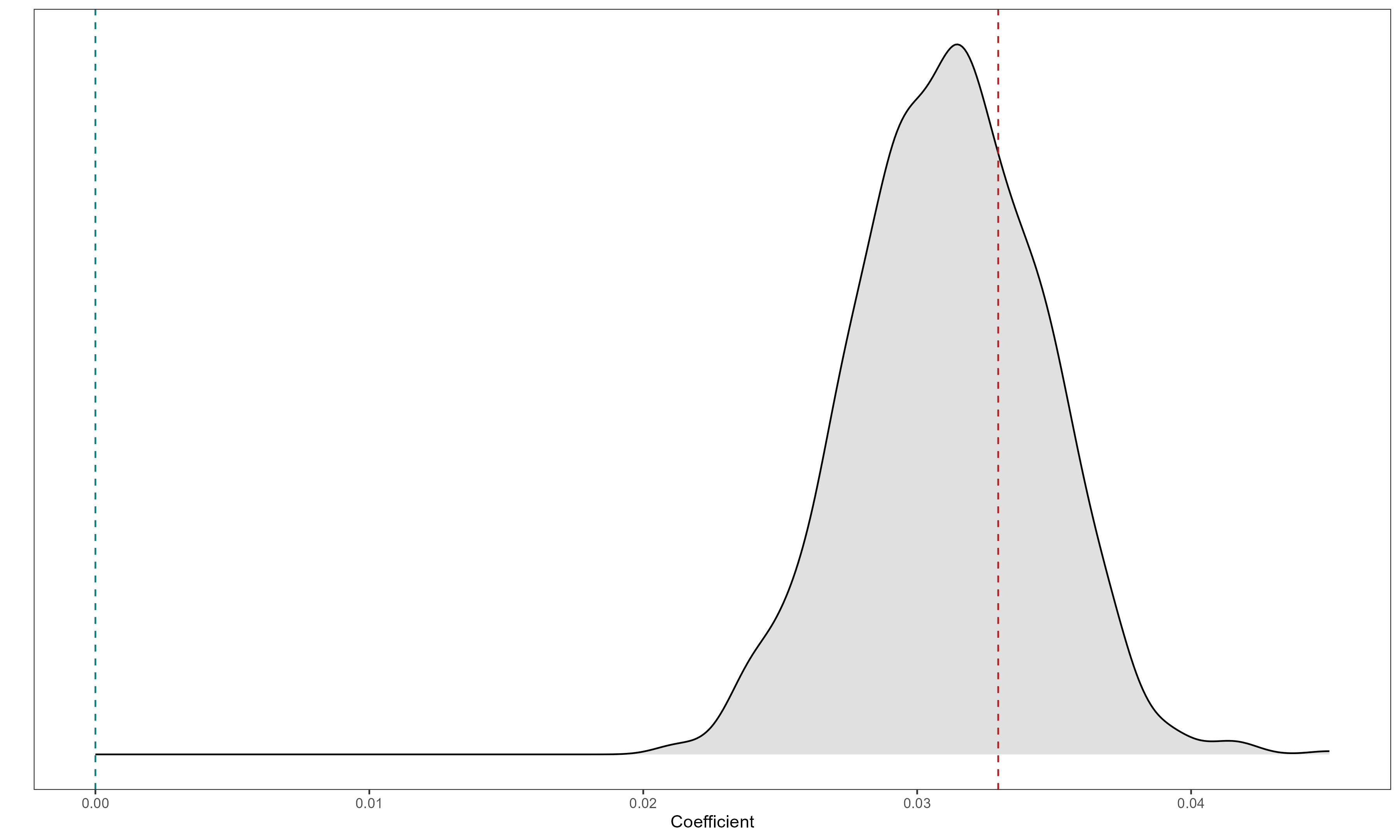}
        \caption{Distribution of coefficient estimates}
        \label{fig:coefficient_plot}
    \end{subfigure}    %
    \begin{subfigure}{0.49\textwidth}
        \centering
        \includegraphics[width=\textwidth]{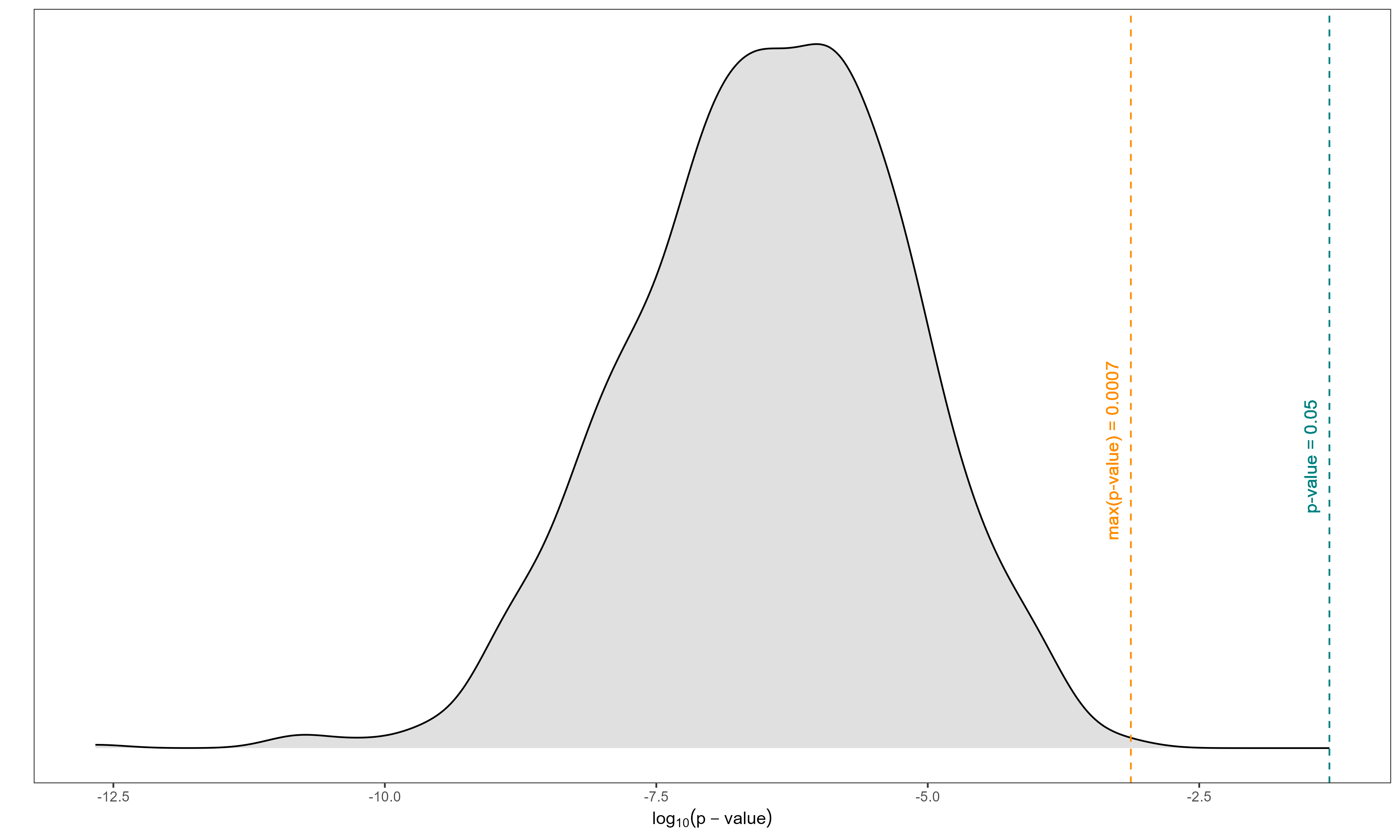}
        \caption{Distribution of log-transformed p-values}
        \label{fig:pval_plot}
    \end{subfigure} %
    \label{fig:bootstrap_plot}
    \begin{minipage}{1.00\textwidth}
    {\footnotesize The figure presents the results of a bootstrapping analysis to test the robustness of the effect of split blocks on firm entry to potential bias from arbitrary post office selection. Panel A shows the density distribution of coefficient estimates from 1,000 bootstrap samples, with the red dashed line indicating the baseline estimate. Panel B displays the distribution of log-transformed p-values from the bootstrap samples. Vertical lines are included to denote the maximum p-value (0.0007) and the conventional significance threshold (0.05)\par}
\end{minipage}
\end{figure}   

Figure \ref{fig:bootstrap_plot} presents the results of the bootstrap analysis. Panel A presents the distribution of coefficient estimates, demonstrating a consistently positive effect of split blocks on firm entry across all samples. The coefficients range from approximately 0.0209 to 0.0450, and none of the 95\% confidence intervals cross zero. Panel B illustrates the distribution of log-transformed p-values, with the maximum p-value around 0.0007, which is significantly below the conventional significance threshold of 0.05 (denoted by the dashed line). These findings indicate that the baseline estimates are stable and reliable, suggesting that they are not artifacts of our custom geocoding method. 

\clearpage
\newpage
\begin{landscape}
\begin{figure}[!ht]
	\centering
	\caption{Geography of Firm Entry and Nightlights}
	\label{map_entry}
	\begin{subfigure}[t]{0.6\textwidth}
		\centering
		\includegraphics[width=\linewidth]{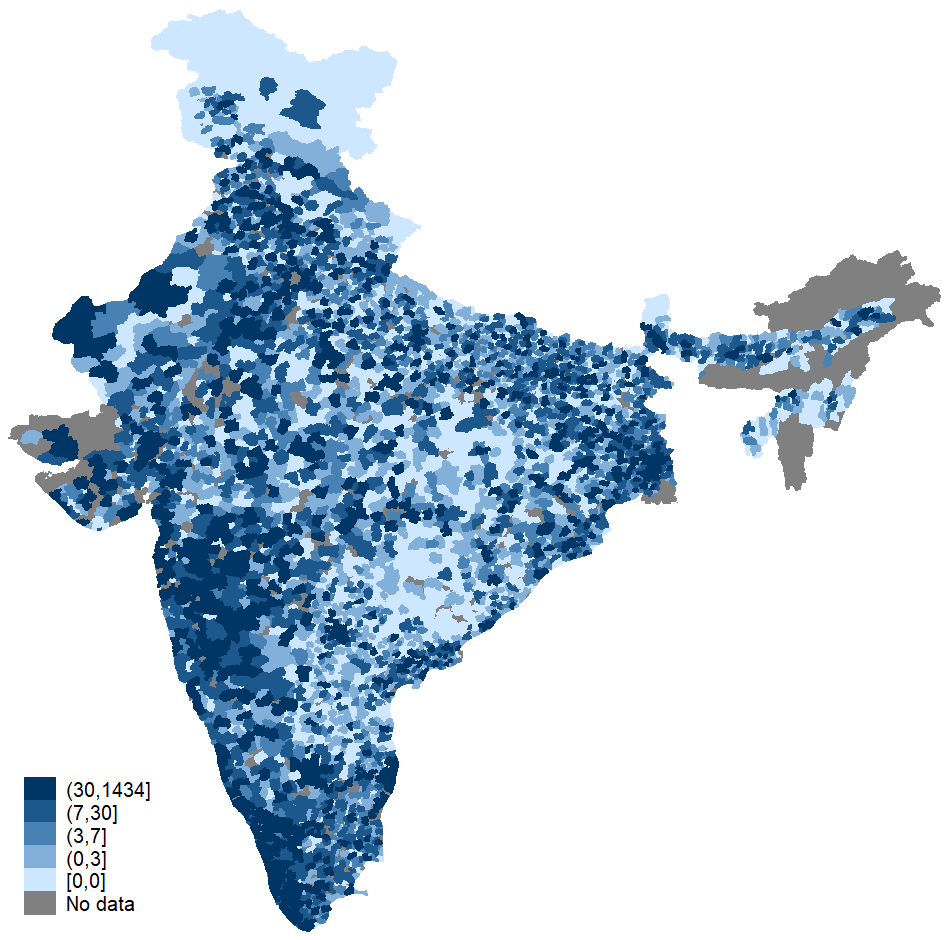}
		\caption{\# New Private Firms} \label{map_entry_a}
	\end{subfigure}
	\begin{subfigure}[t]{0.6\textwidth}
		\centering
		\includegraphics[width=\linewidth]{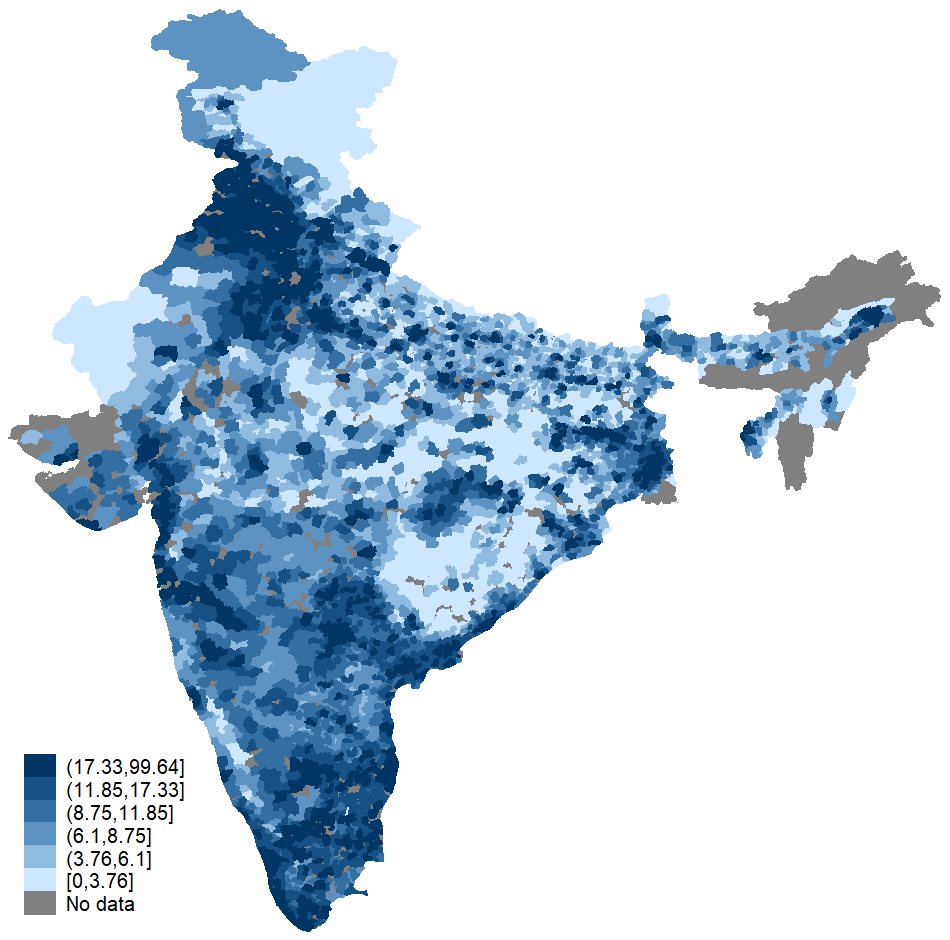} 
		\caption{Average Nightlights} \label{map_entry_b}
	\end{subfigure}
	\begin{minipage}{1.2\textwidth}
		{\footnotesize The figure panel \ref{map_entry_a} plots the geographic distribution of the number of new private firms registered across all blocks in India from 2003 to 2016. The figure panel \ref{map_entry_b} plots the geographic distribution of average night lights across all blocks in India from 2003 to 2016. The sample of private firms includes the universe of all for-profit firms in India registered from 2003 to 2016. The nightlights are standardized to take a minimum value of zero and a maximum value of 100 across blocks for each year. The standardized nightlights in a block are averaged over the period from 2003 to 2016.  Note that we have not verified any boundaries and do not claim authenticity of the same. We do not endorse the geographic boundaries shown here. \par}
	\end{minipage}
\end{figure} 
\end{landscape}

\clearpage
\newpage
\begin{figure}[H]
    \centering
    \caption{Sinuosity of Straight-Line-Like Block Boundary Segment}
    \begin{subfigure}{.4\textwidth}
    \centering
    \includegraphics[width = \textwidth]{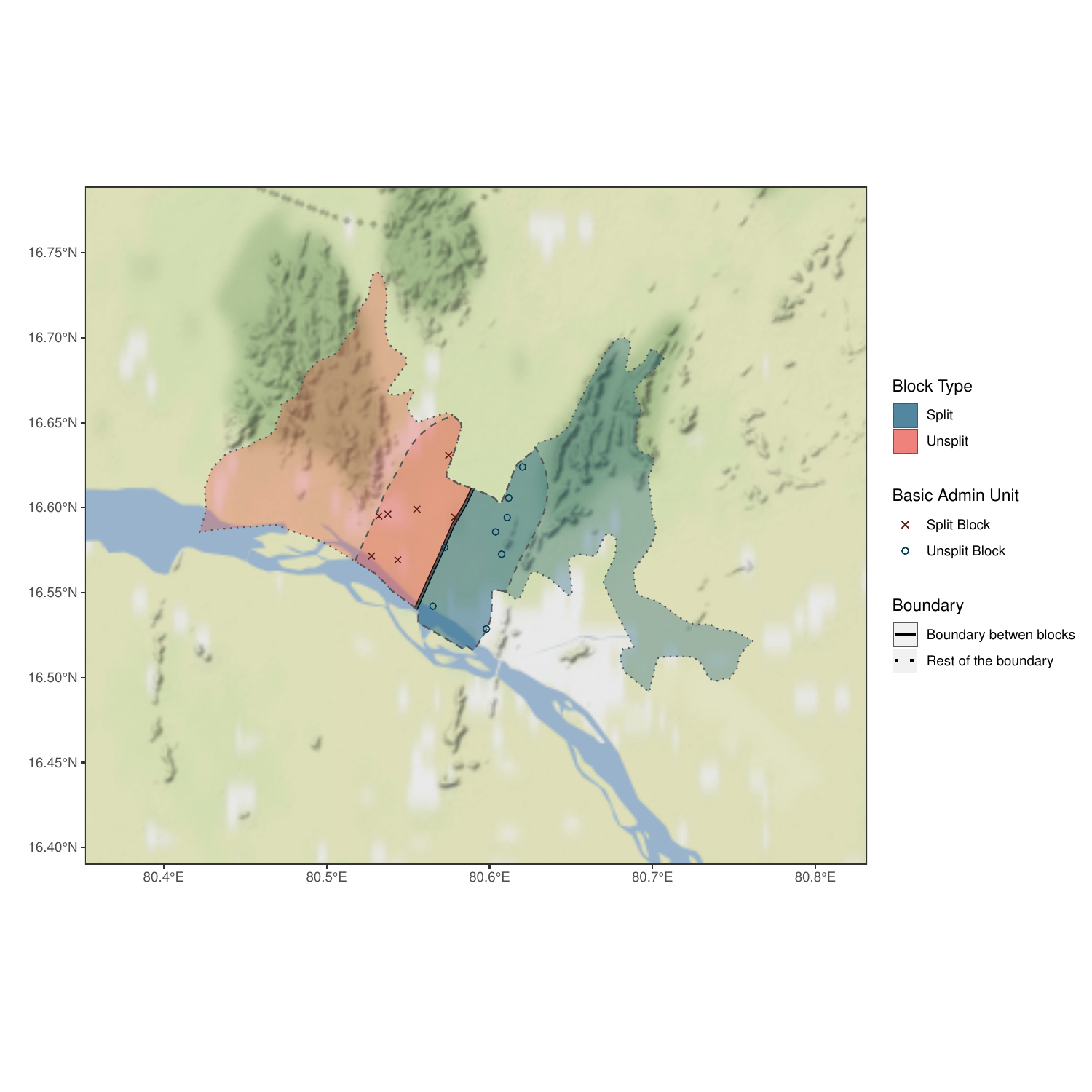}
    \caption{Representative Example}
    \label{straight_rep}
    \end{subfigure} %
    \begin{subfigure}{.4\textwidth}
    \centering
    \includegraphics[width = \textwidth]{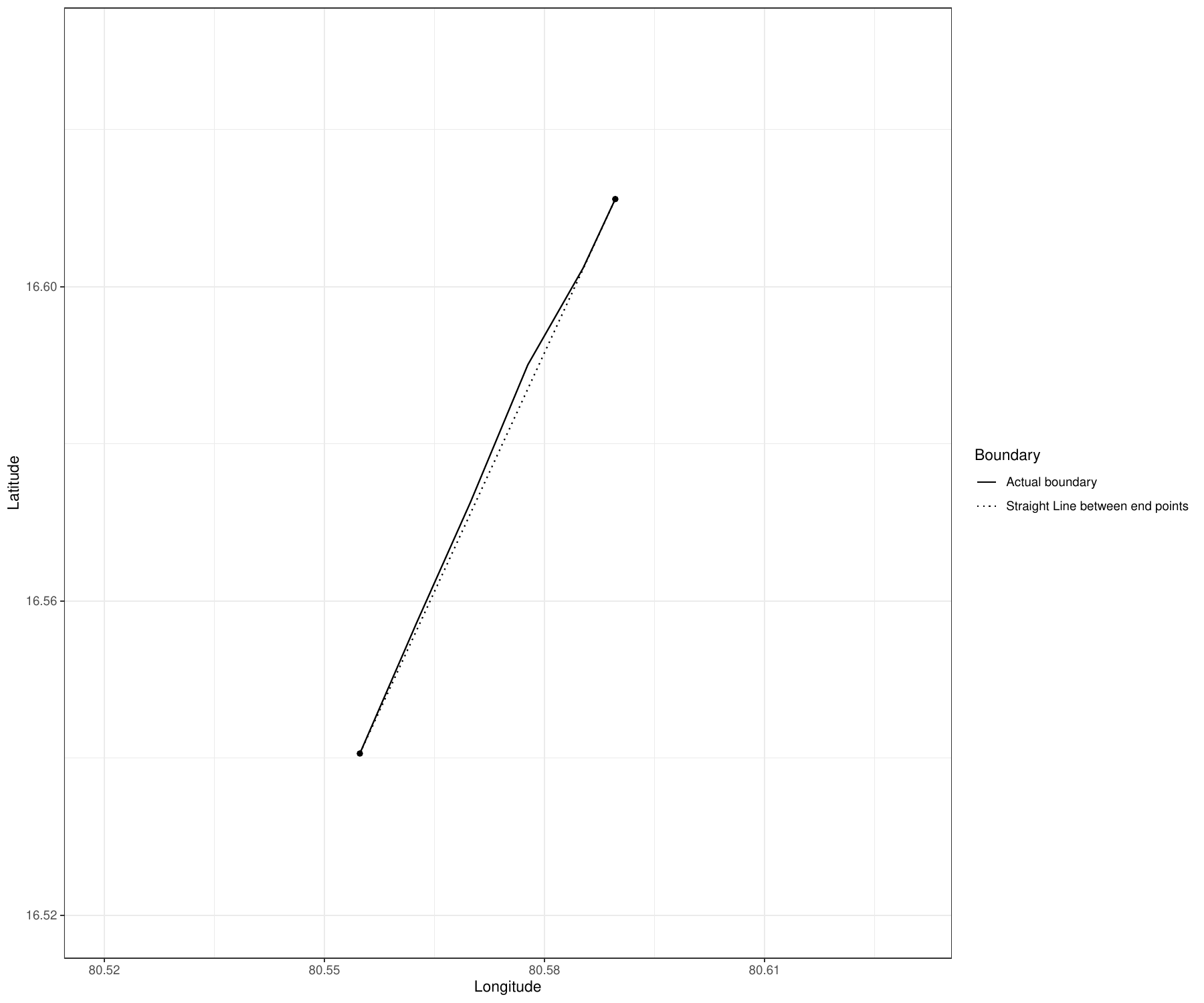}
    \caption{Schematic Diagram}
    \label{straight_sch}
    \end{subfigure} %
    \label{straight_line_ex}
    \begin{minipage}{1\textwidth}
	\begin{center}
		\end{center}
		{\footnotesize Figure \ref{straight_line_ex} illustrates the measure of straightness, \textit{Sinuosity}, using a representative example of straight-line-like boundary. Figure \ref{straight_rep} illustrates the boundary between the blocks, Ibrahimpatnam and Vijayawada (Rural), located in Krishna District, Andhra Pradesh. Figure \ref{straight_sch} compares the block boundary and straight line joining the endpoints of the boundary. The length of the boundary segment is 8.69 km, whereas the shortest distance between the endpoints is 8.68 km. The \textit{Sinuosity} of the segment is 1.000965.   \par}
\end{minipage}
\end{figure}

\begin{figure}[H]
    \centering
    \caption{Sinuosity of a Wiggly Block Boundary Segment}
    \begin{subfigure}{.4\textwidth}
    \centering
    \includegraphics[width = \textwidth]{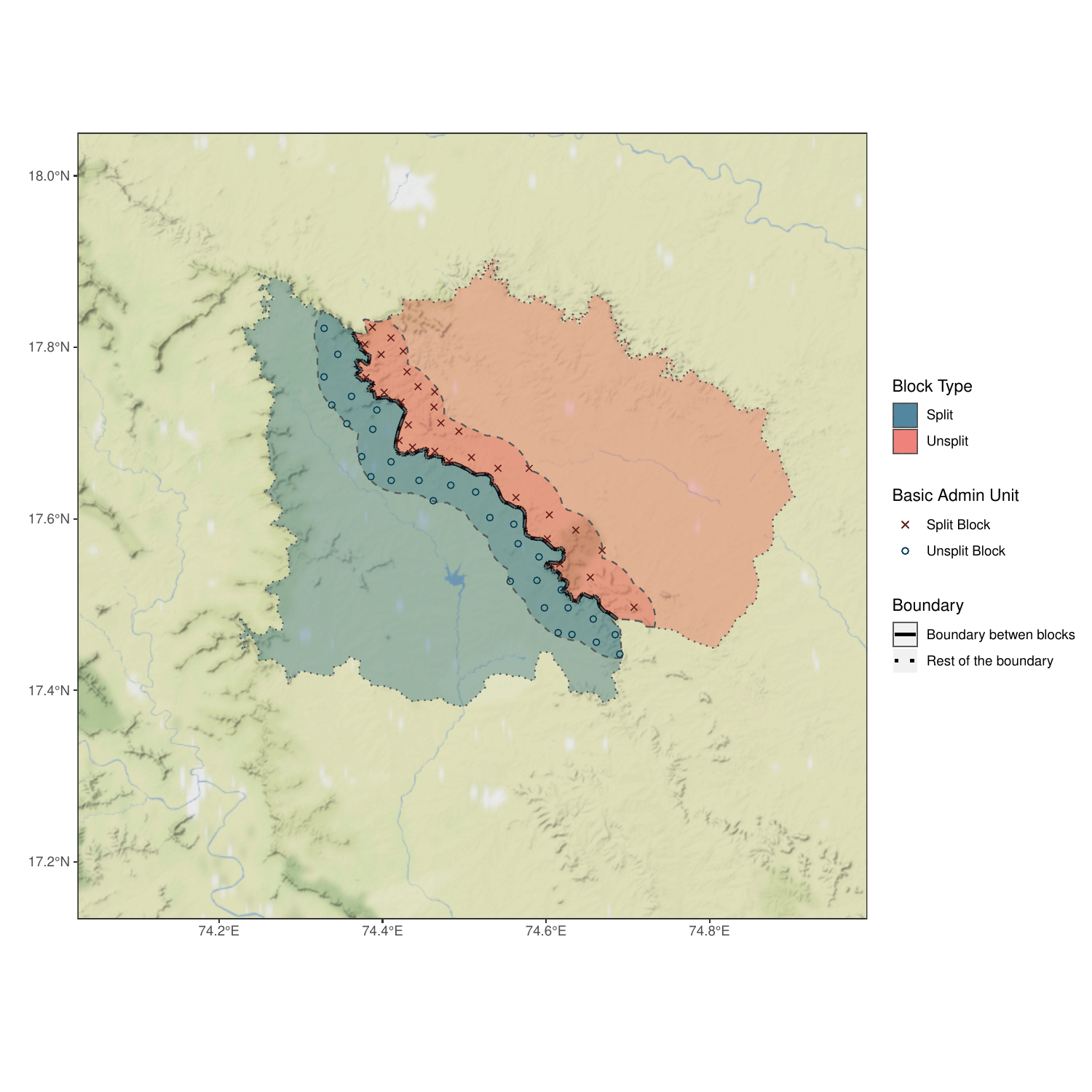}
    \caption{Representative Example}
    \label{wiggly_rep}
    \end{subfigure} %
    \begin{subfigure}{.4\textwidth}
    \centering
    \includegraphics[width = \textwidth]{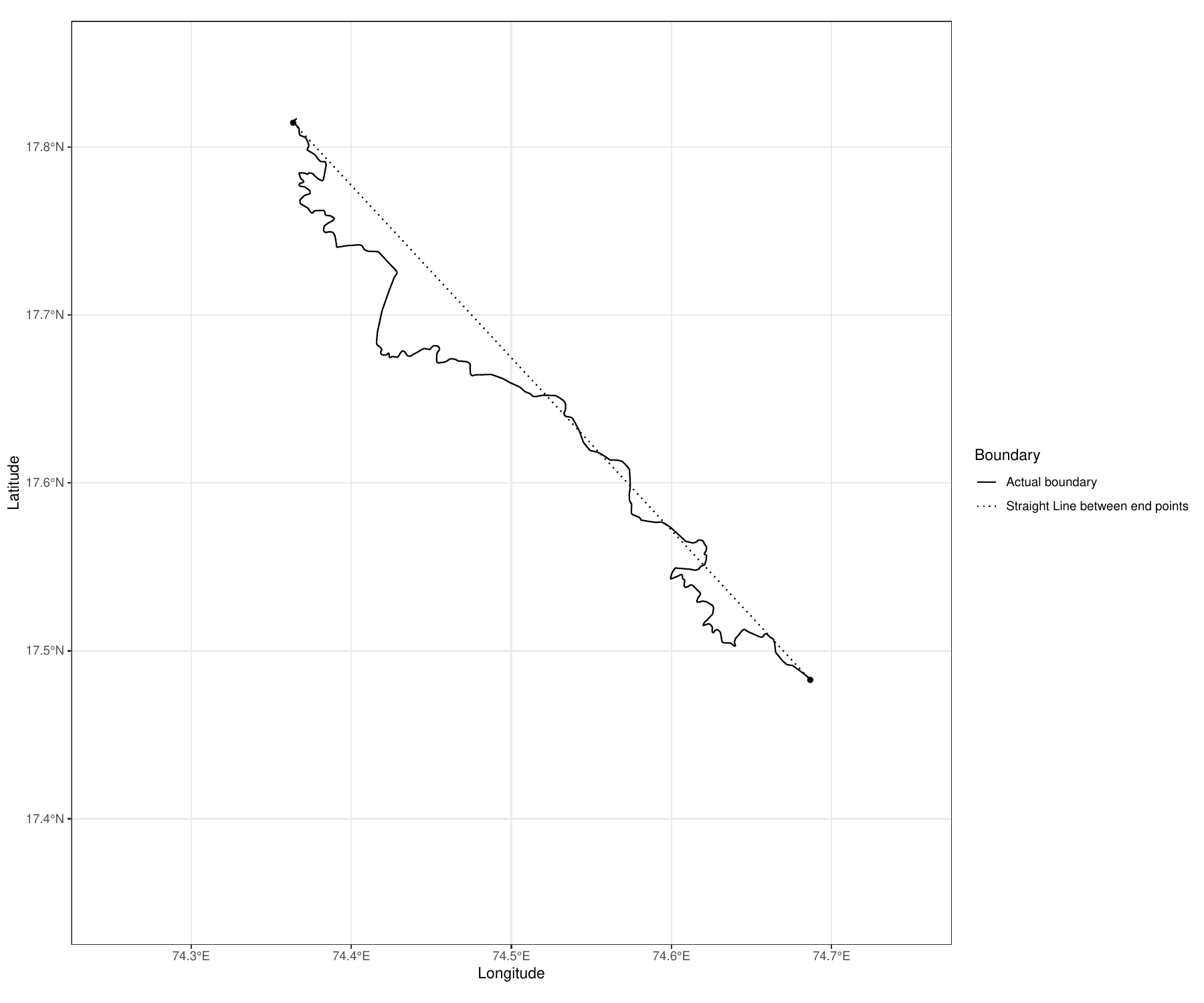}
    \caption{Schematic Diagram}
    \label{wiggly_sch}
    \end{subfigure} %
    \label{wiggly_line_ex}
    \begin{minipage}{1\textwidth}
	\begin{center}
		\end{center}
		{\footnotesize Figure \ref{wiggly_line_ex} illustrates the measure of straightness, \textit{Sinuosity}, using a representative example of wiggly boundary segment. Figure \ref{wiggly_rep} illustrates the boundary between the blocks, Khatav and Phaltan located in Satara District, Maharashtra. Figure \ref{wiggly_sch} compares the block boundary and straight line joining the endpoints of the boundary. The length of the boundary segment is 80.41 km, whereas the shortest distance between the end-points is 50.38 km. The \textit{Sinuosity} of the segment is 1.59. \par}
\end{minipage}
\end{figure}

\clearpage

\begin{table}[htbp]
  \centering
  \caption{Number of Firms}
  \label{app_tab_summary_firms}%
\begin{threeparttable} 
    \begin{tabular}{ccc}
    \toprule
    \toprule
    \multicolumn{1}{l}{Year} & \multicolumn{1}{l}{\# New Firms} & \multicolumn{1}{l}{Nightlights} \\
    \hline 
    2003  & 10,980 &           5.98  \\
    2004  & 15,856 &           7.52  \\
    2005  & 23,051 &           6.72  \\
    2006  & 39,138 &           6.20  \\
    2007  & 58,735 &           7.73  \\
    2008  & 65,299 &           7.39  \\
    2009  & 55,855 &           8.10  \\
    2010  & 80,075 &           8.14  \\
    2011  & 88,833 &           8.10  \\
    2012  & 94,213 &           8.96  \\
    2013  & 83,100 &           9.01  \\
    2014  & 62,858 &         14.64  \\
    2015  & 72,375 &         15.47  \\
    2016  & 75,612 &         15.26  \\
    \hline
    Total/Average & 825,980 & 9.22\\ 
    \bottomrule
    \bottomrule
    \end{tabular}%
 \begin{tablenotes}
\footnotesize	
\item This table presents the number of new firms and the average nightlights index for each year from 2003 and until 2016. \\
\end{tablenotes}
\end{threeparttable}  
\end{table}%

\clearpage
\newpage

\subsection{Caste-Based Patronage Detection}
\label{sec:data_patronage}

\subsubsection{Overview}

This appendix outlines our methodology for identifying firms potentially connected to politicians through shared identity. In India, patronage networks often operate along lines of jati — endogamous groups that shape social organization. Indian surnames frequently carry markers of jati affiliation, providing a basis for detecting such network connections. However, inferring jati from surnames faces substantial challenges arising from heterogeneity in naming conventions and cultural nuances. We address these challenges by implementing a computational pipeline combining machine learning techniques and rich ethnographic documentation. We attempt to go beyond simple surname matching to capture social affiliation through which patronage relationships might realistically operate.

\subsubsection{Data and Sample}

\paragraph{Data Sources} : \\

\noindent\textbf{\textit{Administrative Data.}} The names of politicians and their electoral constituency assignments come from the Socioeconomic High-resolution Rural-Urban Geographic Platform for India (SHRUG) database \citep{almn2021, jv2017}. We obtain the names of firm directors from the Company Register maintained by the Ministry of Corporate Affairs, India. We match firms to politicians based on geographic location, ensuring that our analysis examines potential patronage connections only within the constituencies that politicians represent.\newline

\begin{figure}[H]
    \centering
    \caption{People of India Project Database Structure}
    \label{fig:poi_structure}
    \includegraphics[width=0.9\textwidth]{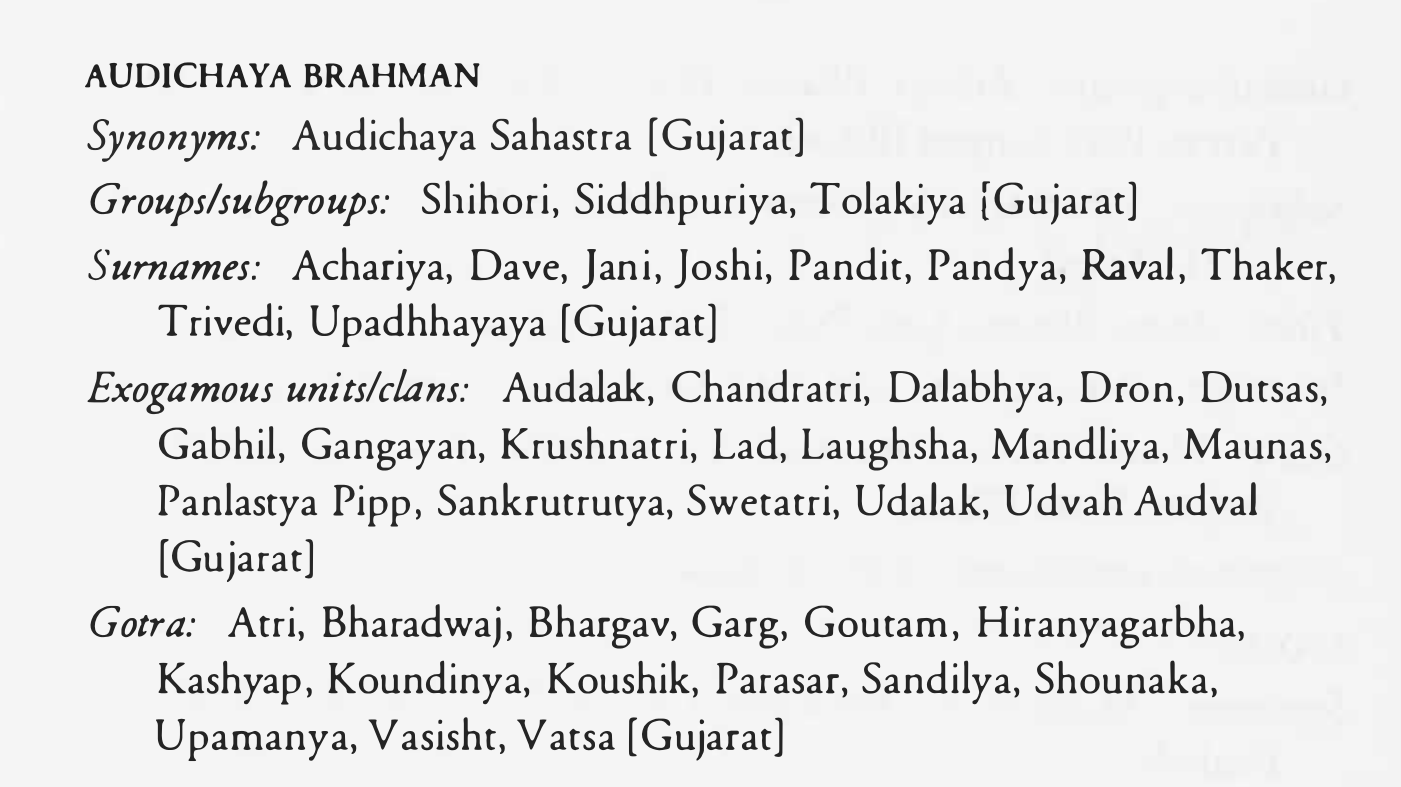}
    \begin{minipage}{0.94\textwidth}
        \begin{center}
        \end{center}
        {\footnotesize The figure displays a representative entry from the People of India Project database for the community (\textit{jati}) ``Audichaya Brahman'' from the state of Gujarat. The entry illustrates the hierarchical structure of ethnographic documentation: \textit{Synonyms} captures alternative community names used regionally; \textit{Groups/subgroups} documents related sub-castes within broader classificatory hierarchies; \textit{Surnames} lists family names associated with the community; \textit{Exogamous units/clans} identifies marriage circles that define permissible unions; and \textit{Gotra} records patrilineal descent groups. The database documents 2,205 such \textit{jatis} across Indian states, with each entry containing multiple surnames and related communities, creating the many-to-many mapping between surnames and \textit{jati} \par}
    \end{minipage}
\end{figure}

\noindent\textbf{\textit{People of India Project Database.}} The People of India Project \citep{singh1996communities} provides ethnographic documentation of Indian communities. Conducted state by state, this project systematically catalogued \textit{jatis}, documenting for each community its associated surnames, regional synonyms, hierarchical sub-groups, exogamous marriage circles, and patrilineal descent groups (gotras). A representative entry for the community (\textit{jati}) ``Audichaya Brahman'', belonging to the state of Gujarat, is illustrated in Figure \ref{fig:poi_structure}.

The relationship between surnames and \textit{jatis} exhibits systematic many-to-many correspondence rather than one-to-one mapping. The methodology followed in the People of India Project documented surnames as communities actually use them, capturing substantial variation within single \textit{jatis} and overlap across related groups. This pattern reflects historical processes—closely related \textit{jatis} often share surnames due to fission from common ancestor communities. Geographic dispersion also produces different surname usage for the same endogamous group, as regional conventions and migration patterns create surname diversity within caste boundaries. Identical surnames may therefore indicate membership in multiple related sub-castes within a given region, while individuals sharing caste identity may carry different surnames.

\clearpage

\paragraph{Sample Construction}: \\ 

\noindent Cultural nuances in Indian naming conventions restrict our analysis to contexts where surname-based \textit{jati} identification operates most reliably.\newline 

\noindent\textbf{\textit{Religious Restriction.}} Surnames carry caste markers primarily for Hindus, where \textit{jati} organization functions most systematically within social structure. We therefore restrict attention to Hindu politicians and firm directors.\newline

\noindent\textbf{\textit{Geographic Restriction.}} North Indian naming practices emphasize hereditary surnames that transmit consistently across generations and carry \textit{jati} affiliation. South Indian conventions prioritize patronymics (father's given name) and village references that change generationally, making surname-based caste identification unreliable. Hence we limit the analysis to eleven North Indian states: Himachal Pradesh, Haryana, Rajasthan, Uttar Pradesh, Bihar, Jharkhand, Orissa, Chhattisgarh, Madhya Pradesh, Gujarat, and Maharashtra.\newline

\noindent\textbf{\textit{Surname-Based Restriction.}} Some surnames (such as Kumar and Singh) function as generic honorifics or titles that diffused across multiple communities, weakening their reliability as caste markers. We exclude politicians with such caste-neutral surnames from the sample. \\

\noindent These restrictions yield a final sample of 915 Hindu politicians from an initial universe of 1,805, matched to 86,767 unique firms operating in their constituencies. This sample provides complete coverage of contexts where our methodology enables reliable patronage detection, avoiding false connections that might arise from uncertain caste assignments or unreliable naming conventions.

\subsubsection{Technical Implementation}

\paragraph{Overview}

In this subsection we detail the methodology used to process politician and director names, following a sequential pipeline to address technical challenges that commonly arise when handling Indian names. Stage 1 classifies religion to identify individuals with Hindu names, for whom \textit{jati} represents a salient identity marker. Stage 2 extracts surnames from complex naming conventions that vary substantially across Indian states. Stage 3 handles transliteration differences and non-standard spellings through embedding-based fuzzy matching with the People of India database, allowing us to infer the \textit{jati} of the politician from their surname despite orthographic variations. Stage 4 implements expanded caste mapping by first building surname inventories for each politician's \textit{jati}, then matching directors against these expanded surname lists to identify potential identity-based connections.

\paragraph{Stage 1: Religion Classification}

We build a classifier to identify individuals with Hindu names for whom surname-based \textit{jati} inference operates systematically. The classifier employs a two-stage approach balancing computational efficiency with accuracy. Initial screening uses character-level Long Short-Term Memory neural networks trained on proprietary matrimonial datasets containing 1.3 million entries on persons' names and religion.\footnote{We implement three specialized binary classifiers—one each for Hindu, Muslim, and Christian identification—rather than a single multi-class model to capture religion-specific linguistic patterns more effectively. Each model processes character sequences through bidirectional LSTM layers with dropout regularization.} We set a high classification standard where only names receiving high-confidence predictions above 0.9 probability proceed automatically. Names with prediction probabilities below this threshold require more sophisticated analysis and are routed to an ensemble of large language models comprising \textit{Qwen 2.5}, \textit{GPT-4o Mini}, and \textit{Gemini 2.0}. These models analyze names using cultural and linguistic knowledge beyond character patterns, with structured output schemas preventing classification errors.\footnote{Final classifications employ majority voting across the ensemble, with validation on out-of-sample matrimonial data ensuring reported performance metrics reflect genuine generalization rather than overfitting.} 

Table \ref{tab:religion_classification} reports classification performance metrics on holdout validation data. High precision (96.3\%) ensures that downstream caste inference operates on reliably identified Hindu names, while high recall (98.2\%) minimizes systematic exclusion of Hindu politicians from the analysis sample. This filtering produces a subsample of Hindu politicians where subsequent surname extraction and caste mapping achieve maximum reliability.

\begin{table}[htbp]
    \centering
    \caption{Religion Classification Performance}
    \label{tab:religion_classification}
    
    \begin{subtable}[t]{0.45\textwidth}
        \centering
        \caption{Confusion Matrix}
        \begin{tabular}{lcc}
            \toprule
            & \multicolumn{2}{c}{\textbf{Actual Religion}} \\
            \cmidrule(l){2-3}
            \textbf{Predicted} & Hindu & Non-Hindu \\
            \midrule
            Hindu & 751 & 29 \\
            Non-Hindu & 14 & 206 \\
            \midrule
            Total & 765 & 235 \\
            \bottomrule
        \end{tabular}
    \end{subtable}
    \hspace{0.5cm}
    \begin{subtable}[t]{0.45\textwidth}
        \centering
        \caption{Performance Metrics}
        \begin{tabular}{lr}
            \toprule
            \textbf{Metric} & \textbf{Value} \\
            \midrule
            Accuracy & 95.7\% \\
            Precision & 96.3\% \\
            Recall & 98.2\% \\
            F1-Score & 97.2\% \\
            \midrule
            Sample Size & 1,000 \\
            \bottomrule
        \end{tabular}
    \end{subtable}
    
    \begin{minipage}{0.94\textwidth}
        \begin{center}
        \end{center}
        {\footnotesize \textbf{Panel A} displays the confusion matrix for binary religion classification where Hindu identification serves as the positive class. Cell entries represent counts of names classified in each category. \textbf{Panel B} reports key performance metrics validated on a holdout sample of 1,000 names from matrimonial data that the LSTM models have not seen during training, ensuring out-of-sample performance assessment. \textit{Accuracy} measures the proportion of correctly classified names across both categories. \textit{Precision} indicates the percentage of names predicted as Hindu that are actually Hindu (type I error control). \textit{Recall} captures the percentage of actual Hindu names correctly identified (type II error control). \textit{F1-Score} provides the harmonic mean of precision and recall, balancing both error types.\par}
    \end{minipage}
\end{table}

\paragraph{Stage 2: Surname Extraction}

We extract surnames from complex Indian names that exhibit substantial variation across regional naming conventions. Different states employ distinct naming patterns—some emphasizing hereditary surnames, others incorporating village names, patronymic elements, or clan affiliations in varying positions within names. Names such as ``Rajendra Pratap Singh alias Moti'' incorporate aliases that may override formal surname conventions, while ``Jadeja Kandhalbhai Sarmanbhai'' reflects state-specific (here, Gujarat) patterns that differ markedly from conventions in other states. Additional complexity arises from honorific titles, professional designations, and administrative variations that obscure genealogically relevant components. Naive extraction of the last name component therefore fails to capture caste-informative elements consistently across India's diverse naming landscape. 

To address this challenge, we leverage large language models that acquire knowledge of cross-cultural naming conventions during pretraining on diverse text corpora, enabling them to distinguish surnames from other name components across different cultural contexts. We provide each model with the politician's home state as contextual information to guide culturally appropriate surname extraction. Our ensemble comprises five models—\textit{Mistral Nemo}, \textit{Grok 3 Mini}, \textit{Gemini 2.0}, \textit{Qwen 2.5}, and \textit{GPT-4o Mini} — pretrained on large-scale datasets \footnote{\citet{huh2024platonic} demonstrate that AI models trained on large corpora develop converging representations across domains, justifying ensemble approaches for cross-cultural tasks.}. Models analyze name structure independently, with consensus achieved through majority voting to identify surname components most likely to carry caste information. This approach enables systematic extraction of caste-relevant surname candidates across India's heterogeneous naming conventions.

\paragraph{Stage 3: Fuzzy Surname Matching}

Extracted surnames must be matched against the People of India database to infer \textit{jati} associations. However, identical surnames exhibit substantial orthographic variation in administrative records. Transliteration from regional scripts to Roman characters produces inconsistent spellings, where the same genealogical identity appears with different spellings due to transcription practices.\footnote{Common examples include ``Chaudhary,'' ``Choudhary,'' and ``Choudhari,'' or ``Agarwal,'' ``Aggarwal,'' and ``Agrawal''—all representing identical surname identities with different romanization conventions.} Exact string matching would systematically miss valid connections between politicians and database entries, undermining inference of \textit{jati}. 

We address this challenge through embedding-based fuzzy matching that captures not just semantic but also contextual similarity across spelling variants. The approach converts surnames to high-dimensional vectors where orthographically different but contextually identical surnames cluster in proximity. We then enable similarity-based retrieval rather than exact correspondence. We rely on vector databases through Pinecone, which indexes People of India surnames as embeddings and retrieves contextual candidate names. We then use \textit{LLMs-as-a-Judge} to validate candidate matches by confirming whether spelling variants represent identical surname identities versus merely similar-sounding but distinct names, preventing false connections while capturing genuine orthographic variations.\footnote{The LLM-as-a-Judge approach employs large language models to evaluate and validate system outputs against semantic criteria, providing more nuanced assessment than simple statistical metrics by capturing contextual appropriateness and semantic equivalence.}

\paragraph{Stage 4: Expanded Caste Mapping}

Given a politician's surname, we seek to identify directors who belong to the same community. A surname maps to multiple related \textit{jatis} that share common ancestry, regional location, and social position. For each politician's surname, we construct a comprehensive surname inventory by extracting all surnames associated with these related \textit{jatis}, creating an expanded list that broadly captures names indicating membership across the interconnected subcommunities. This inventory represents coherent networks for social interaction and mutual support rather than arbitrary surname groupings. Using this expanded surname inventory, we flag potential connections by matching director names against any surname in the comprehensive list. This approach recognizes that patronage operates through broader community-based networks that reflect the actual social structures through which caste identity functions in Indian society.

\paragraph{Illustrative Application: Abhishek Mishra, MLA (Sarojini Nagar, Uttar Pradesh)}: \\

\noindent We illustrate our methology of inferring  of patronage using the example of Abhishek Mishra, an MLA representing the Sarojini Nagar constituency in Uttar Pradesh. Querying the People of India database for ``Mishra'' within Uttar Pradesh reveals that this surname associates with four distinct but related Brahman subcommunities \footnote{External validation from news sources confirms his Brahmin community affiliation. See \url{https://www.news18.com/news/politics/its-bjps-outsider-versus-sps-brahmin-face-in-lucknows-sarojini-nagar-constituency-4729361.html}}:

\begin{itemize}
    \item JUJHAUTIYA BRAHMAN
    \item SAKALDWIPI BRAHMAN  
    \item SANADHYA BRAHMAN
    \item SARAYUPARIA BRAHMAN
\end{itemize}

We extract the complete surname inventory across these four related \textit{jatis} which yields 85 distinct surnames: \textit{Arajariya, Atri, Awasthi, Bharadwaj, Chaube, Dikshit, Dwivedi, Mishra, Pandey, Sharma, Tiwari, Tripathi, Upadhyaya}, among others. Any firm director in Sarojini Nagar constituency carrying any of these 85 surnames would be flagged as a potential patronage connection with Abhishek Mishra. Simple surname matching would identify only directors sharing the ``Mishra'' surname, systematically missing the broader caste network through which preferential treatment might operate. The expanded approach captures more potential connections while trying to remain grounded with ethnographic boundaries that reflect the genuine community relationship.

\clearpage

\subsection{Data Sources and Variables}
\label{app:data_sources}

\subsubsection{Geographic and Political Boundaries}

\begin{enumerate}

\item \textbf{Block Shapefiles}

\textbf{Source:} MLInfomap

\textbf{Description:} Administrative block boundaries as of the 2001 Census of India. Blocks are sub-district administrative units that serve as the primary level for rural development program implementation. The dataset contains polygon geometries for 5,823 blocks across India. We use these boundaries to identify split blocks — blocks governed by multiple MLAs due to overlap with assembly constituency boundaries — and to construct our primary treatment variable.

\item \textbf{Assembly Constituency Shapefiles}

\textbf{Source:} DataMeet

\textbf{Description:} Electoral boundaries for state assembly constituencies corresponding to the 2008 and 1977 delimitations. Assembly constituencies define the electoral districts for Members of Legislative Assembly (MLAs) in India's state legislatures. The 2008 delimitation covers 4,182 constituencies across Indian states. The 1977 delimitation covers 4,109 constituencies. We overlay these electoral boundaries with block boundaries to identify split blocks and to construct our primary treatment variable.

\item \textbf{Village Shapefiles}

\textbf{Source:} Socioeconomic Data and Applications Center (SEDAC) \citep{sedac2018}

\textbf{Description:} Village-level polygon boundaries corresponding to the 2001 Census of India, covering approximately 640,000 villages. We use these boundaries for two purposes: (1) aggregate outcomes for firms and nightlights, and (2) construction of distance-to-boundary measures (running variables) for villages near block boundaries in our regression discontinuity design.

\item \textbf{River Segment Data}

\textbf{Source:} HydroATLAS Global River Database \citep{linke2019hydroatlas}

\textbf{Description:} Hydrological data on river courses across India, including river segment geometries and average discharge measurements. We use this data to classify block boundaries delimited by major water bodies — those that align with major rivers (discharge $\geq$ 200 $m^3/s$) for at least 80\% of their length. This classification enables constructing subsamples where boundary placement is plausibly exogenous to local economic and political conditions.

\end{enumerate}

\subsubsection{Firm Registration Data}

\begin{enumerate}

\item \textbf{Firm Registration Records}

\textbf{Source:} Ministry of Corporate Affairs (MCA)

\textbf{Description:} Statutory records for all companies incorporated under the Companies Act, covering 2003-2016 with approximately 825,980 firms. The data include incorporation dates, operational addresses, and industry classification codes. We aggregate firm entry to the village level, constructing both cross-sectional and time-varying panel measures of entrepreneurial activity.

\item \textbf{Director Records}

\textbf{Source:} Ministry of Corporate Affairs (MCA)

\textbf{Description:} Director names and Director Identification Numbers (DINs) from company incorporation documents. We use director surnames to infer social identity connections between firms and politicians.

\end{enumerate}

\subsubsection{Economic Activity Measures}

\begin{enumerate}
\item \textbf{Nightlight Data}

\textbf{Source:} Harmonized DMSP and VIIRS \citep{li2020harmonized}

\textbf{Description:} Harmonized nighttime light intensity data combining Defense Meteorological Satellite Program (DMSP, 1992-2013) and Visible Infrared Imaging Radiometer Suite (VIIRS, 2012-2020) satellite imagery. We compute mean nightlight radiance within each village boundary, normalize the values within each year to a 0-100 scale. The nightlight index serves as a proxy for local economic activity and development.

\item \textbf{NREGA Application Data}

\textbf{Source:} National Rural Employment Guarantee Act (NREGA) Program

\textbf{Description:} Application-level records from the NREGA employment guarantee program covering seven states (Bihar, West Bengal, Jharkhand, Chhattisgarh, Gujarat, Maharashtra, Karnataka) from 2016-2020. Each record contains individual work applications with days of work demanded for every village. The number of days for which applicants seek work through NREGA provides a measure of demand for unemployment assistance.

\item \textbf{CMIE CapEx}

\textbf{Source:} Centre for Monitoring Indian Economy (CMIE)

\textbf{Description:} Project-level capital expenditure database tracking industrial and infrastructure projects announced between 2003-2016. The data include project announcement dates, implementation commencement dates, and geographic coordinates. We compute distance to split block boundaries (running variable) using spatial coordinates and construct project delay (days from announcement to commencement) as a proxy for efficiency regulatory processing.

\item \textbf{Economic Census}

\textbf{Source:} Government of India, 6th Economic Census 2013 \citep{almn2021, ecindia}

\textbf{Description:} Census of all non-agricultural establishments in India. The data are available at the village level. We use total non-farm employment as a measure of local economic activity.
\end{enumerate}

\subsubsection{Political and Electoral Data}

\begin{enumerate}

\item \textbf{Politician Data}

\textbf{Source:} Trivedi Center for Political Data \citep{jensenius2017}, accessed through SHRUG \citep{almn2021}

\textbf{Description:} Election results and candidate information for state legislative assembly elections across India, including winner names, party affiliations, constituencies, and victory margins. We use party affiliations to construct measures of political non-alignment in split blocks (whether multiple MLAs governing the same block belong to different parties). We also use candidate names to infer social identities and identify patronage connections to entrepreneurs.

\item \textbf{Electoral Affidavit Data}

\textbf{Source:} Association for Democratic Reforms (ADR), accessed through MyNeta.info, scraped data available at \url{https://github.com/Vonter/india-election-affidavits}

\textbf{Description:} Digitized sworn affidavits filed by candidates with the Election Commission of India, covering state legislative assembly elections. The affidavits include self-reported assets and liabilities. We use asset declarations at the start and end of electoral terms to compute percentage changes in politician net wealth as a measure of private returns to office.

\end{enumerate}

\subsubsection{Public Goods Provision}

\begin{enumerate}

\item \textbf{PMGSY Road Construction Data}

\textbf{Source:} Pradhan Mantri Gram Sadak Yojana (PMGSY), accessed through SHRUG \citep{asher2020, almn2021}

\textbf{Description:} Administrative data from the rural road construction program covering projects from 2001-2014. The data include estimated costs and actual costs for road construction projects. We construct cost overruns as the log difference between actual and estimated costs to measure efficiency in public goods provision.

\item \textbf{Population Census}

\textbf{Source:} Government of India Census 2001 \citep{pcindia}, accessed through SHRUG \citep{almn2021}

\textbf{Description:} Village-level demographic, infrastructure, and geographic data from the 2001 Population Census. Demographic variables include total population, female population, and marginalized group (SC/ST) population. Infrastructure measures include health facilities (hospitals and dispensaries), educational facilities (schools and colleges), and other amenities (banks, cinema halls, auditoriums). Geographic characteristics include land area and forest coverage. We use these variables to assess balance on pre-existing characteristics between split and unsplit blocks.

\item \textbf{Swachh Bharat Mission}

\textbf{Source:} Ministry of Drinking Water and Sanitation

\textbf{Description:} Village-level disbursement data from the Swachh Bharat Mission sanitation program covering 2015-2018. We use disbursement amounts as a proxy for implementation efficacy in public goods provision.

\item \textbf{IHDS Survey}

\textbf{Source:} India Human Development Survey (IHDS), 2001 and 2011 rounds

\textbf{Description:} Nationally representative household survey measuring confidence in various institutions including local politicians, police, panchayats, and state government. Respondents rate their confidence on a three-point scale (no confidence, intermediate, high confidence). We aggregate responses to the district level using survey weights to construct district-average measures of institutional confidence.

\end{enumerate}

\subsubsection{Auxiliary Data}

\begin{enumerate}

\item \textbf{People of India Database}

\textbf{Source:} People of India Project \citep{singh1996communities}

\textbf{Description:} Ethnographic documentation of Indian communities (jatis) conducted state by state, cataloguing surnames, regional synonyms, hierarchical sub-groups, marriage circles, and patrilineal descent groups (gotras). The database exhibits many-to-many correspondence between surnames and jatis, reflecting historical fission processes and geographic dispersion. We use this database to map director and politician surnames to social identity groups.

\item \textbf{Matrimony Website Data}

\textbf{Source:} Proprietary data from Indian matrimony websites

\textbf{Description:} User profiles containing names (first and last) and self-reported religion, comprising approximately 2.8 million records. We use this data as training data for LSTM models that learn to predict religious identity from naming patterns. The trained models enable inference of social identity connections between firm directors and politicians.

\end{enumerate}

\clearpage

\subsection{Industry Classification}
\begin{table}[htbp]
\setlength{\tabcolsep}{0.6cm}
  \centering
  \caption{List of Industries with High Regulatory Costs in the Sample}
  \label{tab_app_regulated_classification}%
  \resizebox{1.0\textwidth}{!}{
  \begin{threeparttable}
\begin{tabular}{cll}
\toprule
    \toprule
NIC Code & Industry Name                                                                                                             & Broad Industry                                                       \\ \hline \\
1        & Crop and animal production, hunting and related service activities                                                        & Agriculture, forestry and fishing                                                                               \\
2        & Forestry and Logging                                                                                                      & Agriculture, forestry and fishing                                                                               \\
5        & Mining of coal and lignite                                                                                                & Mining and quarrying                                                                                            \\
19       & Manufacture of coke and refined petroleum products                                                                        & Manufacturing                                                                                                   \\
35       & Electricity, gas, steam and air conditioning supply                                                                       & Electricity, gas, steam and air conditioning supply                                                             \\
36       & Water collection, treatment and supply                                                                                    & \begin{tabular}[c]{@{}l@{}}Water supply; sewerage, waste management\\ and remediation activities\end{tabular}   \\
41       & Construction of buildings                                                                                                 & Construction                                                                                                    \\
45       & \begin{tabular}[c]{@{}l@{}}Wholesale and retail trade and repair of \\ motor vehicles and motorcycles\end{tabular}        & \begin{tabular}[c]{@{}l@{}}Wholesale and retail trade; \\ repair of motor vehicles and motorcycles\end{tabular} \\
50       & Water transport                                                                                                           & Transportation and storage                                                                                      \\
51       & Air transport                                                                                                             & Transportation and storage                                                                                      \\
52       & Warehousing and support activities for transportation                                                                     & Transportation and storage                                                                                      \\
61       & Telecommunications                                                                                                        & Information and communication                                                                                   \\
64       & Financial service activities, except insurance and pension funding                                                        & Financial and insurance activities                                                                              \\
65       & \begin{tabular}[c]{@{}l@{}}Insurance, reinsurance and pension funding, \\ except compulsory social  security\end{tabular} & Financial and insurance activities                                                                              \\
66       & Other financial activities                                                                                                & Financial and insurance activities                                                                              \\
70       & Activities of head offices; management consultancy activities                                                             & Professional, scientific and technical activities                                                               \\
72       & Scientific research and development                                                                                       & Professional, scientific and technical activities                                                               \\
73       & Advertising and market research                                                                                           & Professional, scientific and technical activities                                                               \\
74       & Other professional, scientific and technical activities                                                                   & Professional, scientific and technical activities            \\ 
\bottomrule
\bottomrule
\end{tabular}
    \begin{tablenotes}
\footnotesize	
\item This table presents the list of industries with high regulatory costs in our sample. We define industries with a high regulatory cost following the methodology for the classification of regulated industries in the United States based on \cite{pittman1977market} augmented with India specific regulations and factors as discussed in \cite{awasthi2019cl}. \\
\end{tablenotes}
\end{threeparttable}
}
\end{table}%

\begin{table}[ht!]
\setlength{\tabcolsep}{0.6cm}
  \centering
  \caption{List of Industries with High Rent Seeking or Crony-Capitalism}
  \label{tab_app_crony_classification}%
  \resizebox{1.0\textwidth}{!}{
  \begin{threeparttable}
\begin{tabular}{cll} \toprule
    \toprule
NIC C ode & Industry Name                                                                                                             & Broad Industry                                                                                                \\ \hline \\
1         & Crop and animal production, hunting and related service activities                                                        & Agriculture, forestry and fishing                                                                             \\
2         & Forestry and Logging                                                                                                      & Agriculture, forestry and fishing                                                                             \\
5         & Mining of coal and lignite                                                                                                & Mining and quarrying                                                                                          \\
19        & Manufacture of coke and refined petroleum products                                                                        & Manufacturing                                                                                                 \\
20        & Manufacture of chemicals and chemical products                                                                            & Manufacturing                                                                                                 \\
36        & Water collection, treatment and supply                                                                                    & \begin{tabular}[c]{@{}l@{}}Water supply; sewerage, waste management\\ and remediation activities\end{tabular} \\
41        & Construction of buildings                                                                                                 & Construction                                                                                                  \\
52        & Warehousing and support activities for transportation                                                                     & Transportation and storage                                                                                    \\
64        & Financial service activities, except insurance and pension funding                                                        & Financial and insurance activities                                                                            \\
65        & \begin{tabular}[c]{@{}l@{}}Insurance, reinsurance and pension funding,\\ except compulsory social   security\end{tabular} & Financial and insurance activities                                                                            \\
66        & Other financial activities                                                                                                & Financial and insurance activities                                                                            \\
93        & Sports activities and amusement and recreation activities                                                                 & Arts, entertainment and recreation \\ 
\bottomrule
\bottomrule
\end{tabular}
    \begin{tablenotes}
\footnotesize	
\item This table presents the list of industries with high rent-seeking or high degree of crony-capitalism in our sample. We use the index of industry-level cronyism created by \emph{Economist} using the methodology developed by \emph{Transparency International} to classify firms as high crony and low crony firms. The detailed description of the method to classify industries as crony can be found in The Economist: \href{https://www.economist.com/international/2014/03/15/planet-plutocrat}{Planet Plutocrat}. \\
\end{tablenotes}
\end{threeparttable}
}
\end{table}%

\clearpage
\newpage

\clearpage
\newpage
\section{Robustness Tests}
\label{app_robust}
\begin{table}[!htbp]
\setlength{\tabcolsep}{0.5cm}
  \centering
  \caption{RD Estimate on Firm Entry with Polynomial of Degree 2}
  \label{appendix_tab_poly2}%
  \resizebox{\textwidth}{!}{
  \begin{threeparttable}
    \begin{tabular}{lcccc}
    \toprule
    \toprule
    Dep Var:  LN(0.001+\# New Firms) & (1)   & (2)   & (3)   & (4) \\
    \hline \\
    Split (=1) & 0.0282*** & 0.0322*** & 0.0297*** & 0.0281*** \\
          & (0.0101) & (0.0080) & (0.0070) & (0.0064) \\  \\
    \midrule
    \# Obs & {263,307 } & {356,260 } & {416,694 } & {436,980 } \\
    $R^2$  & {0.1537} & {0.1293} & {0.1147} & {0.1062} \\
    Bandwidth &  5 KM  &  10 KM  &  20 KM  &  50 KM  \\
    Boundary FE & Yes & Yes & Yes & Yes \\
    Polynomial Degree & {2} & {2} & {2} & {2} \\
    \bottomrule
    \bottomrule
    \end{tabular}%
\begin{tablenotes}
\footnotesize	
\item This table presents estimates for specification (\ref{rd}) using the natural logarithm of 0.001 plus firm entry as the dependent variable. We use different bandwidths of 5, 10, 20, and 50 km on either side of the boundary, separating a split block from an unsplit block in columns (1), (2), (3), and (4), respectively.  The unit of observation is a village that lies within the narrow bandwidth of the boundary separating a split and an unsplit block. A block is defined as a split block based on the haphazard overlap of block boundaries with electoral boundaries as per the 2008 delimitation. Firm Entry is defined as the total number of firms that have entered a village from 2008 to 2016.  All regressions include boundary fixed effect, a local quadratic specification estimated separately on each side of the boundary, and use a triangular kernel. Standard errors reported in parentheses are clustered at the block level. \sym{*} \(p<0.1\), \sym{**} \(p<0.05\), \sym{***} \(p<0.01\). \\
\end{tablenotes}
\end{threeparttable}
}
\end{table}%

\begin{table}[!htbp]
\setlength{\tabcolsep}{0.5cm}
  \centering
  \caption{RD Estimate with Uniform Kernel}
  \label{appendix_tab_uni_kernel}%
  \resizebox{\textwidth}{!}{
  \begin{threeparttable}
    \begin{tabular}{lcccc}
    \toprule
    \toprule 
    Dep Var:  LN(0.001+\# New Firms) & (1)   & (2)   & (3)   & (4) \\
    \midrule \\
    Split(=1) & 0.0334*** & 0.0294*** & 0.0308*** & 0.0307*** \\
          & (0.0073) & (0.0062) & (0.0057) & (0.0056) \\ \\
    \midrule
    \# Obs & {257,787} & {348,539} & {407,616} & {427,659} \\
    $R^2$ & {0.6054} & {0.5736} & {0.5453} & {0.5239} \\
    Bandwidth & 5 KM  & 10 KM & 20 KM & 50 KM \\
    Boundary FE & Yes & Yes & Yes & Yes \\
    Kernel & Uniform & Uniform & Uniform & Uniform  \\
    \bottomrule
    \bottomrule
    \end{tabular}%
\begin{tablenotes}
\footnotesize	
\item This table presents estimates for specification (\ref{rd}) using the natural logarithm of 0.001 plus firm entry as the dependent variable. We use different bandwidths of 5, 10, 20, and 50 km on either side of the boundary, separating a split block from an unsplit block in columns (1), (2), (3), and (4), respectively.  The unit of observation is a village that lies within the narrow bandwidth of the boundary separating a split and an unsplit block. A block is defined as a split block based on the haphazard overlap of block boundaries with electoral boundaries as per the 2008 delimitation. Firm Entry is defined as the total number of firms that have entered a village from 2008 to 2016.  All regressions include boundary fixed effect, a local linear specification estimated separately on each side of the boundary, and use a uniform kernel. Standard errors reported in parentheses are clustered at the block level. \sym{*} \(p<0.1\), \sym{**} \(p<0.05\), \sym{***} \(p<0.01\). \\
\end{tablenotes}
\end{threeparttable}  
}
\end{table}%

\clearpage
\newpage
\begin{table}[!htbp]
\setlength{\tabcolsep}{0.5cm}
  \centering
  \caption{RD Estimates with Spatial Polynomial}
  \label{tab:appendix_tab_spat}%
  \resizebox{\textwidth}{!}{
  \begin{threeparttable}
    \begin{tabular}{lcccc}
    \toprule
    \toprule
    Dep Var:  LN(0.001+\# New Firms) & (1)   & (2)   & (3)   & (4) \\
    \midrule \\
    Split(=1) & 0.2456*** & 0.2867*** & 0.3184*** & 0.3314*** \\
          & (0.0821) & (0.0742) & (0.0707) & (0.0706) \\ \\
    \hline 
    \# Obs &  263,307  &  356,260  &  416,694  &  436,980  \\
    $R^2$ & 0.1537 & 0.1293 & 0.1146 & 0.1061 \\
    Bandwidth & 5 KM  & 10 KM & 20 KM & 50 KM \\
    Boundary FE & Yes & Yes & Yes & Yes \\
    Spatial Polynomial & Yes & Yes & Yes & Yes\\
    \bottomrule
    \bottomrule
    \end{tabular}%
\begin{tablenotes}
\footnotesize	
\item This table presents estimates for specification (\ref{rd}) using the natural logarithm of 0.001 plus firm entry as the dependent variable. We use different bandwidths of 5, 10, 20, and 50 km on either side of the boundary, separating a split block from an unsplit block in columns (1), (2), (3), and (4), respectively.  The unit of observation is a village that lies within the narrow bandwidth of the boundary separating a split and an unsplit block. A block is defined as a split block based on the haphazard overlap of block boundaries with electoral boundaries as per the 2008 delimitation. Firm Entry is defined as the total number of firms that have entered a village from 2008 to 2016.  All regressions include boundary fixed effect, a local linear specification with latitude and longitude estimated separately on each side of the boundary, and use a triangular kernel. Standard errors reported in parentheses are clustered at the block level. \sym{*} \(p<0.1\), \sym{**} \(p<0.05\), \sym{***} \(p<0.01\). \\
\end{tablenotes}
\end{threeparttable}
}
\end{table}%

\begin{table}[ht!]
\setlength{\tabcolsep}{0.7cm}
  \centering
  \caption{RD Estimates with \cite{conley1999gmm} Standard Errors}
  \label{appendix_tab_coneley}%
  \resizebox{\textwidth}{!}{
  \begin{threeparttable}
    \begin{tabular}{lcccc}
    \toprule
    \toprule 
    Bandwidth $\downarrow$ & (1)   & (2)   & (3)   & (4) \\
    \midrule \\
    5 KM  & 0.0331*** & 0.0331*** & 0.0331*** & 0.0331*** \\
          & (0.0077) & (0.0077) & (0.0078) & (0.0079) \\
    10 KM & 0.0326*** & 0.0326*** & 0.0326*** & 0.0326*** \\
          & (0.0063) & (0.0063) & (0.0064) & (0.0067) \\
    20 KM & 0.0312*** & 0.0312*** & 0.0312*** & 0.0312*** \\
          & (0.0056) & (0.0056) & (0.0057) & (0.0060) \\
    50 KM & 0.0309*** & 0.0309*** & 0.0309*** & 0.0309*** \\
          & (0.0053) & (0.0053) & (0.0054) & (0.0058) \\ \\
    \midrule
    Conley Cutoff $\rightarrow$ & 2 KM  & 5KM   & 10 KM & 20 KM \\
    \bottomrule
    \bottomrule
    \end{tabular}%
\begin{tablenotes}
\footnotesize	
\item This table presents estimates for specification (\ref{rd}) using the natural logarithm of 0.001 plus firm entry as the dependent variable. We use different bandwidths of 5, 10, 20, and 50 km on either side of the boundary, separating a split block from an unsplit block in rows (1), (2), (3), and (4), respectively.  The unit of observation is a village that lies within the narrow bandwidth of the boundary separating a split and an unsplit block. A block is defined as a split block based on the haphazard overlap of block boundaries with electoral boundaries as per the 2008 delimitation. Firm Entry is defined as the total number of firms that have entered a village from 2008 to 2016.  All regressions include boundary fixed effect, a local linear specification estimated separately on each side of the boundary, and use a triangular kernel. Standard errors reported in parentheses are \cite{conley1999gmm} standard errors, estimated for cutoff values of 2, 5, 10, and 20 km in columns (1), (2), (3), and (4), respectively. \sym{*} \(p<0.1\), \sym{**} \(p<0.05\), \sym{***} \(p<0.01\). \\
\end{tablenotes}
\end{threeparttable}    
}
\end{table}%

\clearpage
\newpage
\begin{table}[ht!]
\setlength{\tabcolsep}{0.7cm}
  \centering
  \caption{Standard Errors with Alternative Clustering}
  \label{appendix_tab_alt_cluster}%
  \resizebox{\textwidth}{!}{
  \begin{threeparttable}
    \begin{tabular}{lcccc}
    \toprule
    \toprule 
    Bandwidth $\downarrow$ & (1)   & (2)   & (3)   & (4) \\
    \midrule \\
5 KM  & 0.0331*** & 0.0331*** & 0.0331*** & 0.0331*** \\
      & (0.0078)  & (0.0081)  & (0.0082)  & (0.0111)  \\
10 KM & 0.0326*** & 0.0326*** & 0.0326*** & 0.0326*** \\
      & (0.0064)  & (0.0072)  & (0.0068)  & (0.0089)  \\
20 KM & 0.0312*** & 0.0312*** & 0.0312*** & 0.0312*** \\
      & (0.0056)  & (0.0068)  & (0.0062)  & (0.0086)  \\
50KM  & 0.0309*** & 0.0309*** & 0.0309*** & 0.0309*** \\
      & (0.0053)  & (0.0066)  & (0.0060)  & (0.0089)  \\ \\
    \midrule
    Clustering $\rightarrow$ & Village  & District   & Boundary & State \\
    \bottomrule
    \bottomrule
    \end{tabular}%
\begin{tablenotes}
\footnotesize	
\item This table presents estimates for specification (\ref{rd}) using the natural logarithm of 0.001 plus firm entry as the dependent variable. We use different bandwidths of 5, 10, 20, and 50 km on either side of the boundary, separating a split block from an unsplit block in rows (1), (2), (3), and (4), respectively.  The unit of observation is a village that lies within the narrow bandwidth of the boundary separating a split and an unsplit block. A block is defined as a split block based on the haphazard overlap of block boundaries with electoral boundaries as per the 2008 delimitation. Firm Entry is defined as the total number of firms that have entered a village from 2008 to 2016.  All regressions include boundary fixed effect, a local linear specification estimated separately on each side of the boundary, and use a triangular kernel. Standard errors reported in parentheses are based on alternative clusters, with clustering at the village, district, boundary, and state level in columns (1), (2), (3), and (4), respectively. \sym{*} \(p<0.1\), \sym{**} \(p<0.05\), \sym{***} \(p<0.01\). \\
\end{tablenotes}
\end{threeparttable}    
}
\end{table}%

\begin{table}[ht!]
\setlength{\tabcolsep}{0.50cm}
  \centering
  \caption{RD Estimate with Alternative Sample}
  \label{appendix_tab_alt_sample_ihs}%
  \resizebox{\textwidth}{!}{
  \begin{threeparttable}
    \begin{tabular}{lcccccc}
    \toprule
    \toprule
    Bandwidth $\downarrow$ & (1)   & (2)   & (3)   & (4)   & (5)   & (6) \\
    \midrule \\
    5 KM  & 0.0331*** & 0.0462*** & 0.1113*** & 0.1606*** & 0.1824*** & 0.0270*** \\
          & (0.0071) & (0.0108) & (0.0220) & (0.0322) & (0.0464) & (0.0068) \\
    10 KM & 0.0326*** & 0.0443*** & 0.1107*** & 0.1575*** & 0.1825*** & 0.0245*** \\
          & (0.0060) & (0.0092) & (0.0191) & (0.0283) & (0.0411) & (0.0057) \\
    20 KM & 0.0312*** & 0.0421*** & 0.1031*** & 0.1464*** & 0.1727*** & 0.0229*** \\
          & (0.0056) & (0.0086) & (0.0181) & (0.0265) & (0.0389) & (0.0052) \\
    50 KM & 0.0309*** & 0.0411*** & 0.1009*** & 0.1428*** & 0.1726*** & 0.0226*** \\
          & (0.0055) & (0.0084) & (0.0177) & (0.0258) & (0.0381) & (0.0050) \\ \\
    \midrule
    Sample & Full Sample & Firms > 0 & Firms > 5 & Firms > 10 & Firms > 20 & Firms <= 1000 \\
    \bottomrule
    \bottomrule
    \end{tabular}%
    \begin{tablenotes}
\footnotesize	
\item This table presents estimates for specification (\ref{rd}) using the natural logarithm of 0.001 plus firm entry as the dependent variable in panel A, and the inverse hyperbolic sine (IHS) transformation of the number of firms in panel B. We use different bandwidths of 5, 10, 20, and 50 km on either side of the boundary, separating a split block from an unsplit block in rows (1), (2), (3), and (4), respectively.  The unit of observation is a village that lies within the narrow bandwidth of the boundary separating a split and an unsplit block. A block is defined as a split block based on the haphazard overlap of block boundaries with electoral boundaries as per the 2008 delimitation. Firm Entry is defined as the total number of firms that have entered a village from 2008 to 2016.  All regressions include boundary fixed effect, a local linear specification estimated separately on each side of the boundary, and use a triangular kernel. Column (1) includes the baseline sample. Column (2) drops all blocks with no firm entry. Columns (3), (4), and (5) drops blocks where less than five, ten, and twenty firms, respectively, entered during the sample period. Column (6) drops all large blocks with number of new firms greater than 1000. Standard errors reported in parentheses are clustered at the block level. \sym{*} \(p<0.1\), \sym{**} \(p<0.05\), \sym{***} \(p<0.01\). \\
\end{tablenotes}
\end{threeparttable}
    }
\end{table}%

\clearpage
\newpage
\begin{table}[ht!]
  \centering
  \caption{RD Estimate with Alternative Transformation}
  \label{appendix_tab_alt_sample_ihs_ppml}%
  \resizebox{\textwidth}{!}{
  \begin{threeparttable}
    \begin{tabular}{lcccccc}
    \toprule
    \toprule 
    \multicolumn{7}{c}{\textit{Panel A : Poisson Pseudo Maximum Likelihood }} \\
    \midrule
    Bandwidth $\downarrow$ & (1)   & (2)   & (3)   & (4)   & (5)   & (6) \\
    \midrule \\
    \multicolumn{1}{l}{5 KM} & 0.2117*** & 0.1667*** & 0.2545*** & 0.2653*** & 0.2487*** & 0.1731*** \\
          & (0.0515) & (0.0525) & (0.0649) & (0.0728) & (0.0835) & (0.0523) \\
    \multicolumn{1}{l}{10 KM} & 0.1856*** & 0.1332*** & 0.1912*** & 0.2015*** & 0.1606** & 0.1411*** \\
          & (0.0430) & (0.0443) & (0.0551) & (0.0624) & (0.0713) & (0.0435) \\
    \multicolumn{1}{l}{20 KM} & 0.2003*** & 0.1444*** & 0.2046*** & 0.1938*** & 0.1589** & 0.1555*** \\
          & (0.0385) & (0.0401) & (0.0505) & (0.0568) & (0.0659) & (0.0389) \\
    \multicolumn{1}{l}{50 KM} & 0.2011*** & 0.1424*** & 0.2016*** & 0.1978*** & 0.1741*** & 0.1535*** \\
          & (0.0369) & (0.0387) & (0.0492) & (0.0560) & (0.0654) & (0.0370) \\
    \midrule
    Sample & Full Sample & Firms > 0 & Firms > 5 & Firms > 10 & Firms > 20 & Firms <= 1000 \\
    \midrule 
    \multicolumn{7}{c}{\textit{Panel B : Outcome - IHS(\#New Firms) }} \\
    \midrule
    Bandwidth $\downarrow$ & (1)   & (2)   & (3)   & (4)   & (5)   & (6) \\
    \hline \\
    5 KM  & 0.0126*** & 0.0196*** & 0.0440*** & 0.0628*** & 0.0757*** & 0.0102*** \\
          & (0.0022) & (0.0035) & (0.0078) & (0.0119) & (0.0178) & (0.0021) \\
    10 KM & 0.0128*** & 0.0196*** & 0.0432*** & 0.0616*** & 0.0753*** & 0.0096*** \\
          & (0.0020) & (0.0031) & (0.0069) & (0.0107) & (0.0160) & (0.0017) \\
    20 KM & 0.0121*** & 0.0186*** & 0.0402*** & 0.0571*** & 0.0704*** & 0.0088*** \\
          & (0.0019) & (0.0030) & (0.0067) & (0.0103) & (0.0155) & (0.0016) \\
    50 KM & 0.0120*** & 0.0183*** & 0.0396*** & 0.0562*** & 0.0709*** & 0.0087*** \\
          & (0.0019) & (0.0030) & (0.0067) & (0.0102) & (0.0155) & (0.0016) \\ \\
    \midrule
    Sample & Full Sample & Firms > 0 & Firms > 5 & Firms > 10 & Firms > 20 & Firms <= 1000 \\
    \bottomrule
    \bottomrule
    \end{tabular}%
    \begin{tablenotes}
\footnotesize	
\item This table presents estimates for specification (\ref{rd}) using the natural logarithm of 0.001 plus firm entry as the dependent variable. We use different bandwidths of 5, 10, 20, and 50 km on either side of the boundary, separating a split block from an unsplit block in rows 1, 2, 3, and 4, respectively.  The unit of observation is a village that lies within the narrow bandwidth of the boundary separating a split and an unsplit block. A block is defined as a split block based on the haphazard overlap of block boundaries with electoral boundaries as per the 2008 delimitation. Firm Entry is defined as the total number of firms that have entered a village from 2008 to 2016.  All regressions include boundary fixed effect, a local linear specification estimated separately on each side of the boundary, and use a triangular kernel. Column (1) includes the baseline sample. Column (2) drops all blocks with no firm entry. Columns (3), (4) and (5) drops blocks where less than 5, 10, and 20 firms, respectively, entered during the sample period. Column (6) drops all large blocks with more than 1000 firms. Standard errors reported in parentheses are clustered at the block level. \sym{*} \(p<0.1\), \sym{**} \(p<0.05\), \sym{***} \(p<0.01\). \\
\end{tablenotes}
\end{threeparttable}
    }
\end{table}%

\clearpage
\newpage
\begin{table}[!htbp]
\setlength{\tabcolsep}{0.6cm}
  \centering
  \caption{RD Estimates with Non-Pooled Data}
  \label{tab:appendix_tab_nonpool}%
  \resizebox{\textwidth}{!}{
  \begin{threeparttable}
    \begin{tabular}{lcccc}
    \toprule
    \toprule 
    Dep Var:  LN(0.001+\# New Firms) & (1)   & (2)   & (3)   & (4) \\
    \midrule \\
    Split(=1) & 0.0215*** & 0.0214*** & 0.0199*** & 0.0198*** \\
          & (0.0036) & (0.0032) & (0.0031) & (0.0031) \\ \\
    \hline 
    \# Obs &  1,813,190  &  2,497,757  &  2,958,673  &  3,120,412  \\
    $R^2$ & 0.1219 & 0.0997 & 0.0873 & 0.0805 \\
    Bandwidth & 5 KM  & 10 KM & 20 KM & 50 KM \\
    Boundary $\times$ Year FE & Yes & Yes & Yes & Yes \\
    Spatial Polynomial & Yes & Yes & Yes & Yes\\
    \bottomrule
    \bottomrule
    \end{tabular}%
\begin{tablenotes}
\footnotesize	
\item This table presents estimates for specification (\ref{rd}) using the natural logarithm of 0.001 plus firm entry as the dependent variable. We use different bandwidths of 5, 10, 20, and 50 km on either side of the boundary, separating a split block from an unsplit block in columns (1), (2), (3), and (4), respectively.  The unit of observation is a village-year that lies within the narrow bandwidth of the boundary separating a split and an unsplit block from 2008 to 2016. A block is defined as a split block based on the haphazard overlap of block boundaries with electoral boundaries as per the 2008 delimitation. Firm Entry is defined as the total number of firms that have entered a village during the year.  All regressions include boundary fixed effect, a local linear specification with latitude and longitude estimated separately on each side of the boundary, and use a triangular kernel. Standard errors reported in parentheses are clustered at the block level. \sym{*} \(p<0.1\), \sym{**} \(p<0.05\), \sym{***} \(p<0.01\). \\
\end{tablenotes}
\end{threeparttable} 
}
\end{table}%

\begin{table}[!htbp]
\setlength{\tabcolsep}{1.2cm}
  \centering
  \caption{RD Estimates with Alternative Grids}
  \label{tab:appendix_tab_grid}%
  \resizebox{\textwidth}{!}{
  \begin{threeparttable}
\begin{tabular}{cccc}
  \toprule
      \toprule
  \multirow{2}{*}{Bandwidth $\downarrow$} & \multicolumn{3}{c}{Grid Size}     \\ \cline{2-4}
   & (1) & (2) & (3) \\
                             & 1 $\times$ 1       & 2 $\times$ 2       & 5 $\times$ 5       \\ \hline \\
  5 KM                       & 0.0139*** & 0.0450*** & 0.1796*** \\
                             & (0.0025)  & (0.0084)  & (0.0423)  \\
  10 KM                      & 0.0142*** & 0.0486*** & 0.1621*** \\
                             & (0.0022)  & (0.0074)  & (0.0309)  \\
  20 KM                      & 0.0155*** & 0.0535*** & 0.1833*** \\
                             & (0.0022)  & (0.0073)  & (0.0262)  \\
  50 KM                      & 0.0189*** & 0.0655*** & 0.2248*** \\
                             & (0.0019)  & (0.0064)  & (0.0222) \\ \bottomrule
                             \bottomrule
  \end{tabular}
\begin{tablenotes}
\footnotesize	
\item This table presents estimates for specification (\ref{rd}) using the natural logarithm of 0.001 plus firm entry as the dependent variable. We use different bandwidths of 5, 10, 20, and 50 km on either side of the boundary, separating a split block from an unsplit block in rows 1, 2, 3, and 4, respectively.  The unit of observation is an arbitary grid of size $x \times x$ that lies within the narrow bandwidth of the boundary separating a split and an unsplit block from 2008 to 2016. Columns 1, 2 and 3 use a grid size of $1 \times 1$, $2 \times 2$ and $5 \times 5$, respectively. A block is defined as a split block based on the haphazard overlap of block boundaries with electoral boundaries as per the 2008 delimitation. Firm Entry is defined as the total number of firms that have entered in the cell during the period.  All regressions include boundary fixed effect, a local linear specification estimated separately on each side of the boundary, and use a triangular kernel. Standard errors reported in parentheses are clustered at the block level. \sym{*} \(p<0.1\), \sym{**} \(p<0.05\), \sym{***} \(p<0.01\). \\
\end{tablenotes}
\end{threeparttable} 
}
\end{table}%

\clearpage
\newpage
\begin{table}[htbp]
  \centering
  \caption{RD Estimates before the 2008 Delimitation}
  \label{tab:appendix_tab_balance_firm_nl}%
  \begin{threeparttable}
\begin{tabular}{lcccc}
\toprule
    \toprule
Dep Var $\downarrow$               & (1)       & (2)       & (3)       & (4)       \\ \hline \\
LN(0.001+\# New Firms) & 0.0142    & 0.0114    & 0.0064    & 0.0046    \\
                       & (0.01310) & (0.01170) & (0.01120) & (0.01090) \\ \\ 
LN(0.001+Nightlights)  & 0.0059    & 0.0442    & -0.0408   & -0.0763   \\
                       & (0.13170) & (0.15160) & (0.16380) & (0.16900) \\ \\ \hline 
Bandwidth  $\rightarrow$            & 5 KM      & 10 KM     & 20 KM     & 50 KM    \\ 
\bottomrule
\bottomrule
\end{tabular}
\begin{tablenotes}
\footnotesize	
\item This table presents estimates for specification (\ref{rd}) using the natural logarithm of 0.001 plus firm entry and nightlights as the dependent variable. We use different bandwidths of 5, 10, 20, and 50 km on either side of the boundary, separating a split block from an unsplit block in columns (1), (2), (3), and (4), respectively.  The unit of observation is a village-year that lies within the narrow bandwidth of the boundary separating a split and an unsplit block from 2003 until 2007. A block is defined as a split block based on the haphazard overlap of block boundaries with electoral boundaries as per the 2008 delimitation. Firm Entry is defined as the total number of firms that have entered a village during the year.  All regressions include boundary fixed effect, a local linear specification estimated separately on each side of the boundary, and use a triangular kernel. Standard errors reported in parentheses are clustered at the block level. \sym{*} \(p<0.1\), \sym{**} \(p<0.05\), \sym{***} \(p<0.01\). \\
\end{tablenotes}
\end{threeparttable} 
\end{table}%

\begin{table}[ht!]
  \centering
  \caption{RD Estimate with Additional Controls -- Dep Var:  LN(0.001+\# New Firms)}
  \label{appendix_tab_controls}%
  \resizebox{\textwidth}{!}{
  \begin{threeparttable}
    \begin{tabular}{lcccccc}
    \toprule
    \toprule 
    Bandwidth $\downarrow$ & (1)   & (2)   & (3)   & (4)   & (5)   & (6) \\
    \midrule \\
    5 KM  & 0.0331*** & 0.0291*** & 0.0292*** & 0.0474*** & 0.0317*** & 0.0416*** \\
          & (0.0071) & (0.0069) & (0.0069) & (0.0097) & (0.0070) & (0.0093) \\
    10 KM & 0.0326*** & 0.0277*** & 0.0283*** & 0.0464*** & 0.0308*** & 0.0382*** \\
          & (0.0060) & (0.0058) & (0.0058) & (0.0084) & (0.0060) & (0.0080) \\
    20 KM & 0.0312*** & 0.0264*** & 0.0273*** & 0.0424*** & 0.0293*** & 0.0337*** \\
          & (0.0056) & (0.0054) & (0.0054) & (0.0083) & (0.0056) & (0.0078) \\
    50 KM & 0.0309*** & 0.0263*** & 0.0269*** & 0.0401*** & 0.0289*** & 0.0325*** \\
          & (0.0055) & (0.0052) & (0.0052) & (0.0082) & (0.0054) & (0.0078) \\ \\
    \midrule
    Controls & None  & Population & Area  & Distance to HQ & Compactness & All \\
    \bottomrule
    \bottomrule
    \end{tabular}%
    \begin{tablenotes}
\footnotesize	
\item This table presents estimates for specification (\ref{rd}), augmented for controls, using the natural logarithm of 0.001 plus firm entry as the dependent variable. We use different bandwidths of 5, 10, 20, and 50 km on either side of the boundary, separating a split block from an unsplit block in rows 1, 2, 3, and 4, respectively.  The unit of observation is a village that lies within the narrow bandwidth of the boundary separating a split and an unsplit block. A block is defined as a split block based on the haphazard overlap of block boundaries with electoral boundaries as per the 2008 delimitation. Firm Entry is defined as the total number of firms that have entered a village from 2008 to 2016.  All regressions include boundary fixed effect, a local linear specification estimated separately on each side of the boundary, and use a triangular kernel. Column (1) includes no controls. Columns (2), (3), and (4) include a linear and quadratic term for the 2001 population, geographic area, and the distance to district headquarters for a village. Column (5) controls for compactness of the block measured as in \cite{harari2020}. Column (6) includes all controls included in columns (2)-(5). Standard errors reported in parentheses are clustered at the block level. \sym{*} \(p<0.1\), \sym{**} \(p<0.05\), \sym{***} \(p<0.01\). \\
\end{tablenotes}
\end{threeparttable}
}
\end{table}%

\clearpage
\newpage
\begin{table}[ht!]
\setlength{\tabcolsep}{0.5cm}
  \centering
  \caption{RD Estimates Controlling for Politician Quality}
  \label{tab:appendix_ac_fe}%
  \resizebox{\textwidth}{!}{
  \begin{threeparttable}
    \begin{tabular}{lcccc}
    \toprule
    \toprule
    Dep Var:  LN(0.001+\# New Firms) & (1)   & (2)   & (3)   & (4) \\ \hline \\
    Split (=1) & 0.0239*** & 0.0164*** & 0.0130** & 0.0121**  \\
          & (0.0073) & (0.0062) & (0.0057) & (0.0056) \\ \\
    \midrule
    \#Obs & 263,156  & 356,101  & 416,528 & 436,814 \\
    $R^2$  & 0.1965 & 0.1670 & 0.1497 & 0.1394 \\
    Bandwidth & 5 KM  & 10 KM & 20 KM & 50 KM \\
    Boundary FE & Yes   & Yes   & Yes   & Yes \\
    Constituency FE & Yes   & Yes   & Yes   & Yes \\
    \bottomrule
    \bottomrule
    \end{tabular}%
\begin{tablenotes}
\footnotesize	
\item This table presents estimates for specification (\ref{rd}) using the natural logarithm of 0.001 plus firm entry as the dependent variable. We use different bandwidths of 5, 10, 20, and 50 km on either side of the boundary, separating a split block from an unsplit block in columns (1), (2), (3), and (4), respectively.  The unit of observation is a village that lies within the narrow bandwidth of the boundary separating a split and an unsplit block. A block is defined as a split block based on the haphazard overlap of block boundaries with electoral boundaries as per the 2008 delimitation. Firm Entry is defined as the total number of firms that have entered a village from 2008 to 2016.  All regressions include boundary fixed effect and constituency fixed effect.a local linear specification estimated separately on each side of the boundary, and use a triangular kernel. Standard errors reported in parentheses are clustered at the block level. \sym{*} \(p<0.1\), \sym{**} \(p<0.05\), \sym{***} \(p<0.01\). \\
\end{tablenotes}
\end{threeparttable} 
}
\end{table}%

\begin{table}[ht!]
\setlength{\tabcolsep}{0.5cm}
  \centering
  \caption{Probability of Firm Exit}
  \label{tab:appendix_prob_firm_exit}%
  \resizebox{\textwidth}{!}{
 \begin{threeparttable}
\begin{tabular}{lcccc}
\toprule
    \toprule
Dep   Var:  Firm exits within x time & (1)      & (2)      & (3)      & (4)      \\ \hline \\
6 months                             & -0.0008  & -0.0004  & -0.0003  & -0.0002  \\
                                     & (0.0007) & (0.0004) & (0.0003) & (0.0002) \\ \\
1 year                               & -0.0017  & -0.0003  & -0.0003  & -0.0005  \\
                                     & (0.0014) & (0.0008) & (0.0007) & (0.0006) \\ \\
2 year                               & -0.0044  & -0.0008  & -0.0010  & -0.0022 \\
                                     & (0.0028) & (0.0019) & (0.0017) & (0.0016) \\ \\ \hline 
Bandwidth                            & 5 KM     & 10 KM    & 20 KM    & 50 KM    \\
Boundary X Cohort Year FE            & Yes      & Yes      & Yes      & Yes     \\ 
\bottomrule
\bottomrule
\end{tabular}
\begin{tablenotes}
\footnotesize	
\item This table presents results from a firm-level regression, estimating the relationship between the probability of firm exit and split variable. It uses an RD specification similar to equation  (\ref{rd}), but the data is at the firm level. For each firm that started between 2008 and 2016, we estimate the probability of it closing within six months, one year, and two years of its start as a function of the split status of the block. We use different bandwidths of 5, 10, 20, and 50 km on either side of the boundary, separating a split block from an unsplit block in columns (1), (2), (3), and (4), respectively. The unit of observation is a firm that lies within the narrow bandwidth of the boundary separating a split and an unsplit block. A block is defined as a split block based on the haphazard overlap of block boundaries with electoral boundaries as per the 2008 delimitation. All regressions include boundary $\times$ cohort fixed effects. All firms that entered during the same year are considered part of the same cohort. Each row of the table comes from a different regression. Rows 1, 2, and 3 estimate the probability of firm exit within six months, one year, and two years of firm entry, respectively. a local linear specification estimated separately on each side of the boundary, and use a triangular kernel. Standard errors reported in parentheses are clustered at the block level. \sym{*} \(p<0.1\), \sym{**} \(p<0.05\), \sym{***} \(p<0.01\). \\
\end{tablenotes}
\end{threeparttable} 
}
\end{table}%

\clearpage
\newpage
\begin{figure}[!htbp]
	\centering
	\caption{RD Estimates for various Bandwidth choices}
	\includegraphics[width=\textwidth]{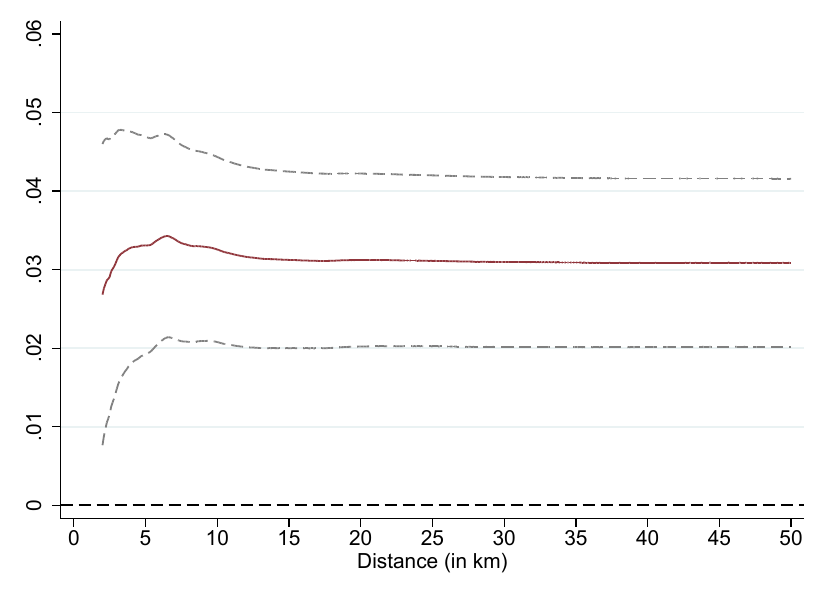}
	\label{fig_cont_bw}
	\begin{minipage}{\textwidth}
{\fontsize{8pt}{9pt}\selectfont This table presents RD estimates and 95\% confidence interval for specification (\ref{rd}) using the natural logarithm of 0.001 plus firm entry as the dependent variable. We estimate specification (\ref{rd}) for several bandwidths between 2 km and 50 km in increments of 0.1 km. The Y-axis reports the RD estimate, and the X-axis reports the bandwidth used to estimate the RD coefficient. The unit of observation is a village that lies within the narrow bandwidth of the boundary separating a split and an unsplit block. A block is defined as a split block based on the haphazard overlap of block boundaries with electoral boundaries as per the 2008 delimitation. Firm Entry is defined as the total number of firms that have entered a village from 2008 to 2016. All regressions include boundary fixed effect, a local linear specification estimated separately on each side of the boundary, and use a triangular kernel. The solid red line reports the RD estimate, and the dashed grey lines indicate the 95\% confidence bands calculated based on standard errors clustered at the block level.
\par}
	\end{minipage}
\end{figure}

\clearpage

\begin{table}[htbp]
  \centering
  \caption{Number of Politicians in Split Blocks and Firm Entry}
  \label{sptd_firm_entry_multi}%
  \resizebox{\textwidth}{!}{ 
  \begin{threeparttable}
    \begin{tabular}{lcccc}
    \toprule
    \toprule 
    Dep Var:  LN(0.001+\# New Firms) & {(1)} & {(2)} & {(3)} & {(4)} \\
    \midrule \\
    Politician (=2) & {0.0261***} & {0.0256***} & {0.0236***} & {0.0229***} \\
          & {(0.0072)} & {(0.0060)} & {(0.0056)} & {(0.0054)} \\
    Politician (=3) & {0.0524***} & {0.0472***} & {0.0487***} & {0.0502***} \\
          & {(0.0180)} & {(0.0163)} & {(0.0173)} & {(0.0185)} \\
    Politician (>=4) & {0.3661***} & {0.3956***} & {0.4322***} & {0.4526***} \\
          & {(0.1099)} & {(0.1163)} & {(0.1095)} & {(0.1088)} \\ \\
    \midrule
    \#Obs & {263,307} & {356,260} & {416,694} & {436,980} \\
    $R^2$  & {0.1541} & {0.1297} & {0.1151} & {0.1066} \\
    Bandwidth & {5 KM} & {10 KM} & {20 KM} & {50 KM} \\
    Boundary FE & Yes & Yes & Yes & Yes \\
    \midrule
    Politician (=3) - Politician (=2) & 0.0263 & 0.0216 & 0.0251 & 0.0273 \\
    Politician (=4) - Politician (=2)  & 0.340  & 0.370  & 0.4086 & 0.4297 \\
    F-Stat & 5.80   & 5.89  & 7.98  & 8.85 \\
    Prob > F & 0.0030 & 0.0028 & 0.0003 & 0.0001 \\
    \bottomrule
    \bottomrule
    \end{tabular}%
    \begin{tablenotes}
\footnotesize	
\item This table presents estimates for specification (\ref{rd}) augmented for the number of politicians in split blocks using the natural logarithm of 0.001 plus firm entry as the dependent variable. We use different bandwidths of 5, 10, 20, and 50 km on either side of the boundary, separating a split block from an unsplit block in columns (1), (2), (3), and (4), respectively.  The unit of observation is a village that lies within the narrow bandwidth of the boundary separating a split and an unsplit block. A block is defined as a split block based on the haphazard overlap of block boundaries with electoral boundaries as per the 2008 delimitation. We further segregate split blocks based on the number of politicians governing the block. Politician (=2) is a binary variable taking a value of 1 for split blocks with exactly two politicians. Politician (=3) is a binary variable taking a value of 1 for split blocks with exactly politicians. Politician (>=4) is a binary variable taking a value of 1 for split blocks with four or more politicians. Firm Entry is defined as the total number of firms that have entered a village from 2008 to 2016.  All regressions include boundary fixed effects, a local linear specification estimated separately on each side of the boundary, and use a triangular kernel. Standard errors reported in parentheses are clustered at the block level. \sym{*} \(p<0.1\), \sym{**} \(p<0.05\), \sym{***} \(p<0.01\). \\
\end{tablenotes}
\end{threeparttable}
}
\end{table}%

\clearpage

\begin{table}[ht!]
\centering
\captionsetup{justification=centering}
   \caption{P-value Adjusted for Multiple Hypothesis Testing} 
   \label{tab_mht}
\resizebox{1.00\textwidth}{!}{%
\begin{threeparttable}
\begin{tabular}{lHcccccc}
\toprule \toprule
               \multirow{3}{*}{}             & & (1)            & (2)            & (3)            & (4)            & (5)  & (6)                                                                         \\ \cline{3-8}
                            & & \multicolumn{3}{c}{5km} & \multicolumn{3}{c}{10km} \\
                            & & \# Firms     & Nightlight     & Employment     & \# Firms     & Nightlight     & Employment  \\ \hline \\
Adjusted (List et al) &	p-value	&	0.000	&	0.000	&	0.000	&	0.000	&	0.000	&	0.000	\\
Adjusted (Bonferroni)	&	p-value	&	0.001	&	0.001	&	0.001 &	0.001	&	0.001	&	0.001		\\
Adjusted (Holm)	&	p-value	&	0.001	&	0.000	&	0.000 & 0.001	&	0.001	&	0.000		\\
  \bottomrule \bottomrule
\end{tabular}
\begin{tablenotes}
\scriptsize	
\item This table explores various adjustments due to the multiplicity of outcomes and reports the p-values for $\beta$ from the specification (\ref{rd}) using the natural logarithm of 0.001 plus firm entry, nightlight intensity, and employment as the dependent variables. We use different bandwidths of 5 and 10 kms. The columns of the table reports the various ways to adjust standard errors for the multiplicity of the null hypothesis and reports the p-values post the error correction. In First row, we present p-values based on adjustment discussed in \citet{list2019multiple} and \citet{list2023multiple}. The second row presents p-values adjusted based on  Bonferroni. The third row presents p-value adjusted based on Holm. We use code from \citet{list2023multiple} to implement the adjustment. 

\end{tablenotes}
\end{threeparttable}
}
\end{table}

\clearpage

\newpage
\subsection{Falsification and Placebo Tests}
\begin{figure}[H]
    \centering
    \caption{Falsification Test: Sample Construction}
    \begin{subfigure}{.45\textwidth}
    \centering
    \includegraphics[width = \textwidth]{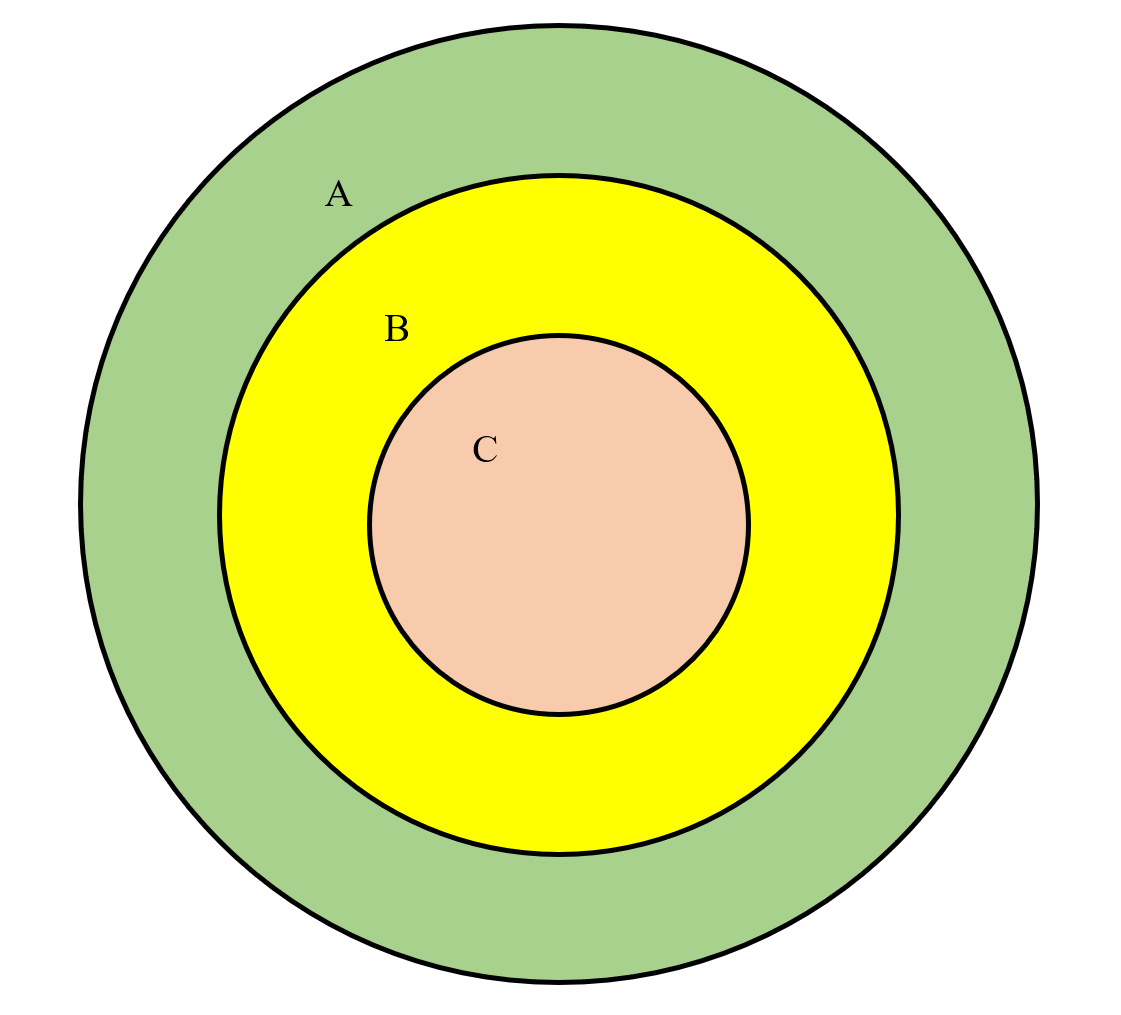}
    \caption{Schematic Diagram}
    \label{app_fig_schematic_1}
    \end{subfigure} %
    \begin{subfigure}{.45\textwidth}
    \centering
    \includegraphics[width = \textwidth]{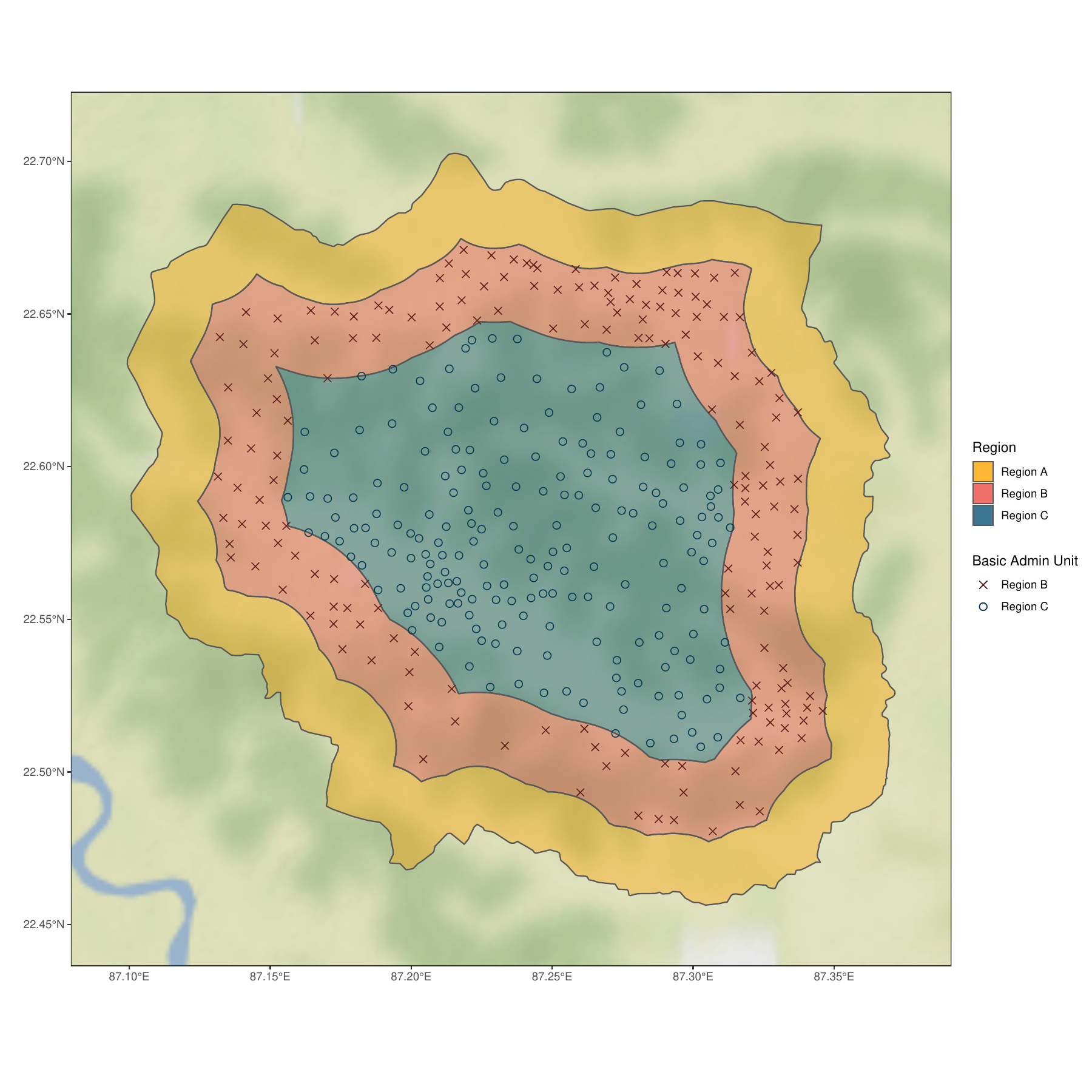}
    \caption{Representative Example}
    \label{app_fig_schematic_2}
    \end{subfigure} %
    \label{app_fig_schematic}
    \begin{minipage}{1.00\textwidth}
	\begin{center}
		\end{center}
		{\footnotesize 	Figure \ref{app_fig_schematic_1} presents the schematic diagram of a block that is split into three regions A, B and C. These regions are equal in area and account for one-third of the entire block area. Our falsification exercise compares villages in region B with villages in region C within a narrow bandwidth of the boundary separating the two regions. Figure \ref{app_fig_schematic_2} presents the schematic diagram using a representative example using the block of Salbani in the West Medinipur district in the state of West Bengal.  \par}
\end{minipage}
\end{figure}

We first describe the methodology to construct the falsification sample. We divide each block into three regions A, B and C, as shown in Figure \ref{app_fig_schematic}. These regions are equal in area and account for one-third of the entire block area. Our falsification exercise compares villages in region B with villages in region C within a narrow bandwidth of the boundary separating the two regions. The intuition of this test is that crossing from region B to region C does not change the number of politicians governing the area. Hence, if our baseline results capture the effect of the change in the number of politicians and not spatial correlation, we should observe a null effect when moving from region B to region C. 

\clearpage
\newpage
\begin{table}[htbp]
  \centering
  \caption{Falsification Test: Results}
  \label{tab_app_falsification}%
  \begin{threeparttable}
    \begin{tabular}{lcccc}
    \toprule
    \toprule 
    Dep Var:  LN(0.001+\# New Firms) & (1)   & (2)   & (3)   & (4) \\
    \midrule \\
    Inside Region C (=1) & -0.0007 & -0.0037 & -0.0058 & -0.0047 \\
          & (0.0088) & (0.0067) & (0.0058) & (0.0055) \\ \\
    \midrule
    \# Obs & 200,843 & 278,097 & 320,484 & 332,161 \\
    $R^2$  & 0.1587 & 0.1306 & 0.1190 & 0.1157 \\
    Bandwidth & 5 KM  & 10 KM & 20 KM & 50 KM \\
    Boundary FE & Yes & Yes & Yes & Yes  \\
    \bottomrule
    \bottomrule
    \end{tabular}%
    \begin{tablenotes}
\footnotesize	
\item This table presents estimates for specification (\ref{rd}) for the falsification sample using the natural logarithm of 0.001 plus firm entry as the dependent variable. We use different bandwidths of 5, 10, 20, and 50 km on either side of the boundary in columns (1), (2), (3), and (4), respectively.  The unit of observation is a village that lies within the narrow bandwidth of the boundary separating region B and region C within a block. A block is divided into three regions, A, B, and C, which are equal in area as shown in Figure \ref{app_fig_schematic}. The falsification test compares villages in region B and region C. Firm Entry is defined as the total number of firms that have entered a village from 2008 to 2016.  All regressions include boundary fixed effect, a local linear specification estimated separately on each side of the boundary, and use a triangular kernel. Standard errors reported in parentheses are clustered at the block level. \sym{*} \(p<0.1\), \sym{**} \(p<0.05\), \sym{***} \(p<0.01\). \\
\end{tablenotes}
\end{threeparttable}
\end{table}%

\clearpage
\newpage
\begin{figure}[H]
\centering
	\caption{Placebo Test}
    \centering
    \begin{subfigure}{.49\textwidth}
    \centering
    \includegraphics[width = \textwidth]{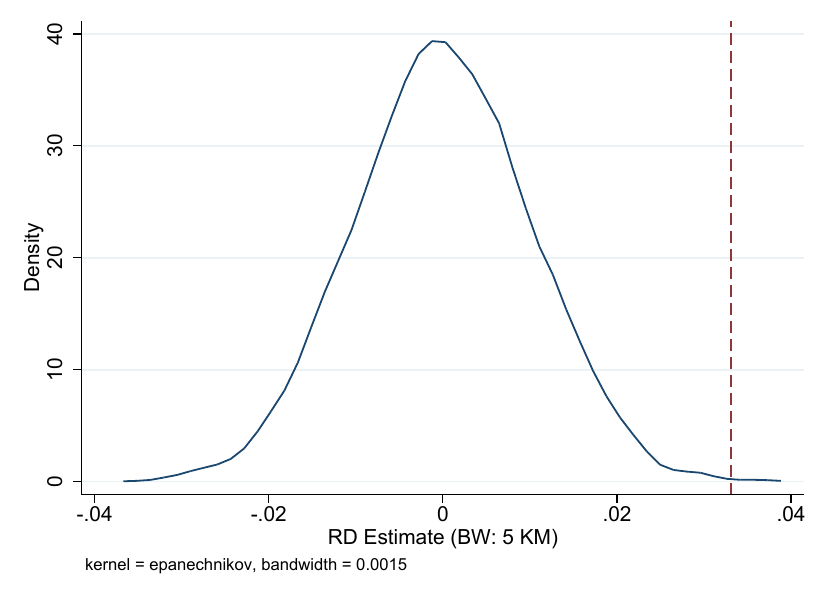}
    \caption{Bandwidth: 5 KM}
    \label{fig_placebo_1}
    \end{subfigure} %
    \begin{subfigure}{.49\textwidth}
    \centering
    \includegraphics[width = \textwidth]{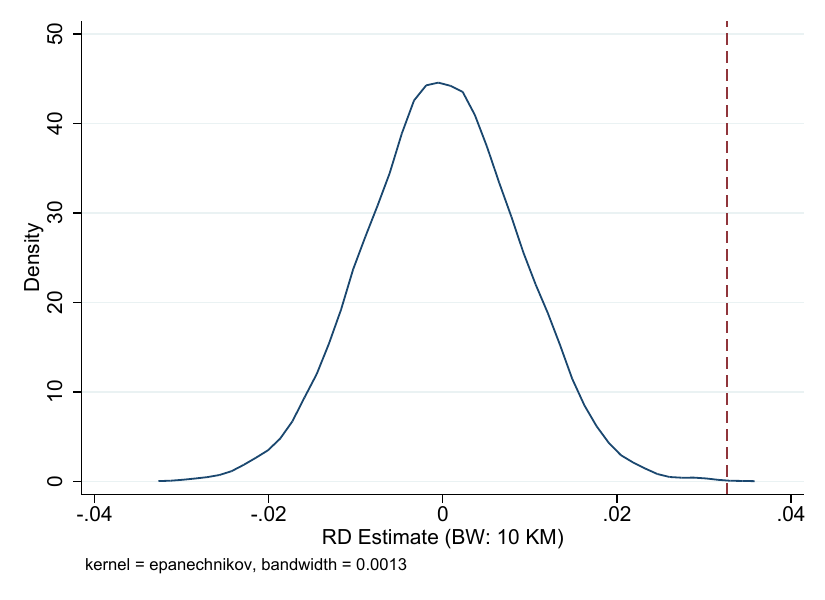}
    \caption{Bandwidth: 10 KM}
    \label{fig_placebo_2}
    \end{subfigure} %
    \begin{subfigure}{.49\textwidth}
    \centering
    \includegraphics[width = \textwidth]{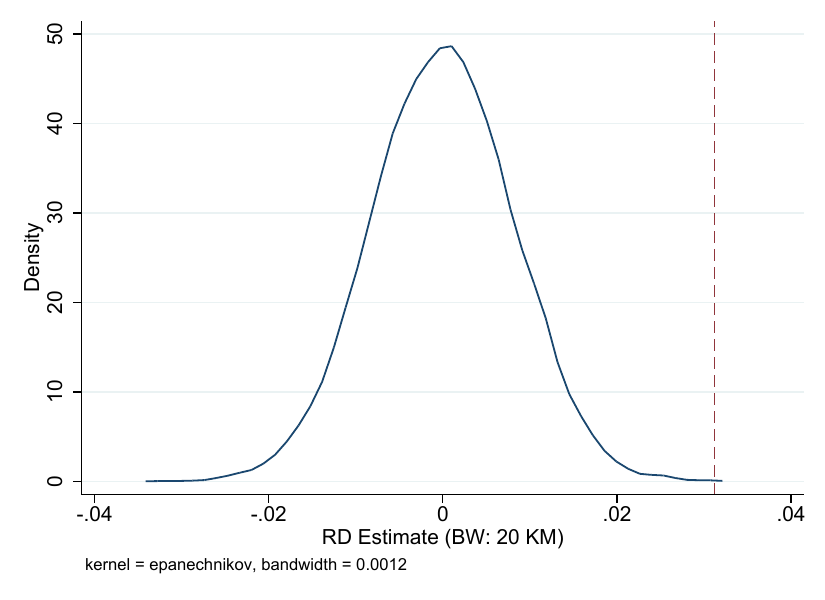}
    \caption{Bandwidth: 20 KM}
    \label{fig_placebo_3}
    \end{subfigure}
    \begin{subfigure}{.49\textwidth}
    \centering
    \includegraphics[width = \textwidth]{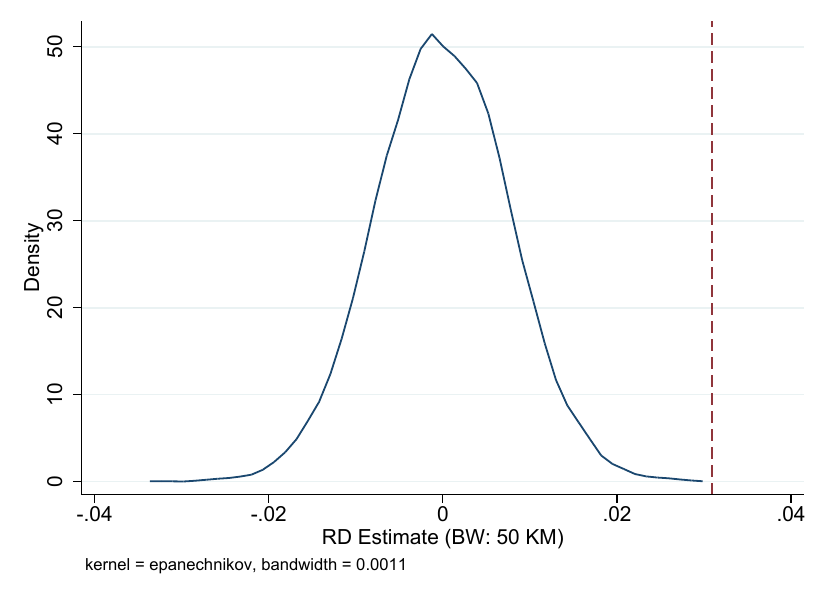}
    \caption{Bandwidth: 50 KM}
    \label{fig_placebo_4}
    \end{subfigure} 
    \label{fig_placebo}
    \begin{minipage}{\textwidth}
{\footnotesize The figure plots the kernel density of the RD estimates for specification (\ref{rd}) for the randomly assigned split status obtained from the 10,000 Monte-Carlo simulations. The dependent variable is the natural logarithm of 0.001 plus firm entry. We randomly assign each block to be a split block or an unsplit block. Hence, a placebo split binary variable is generated for each block. Every block has a probability of 0.411 to be assigned as a split block. This is based on the fraction of total split blocks in the sample. We repeat this process 10,000 times and estimate the RD specification (\ref{rd}) with the placebo split variable for different bandwidths of 5 (Figure \ref{fig_placebo_1}), 10 (Figure \ref{fig_placebo_2}), 20 (Figure \ref{fig_placebo_3}), and 50 km (Figure \ref{fig_placebo_4}) on either side of the boundary. The dashed red line shows the point estimate from the corresponding bandwidth as in Panel A of Table \ref{sptd_firm_entry}.
\par}
	\end{minipage}
\end{figure}

\clearpage
\newpage
\begin{table}[ht!]
  \centering
  \caption{2008 Delimitation and Elections After 2008: Firm Entry and Change in Number of Politicians}
  \label{delim_election}%
   \resizebox{0.9\textwidth}{!}{     
  \begin{threeparttable} 
    \begin{tabular}{lcccc}
    \toprule
    \toprule
    \multicolumn{5}{c}{\textit{Panel A : Unsplit to Split}} \\
    \midrule
    Dep Var:  LN(0.001+\# New Firms) & (1)   & (2)   & (3)   & (4) \\
    \midrule \\
    Treat x Post Delimitation & 0.0103* & 0.0093* & 0.0083* & 0.0079* \\
          & (0.0061) & (0.0051) & (0.0047) & (0.0044) \\ \\
    Treat x Post Election & 0.0140** & 0.0098* & 0.0083* & 0.0075 \\
          & (0.0062) & (0.0054) & (0.0050) & (0.0048) \\ \\
    \midrule
    \# Obs & 687,792 & 975,312 & 1,185,816 & 1,263,456 \\
    $R^2$  & 0.6489 & 0.6329 & 0.6204 & 0.6129 \\
    Bandwidth & 5 KM  & 10 KM & 20 KM & 50 KM \\
    Boundary $\times$ Year FE & Yes   & Yes   & Yes   & Yes \\
    Village FE & Yes   & Yes   & Yes   & Yes \\
    \midrule 
    \multicolumn{5}{c}{\textit{Panel B : Split to Unsplit}} \\
    \midrule
    Dep Var:  LN(0.001+\# New Firms) & (1)   & (2)   & (3)   & (4) \\
    \midrule \\
    Treat x Post Delimitation & -0.0092*** & -0.0103*** & -0.0108*** & -0.0111*** \\
          & (0.0017) & (0.0015) & (0.0015) & (0.0015) \\ \\
    Treat x Post Election & -0.0029 & -0.0023 & -0.0026* & -0.0024* \\
          & (0.0018) & (0.0016) & (0.0015) & (0.0014) \\ \\
    \midrule
    \# Obs & 2,469,408 & 3,297,648 & 3,812,364 & 3,978,144 \\
    $R^2$  & 0.6888 & 0.6861 & 0.6817 & 0.6769 \\
    Bandwidth & 5 KM  & 10 KM & 20 KM & 50 KM \\
    Boundary $\times$ Year FE & Yes   & Yes   & Yes   & Yes \\
    Village FE & Yes   & Yes   & Yes   & Yes \\
    \bottomrule
    \bottomrule
    \end{tabular}%
 \begin{tablenotes}
\footnotesize   
\item This table presents estimates for supplemented specification (\ref{did_1}) and (\ref{did_2}) using the natural logarithm of 0.001 plus firm entry as the dependent variable in Panels A and B, respectively. The table supplements these specifications to include the interaction term of Treat and Post Election. In Panel A the treated group is the set of blocks that switched from being unsplit to split following the 2008 delimitation, whereas the control group comprises of blocks that were always unsplit both before and after the delimitation and border the treated group. In Panel B, the treated group is the set of blocks that switched from being split to unsplit following the 2008 delimitation, whereas the control group comprises blocks that are always split among multiple politicians both before and after the 2008 delimitation and border the treated group. The variable $Post x Delimitation$ takes a value of 1 for all years since the announcement of delimitation in 2008. The $Post x Election$ variable assumes a value of 1 for all years succeeding the initial state legislative election held after the announcement of delimitation. We use different bandwidths of 5, 10, 20, and 50 km on either side of the boundary, separating the treatment and the control groups in columns (1), (2), (3), and (4), respectively.  The unit of observation is a village-year that lies within the narrow bandwidth of the boundary separating the treatment and the control group. Firm entry is defined as the total number of firms that have entered a village during the year. All regressions include village and boundary $\times$ year fixed effects, a local linear specification estimated separately on each side of the boundary, and use a triangular kernel. Standard errors reported in parentheses are clustered at the block level. \sym{*} \(p<0.1\), \sym{**} \(p<0.05\), \sym{***} \(p<0.01\). \\
\end{tablenotes}
\end{threeparttable}}    
\end{table}%

\clearpage
\newpage

\clearpage
\section{Mechanism}

\begin{table}[htbp]
\setlength{\tabcolsep}{0.6cm}

  \centering
  \caption{Multiple Politicians and Power Supply}
  \label{tab_app_efficiency_power}%
    \resizebox{0.95\textwidth}{!}{

  \begin{threeparttable}
    \begin{tabular}{lcccc}
    \toprule
    \toprule 
    Dep Var: High Power Supply & (1)   & (2)   & (3)   & (4) \\
    \midrule \\
    Split (=1) & 0.0083* & 0.0102** & 0.0117** & 0.0109** \\
          & (0.0048) & (0.0046) & (0.0046) & (0.0047) \\ \\
    \hline 
    \# Obs & 105,264 & 143,320 & 168,952 & 177,819 \\
    $R^2$  & 0.7026 & 0.6944 & 0.6854 & 0.6791 \\
    Boundary FE & Yes & Yes & Yes & Yes \\
    \bottomrule
    \bottomrule
    \end{tabular}%
        \begin{tablenotes}
\footnotesize	
\item This table presents estimates for specification (\ref{rd}) using high power supply as the dependent variable. We use different bandwidths of 5, 10, 20, and 50 km on either side of the boundary, separating a split block from an unsplit block in columns (1), (2), (3), and (4), respectively. The unit of observation is a village that lies within the narrow bandwidth of the boundary separating a split and an unsplit block. A block is defined as a split block based on the haphazard overlap of block boundaries with electoral boundaries as per the 2008 delimitation. A village is defined as a high power supply village if it receives commercial electricity supply for more than 70\% of time, both during winter and summer seasons, and is defined as low power supply village otherwise. The data on village level commercial power supply comes from the 2011 Indian Census.  All regressions include boundary fixed effect, a local linear specification estimated separately on each side of the boundary, and use a triangular kernel. Standard errors reported in parentheses are clustered at the block level. \sym{*} \(p<0.1\), \sym{**} \(p<0.05\), \sym{***} \(p<0.01\). \\
\end{tablenotes}
\end{threeparttable}}
\end{table}%

\begin{table}[ht!]
\setlength{\tabcolsep}{0.6cm}

  \centering
  \caption{{Multiple Politicians and Swachh Bharat Mission}}
   \label{tab_sbm}%
     \resizebox{0.95\textwidth}{!}{

   \begin{threeparttable}
    \begin{tabular}{lcccc}
    \toprule
    \toprule
    Dep Var:  LN(1 + Funding Received) & (1)   & (2)   & (3)   & (4) \\
    \midrule\\
    Split (=1) & 0.0931* & 0.1066** & 0.1165** & 0.1163** \\
          & (0.0514) & (0.0485) & (0.0480) & (0.0486) \\ \\
    \midrule 
    \# Obs & 220,665 & 298,546 & 349,219 & 366,325 \\
    $R^2$  & 0.2959 & 0.2912 & 0.2887 & 0.2884 \\
    Bandwidth & 5 KM  & 10 KM & 20 KM & 50 KM \\
    Boundary FE & Yes & Yes & Yes & Yes \\
    \bottomrule
    \bottomrule
    \end{tabular}%
            \begin{tablenotes}
\footnotesize	
\item This table presents estimates for specification (\ref{rd}) using the natural logarithm of funding disbursed under the Swachh Bharat Mission (SBM) as the dependent variable. We use different bandwidths of 5, 10, 20, and 50 km on either side of the boundary, separating a split block from an unsplit block in columns (1), (2), (3), and (4), respectively. The unit of observation is a village that lies within the narrow bandwidth of the boundary separating a split and an unsplit block. A block is defined as a split block based on the haphazard overlap of block boundaries with electoral boundaries as per the 2008 delimitation. The data on village level funding disbursed under SBM is the sum of all funding received from 2015 until 2018.  All regressions include boundary fixed effect, a local linear specification estimated separately on each side of the boundary, and use a triangular kernel. Standard errors reported in parentheses are clustered at the block level. \sym{*} \(p<0.1\), \sym{**} \(p<0.05\), \sym{***} \(p<0.01\). \\
\end{tablenotes}
\end{threeparttable}}
\end{table}%

\clearpage
\newpage
\begin{table}[htbp]
\centering
\caption{Confidence in Local Politician and State Machinery}
\label{state_mach}
\resizebox{0.84\textwidth}{!}{
\begin{threeparttable}
\begin{tabular}{lccGcc}
\toprule
    \toprule
\multirow{3}{*}{\begin{tabular}[c]{@{}l@{}}Dep Var: Confidence Level\end{tabular}} & (1)                                                                         & (2)       & (3)        & (3)        & (4)                                                                         \\ \cline{2-6}
 & \multirow{2}{*}{\begin{tabular}[c]{@{}c@{}}Local\\ Politician\end{tabular}} & \multicolumn{3}{c}{State Machinery} & \multirow{2}{*}{\begin{tabular}[c]{@{}c@{}}State\\  Government\end{tabular}} \\ \cline{3-5}
                                                                                     &                                                                             & Police    & Courts     & Panchayats &                                                                             \\ \midrule \\
Fraction of Split Blocks                                                             & 0.2841**                                                                    & 0.1659**  & 0.2850***  & 0.2112***  & -0.0109                                                                     \\
                                                                                     & (0.1389)                                                                    & (0.0786)  & (0.0952)   & (0.0812)   & (0.1424)                                                                    \\ \\ \midrule
Sample Mean                                                                          & 1.4257                                                                      & 1.9617    & 2.4290     & 2.1107     & 2.0751                                                                      \\
Standard Deviation                                                                   & 0.5249                                                                      & 0.3397    & 0.2895     & 0.3075     & 0.4709                                                                      \\ \midrule
\# Obs                                                                               & 686                                                                         & 686       & 686        & 686        & 686                                                                         \\
$R^{2}$                                                                              & 0.6277                                                                      & 0.7125    & 0.5558     & 0.5592     & 0.5192                                                                      \\

District FE    & Yes    & Yes   & Yes   & Yes       & Yes  \\
Year FE    & Yes    & Yes   & Yes   & Yes       & Yes  \\ 
\bottomrule
\bottomrule
\end{tabular}
            \begin{tablenotes}
\footnotesize	
\item This table presents estimates from the regression of confidence in local politicians, state machinery, and the state government on the fraction of split blocks in the district. The unit of observation is district-year. The data on confidence comes from the two survey waves of the India Human Development Survey (IHDS) conducted in 2001 and 2011. Each respondent is asked, ``How much confidence do you have in X?'', where ``X" denotes local politicians, police, panchayat, and the government at the state level.  A respondent can say - (1) No Confidence coded as 1, (2) Intermediate level of confidence coded as 2, or (3) High level of confidence coded as 3. We collapse the respondent level data at the village level using survey weights to create a new variable that measures the district-level average confidence in ``X." For each district, we create a variable - fraction of split blocks. It is calculated as the ratio of the number of split blocks to the total number of blocks in the district. We define a block as split in 2001 and 2011 based on 1977 and 2008 delimitation. Columns (1), (2), (3), and (4) use confidence in the local politician, police, panchayat, and the state government as the dependent variable. We also present the average and the standard deviation of the dependent variables. Robust standard errors are reported in parentheses. \sym{*} \(p<0.1\), \sym{**} \(p<0.05\), \sym{***} \(p<0.01\). \\
\end{tablenotes}
\end{threeparttable}
}
\end{table}

\clearpage
\begin{table}[htbp]
\centering
\caption{Difference in Confidence among Government and Private Schools and Hospitals}
\label{conf_school}%
\begin{threeparttable}
\begin{tabular}{lcc}
\toprule
    \toprule
\multirow{2}{*}{\begin{tabular}[c]{@{}l@{}}Dep Var: Confidence Level\end{tabular}} & (1)       & (2)       \\ \cline{2-3}
                                                                                     & Schools    & Hospitals \\ \midrule \\  \\
Govt $\times$ Fraction of Split Blocks                                                & 0.2061**  & 0.2089**  \\
                                                                                     & (0.0956)  & (0.1022)  \\
Govt                                                                                  & -0.2456** & -0.3635** \\
                                                                                     & (0.1178)  & (0.1456)  \\ \\ \midrule 
Sample Mean                                                                          & 2.5856    & 2.5535    \\
Standard Deviation                                                                   & 0.2646    & 0.2937    \\ \midrule
District FE                                                                         & Yes     & Yes       \\ \midrule
\# Obs                                                                               & 688       & 688       \\
$R^{2}$                                                                              & 0.6986    & 0.6950    \\ \bottomrule
\bottomrule
\end{tabular}
            \begin{tablenotes}
\footnotesize	
\item This table presents estimates from the regression of confidence in schools and hospitals on the interaction term of schools and hospitals provided by the government and the fraction of split blocks in the district. The unit of observation is district-provider; that is, for each district, we have two observations, one indicating the confidence in the government good and another in the private sector good. The data on confidence comes from the 2011 survey wave of the India Human Development Survey (IHDS). Each respondent is asked - ``How much confidence do you have in X?'', where ``X" denotes government and private schools and hospitals.  A respondent can say - (1) No Confidence coded as 1, (2) Intermediate level of confidence coded as 2, or (3) High level of confidence coded as 3. We collapse the respondent level data at the village-provider level using survey weights to create a new variable that measures the district-provider-level average confidence in ``X." Provider is either the private sector or the government. For each district, we create a variable - fraction of split blocks. It is calculated as the ratio of the number of split blocks to the total number of blocks in the district. We define a block as a split block based on the 2008 delimitation. Columns (1) and (2) use confidence in schools and hospitals as the dependent variable. We also present the average and the standard deviation of the dependent variables. Robust standard errors are reported in parentheses. \sym{*} \(p<0.1\), \sym{**} \(p<0.05\), \sym{***} \(p<0.01\). \\
\end{tablenotes}
\end{threeparttable}
\end{table}%

\clearpage
\begin{table}[htbp]
\centering
\caption{Likelihood of Incumbent's Re-election}
\label{prob_reelect}
\resizebox{0.84\textwidth}{!}{
\begin{threeparttable}
\begin{tabular}{lccc}
\toprule
    \toprule
Dep Var: Incumbent Re-elected & (1)       & (2)       & (3)       \\ \hline \\ 
Fraction of Split Blocks      & 0.0545*** & 0.0518*** & 0.0517*** \\
                              & (0.0183)  & (0.0184)  & (0.0184)  \\ \\ \hline 
State $\times$ Year FE        & Yes       & Yes       & Yes       \\ 
Reservation FE                &           & Yes       & Yes       \\
By-Election FE               &           &           & Yes       \\ \hline 
\# Obs                        & 4,667     & 4,667     & 4,667     \\
$R^{2}$                       & 0.0798    & 0.0809    & 0.0812    \\ \bottomrule
\bottomrule
\end{tabular}
            \begin{tablenotes}
\footnotesize	
\item This table presents estimates from the regression of the likelihood of the incumbent's reelection and the fraction of split blocks in the assembly constituency (AC). The unit of observation is AC-year from 2008 until 2018. The data on election outcomes comes from the Election Commission of India. For each AC, we create a variable - fraction of split blocks. It is calculated as the ratio of the number of split blocks to the total number of blocks in the AC. We define a block as a split block based on the 2008 delimitation. Column (1) includes state $\times$ year fixed effect. Column (2) adds reservation fixed effect, which controls for whether the AC is reserved for lower castes. Column (3) adds a by-election fixed effect and controls for the election being outside the normal election cycle. Standard errors reported in parentheses are clustered at the AC level. \sym{*} \(p<0.1\), \sym{**} \(p<0.05\), \sym{***} \(p<0.01\). \\
\end{tablenotes}
\end{threeparttable}
}
\end{table}

\clearpage

\newpage

\begin{table}[htbp]
  \centering
  \caption{{Effect of Multiple Politicians on Block Development Officers}}
   \label{tab_survey}%
   \resizebox{0.84\textwidth}{!}{ 
   \begin{threeparttable}
    \begin{tabular}{lccccc}
    \toprule \toprule
             & (1)                                                                                              & (2)                                                                       & (3)          & (4)                                                           & (5)                                                          \\ \cline{2-6}
             & \multirow{2}{*}{\begin{tabular}[c]{@{}c@{}}Greater \\ Management \\ by Politicians\end{tabular}} & \multirow{2}{*}{\begin{tabular}[c]{@{}c@{}}Job\\ Difficulty\end{tabular}} & \multicolumn{3}{c}{Resources}       \\ \cline{4-6}
             &                                                                                                  &                                                                           & \# Computers & \begin{tabular}[c]{@{}c@{}}\# Permanent \\ Staff\end{tabular} & \begin{tabular}[c]{@{}c@{}}\# Temporary\\ Staff\end{tabular} \\ \hline 
Split(=1)    & 0.6574**                                                                        & 0.8014**                                                 & 0.0559       & 0.0047                                                        & -0.0495                                                      \\
             & (0.3144)                                                                        & (0.3333)                                                 & (0.0711)     & (0.0294)                                                      & (0.1562)                                                     \\  \hline 
Assembly FE  & Yes                                                                             & Yes                                                      & Yes          & Yes                                                           & Yes                                                          \\
Controls     & Yes                                                                             & Yes                                                      & Yes          & Yes                                                           & Yes       \\
Model        & Ordered Probit                                                                                   & Ordered Probit                                                            & Poisson      & Poisson                                                       & Poisson                                                      \\ \hline 
\# Obs       & 360                                                                             & 365                                                      & 354          & 363                                                           & 357                                                          \\
Psuedo $R^2$ & 0.5697                                                                          & 0.3982                                                   & 0.2582       & 0.6799                                                        & 0.6547      \\ \bottomrule
\bottomrule
\end{tabular}%
      \begin{tablenotes}
\footnotesize	
\item This table presents estimates the differences in the management and the resources for Block Development Officers (BDO's) across split and unsplit blocks. A block is defined as a split block based on the haphazard overlap of block boundaries with electoral boundaries as per the 2008 delimitation. Column (1) uses the response of the opinion of BDO's on the management by the politicians. The response is noted on a three point scale 1 to 3 where 1 indicates lower management and 3 indicates greater management. Column (2) uses the response of the opinion of BDO's on their job difficulty. The response is noted on a ten point scale 1 to 10 where 1 indicates low difficulty and 10 indicates high difficulty. Column (3)-(5) uses the BDO's response on the availability of resources. Columns (3)-(5) use the number of computers, number of permanent staff, and the number of temporary staff as the dependent variable. Columns (1) and (2) estimate the relationship using an ordered probit model. Columns (3)-(5) estimate the relationship using a Poisson model. All regressions include assembly constituency fixed effects and BDO specific controls. Controls include education fixed effects, gender fixed effects, age, and experience. Robust standard errors reported in parentheses. \sym{*} \(p<0.1\), \sym{**} \(p<0.05\), \sym{***} \(p<0.01\). \\
\end{tablenotes}
\end{threeparttable}
}
\end{table}%

\end{appendices}
\end{document}